\documentclass[a4paper, 11pt]{article}
\usepackage{amsmath, amssymb, color} 
\usepackage{braket}
\usepackage{slashbox}
\usepackage{wrapfig}
\def\hsymbl#1{\smash{\hbox{\huge$#1$}}}
\def\hsymbu#1{\smash{\lower1.7ex\hbox{\huge$#1$}}}
\newtheorem{Def}{Definition}[section]
\newtheorem{Prop}[Def]{Proposition}
\newtheorem{Lem}[Def]{Lemma}
\newtheorem{Cor}[Def]{Corollary}
\newtheorem{prop}{Proposition}
\newenvironment{Propbis}[1]
  {\renewcommand{\theprop}{\ref{#1} bis}
   \addtocounter{prop}{-1}
   \begin{prop}}
  {\end{prop}}
\DeclareMathOperator \bnol 
{\begin{picture}(2.5,9)(0,0) \put(-1,-0.5)
{$\stackrel{\scriptscriptstyle \star}{\scriptscriptstyle \star}$} \end{picture}}
\DeclareMathOperator \bnor 
{\begin{picture}(4,9)(0,0) \put(-.5,-0.5)
{$\stackrel{\scriptscriptstyle \star}{\scriptscriptstyle \star}$}\end{picture}}
\newcommand{\bs}{\begin{subequations}}
\newcommand{\es}{\end{subequations}}
\newcommand{\bbm}{\begin{bmatrix}}
\newcommand{\ebm}{\end{bmatrix}}
\newcommand{\mrm}{\mathrm}
\newcommand{\bsymb}{\boldsymbol}
\newcommand{\p}{\partial}
\newcommand{\diff}{\mathrm{d}}
\newcommand{\iu}{\mathrm{i}}
\newcommand{\e}{{\hspace{.1em}{\mathrm e}}}
\newcommand{\GF}{\mathrm{GF}}
\newcommand{\FP}{\mathrm{FP}}
\newcommand{\RS}{\mathrm{RS}}
\newcommand{\bpz}[1]{\mathrm{bpz} \hspace{-.1em} \left( #1 \right)}
\newcommand{\gp}[1]{\epsilon \hspace{-.1em} \left( #1 \right)}
\newcommand{\hc}[1]{\mathrm{hc} \hspace{-.1em} \left( #1 \right)}
\newcommand{\dB}{\delta_B}
\newcommand{\Cu}{{C_1}}
\newcommand{\Pcz}{{\Phi}^{\overline{c}\zeta}}
\newcommand{\Pc}{{\Phi}^{\overline{c}-}}
\newcommand{\Pz}{{\Phi}^{-\zeta}}
\newcommand{\Pn}{{\Phi}^{--}}
\newcommand{\Lcz}{{\Lambda}^{\overline{c} \zeta}}
\newcommand{\Lc}{{\Lambda}^{\overline{c}-}}
\newcommand{\Lz}{{\Lambda}^{-\zeta}}
\newcommand{\Ln}{{\Lambda}^{--}}
\newcommand{\tP}{\widetilde{\Phi}}

\newcommand{\tPc}{\widetilde{\Phi}^{\overline{c}-}}

\newcommand{\tPn}{\widetilde{\Phi}^{--}}
\newcommand{\hP}{\widehat{\Phi}}
\newcommand{\bz}{b_0}
\newcommand{\dz}{d_0}
\newcommand{\cz}{c_0}
\newcommand{\xz}{\xi_0}
\newcommand{\zz}{\zeta_0}
\newcommand{\ez}{\eta_0}
\newcommand{\Lxy}{L_{x,y}}
\newcommand{\Lbar}{\overline{L}_0}
\newcommand{\czbar}{\overline{c}_0}
\newcommand{\zzbar}{\overline{\zeta}_0}
\newcommand{\ezbar}{\overline{\eta}_0}
\newcommand{\Xbar}{\overline{X}_0}
\newcommand{\PXb}{P_{X \hspace{-.1em}b}}
\newcommand{\KXb}{K_{X \hspace{-.1em}b}}
\newcommand{\PXbstar}{P^\star_{X \hspace{-.1em}b}}
\newcommand{\Pxistar}{P^\star_\xi}
\newcommand{\tPb}{\widetilde{P}_b}
\newcommand{\tKXb}{\widetilde{K}_{X \hspace{-.1em}b}}
\newcommand{\tp}{\widetilde{p}}
\newcommand{\tm}{\widetilde{m}}
\newcommand{\tk}{\widetilde{k}}
\newcommand{\tB}{\widetilde{B}}
\newcommand{\tK}{\widetilde{K}}
\newcommand{\tM}{\widetilde{M}}
\newcommand{\calB}{\mathcal{B}}
\newcommand{\calC}{\mathcal{C}}
\newcommand{\calD}{\mathcal{D}}
\newcommand{\calE}{\mathcal{E}}
\newcommand{\calF}{\mathcal{F}}
\newcommand{\calL}{\mathcal{L}}
\newcommand{\calM}{\mathcal{M}}
\newcommand{\tcalM}{\widetilde{\mathcal{M}}}
\newcommand{\calN}{\mathcal{N}}
\newcommand{\calO}{\mathcal{O}}
\newcommand{\calX}{\mathcal{X}}
\newcommand{\calZ}{\mathcal{Z}}
\newcommand{\hz}{\hat{z}}
\newcommand{\hb}{\hat{b}}
\newcommand{\hbl}{\hat{b}^\lambda}
\newcommand{\hv}{\hat{v}}
\newcommand{\hvl}{\hat{v}^\lambda}
\newcommand{\hxi}{\hat{\xi}}
\newcommand{\hxil}{\hat{\xi}^\lambda}
\newcommand{\hd}{\hat{d}}
\newcommand{\hdl}{\hat{d}^\lambda}

\newcommand{\zbar}{\overline{z}}
\newcommand{\indenths}{\hspace{1.22em}\ }
\newcommand{\hs}{\hspace{.07em}}

\allowdisplaybreaks
\begin{document}
\begin{titlepage}
\rightline{}
\rightline{\tt arXiv:1201.1762}
\rightline{\tt UT-Komaba/12-2}
\begin{center}
\vskip 2.5cm
{\Large \bf {
Validity of Gauge-Fixing Conditions
}}\\
\vskip 0.5cm
{\Large \bf {and the Structure of Propagators in Open Superstring Field Theory}}\\
\vskip 3.0cm
{\large Shingo {\sc Torii}}
\vskip 1cm
{\it {Institute of Physics, University of Tokyo,}}\\
{\it {Komaba, Meguro-ku, Tokyo 153-8902, Japan}}\\[.5ex]
storii@hep1.c.u-tokyo.ac.jp
\vskip 3.5cm
{\bf Abstract}
\end{center}
\vskip 0.5cm
\baselineskip 16pt

We make a detailed analysis on validity of gauge-fixing conditions and the structure of propagators
in the Wess-Zumino-Witten-type open superstring field theory.
First, we generalize the gauge-fixing conditions considered in JHEP {\bf 03} (2012) 030 [arXiv:1201.1761] by the present author et al.,
and propose a large class of conditions characterized by zero modes of world-sheet oscillators.
Then we demonstrate its validity:
we prove that gauge degrees of freedom allow us to impose the conditions,
and that the conditions fix the gauges completely.
Moreover, we elucidate how the information about the gauge choices is reflected in the structure of propagators.
The results can be readily extended to the case 
in which gauge-fixing conditions involve linear combinations of the world-sheet oscillators, including nonzero modes.
We investigate also such extended gauges, which are the counterpart of linear $b$-gauges in bosonic string field theory, 
and obtain the corresponding propagators.
\end{titlepage}
\baselineskip 17pt
\tableofcontents

\newpage
\section{Introduction}
\setcounter{equation}{0}
\indenths
The open-closed duality is one of the most fundamental properties of string theory.
In the perturbative description, closed strings naturally appear in loop diagrams of open strings.
However, in the framework of open string \emph{field} theory, it is enigmatic 
how closed strings are encoded in the form of open string fields.
In fact, we have not yet known even the answer to the question whether or not closed strings can be 
described in terms of \emph{fields} of open strings.

One natural approach to gaining a key to this question will be to quantize the field theory.
From the viewpoint of the open-closed duality, the quantum effect of open strings inevitably involves closed strings.
Hence, the point of the matter is whether or not open string field theory is consistently quantized without additional dynamical degrees of freedom,
such as closed string fields.
A criterion for the judgment is provided by the Batalin-Vilkovisky (BV) formalism~\cite{BV}.\footnote{
See ref.~\cite{GPS} for a pedagogical review.}
According to this formalism, consistent path-integral quantization of a gauge system is possible only when a certain fundamental equation has a solution.
The equation, which is called the quantum master equation, is an extension of the Ward-Takahashi identity, and
the existence of its solution ensures gauge independence of the partition function.
Moreover, it is known that violation of the quantum master equation corresponds to a gauge anomaly~\cite{TNP}.
Therefore, we expect that the equation will play the role of the touchstone:
if open string field theory lacks some needed information about closed strings, then its gauge symmetry would be anomalous, and
the quantum master equation could not be satisfied.

In bosonic string field theory~\cite{Witten}, an attempt was made by Thorn~\cite{ThornSFT} to construct the solution to the quantum master equation,
but there was a difficulty of divergences from tadpole diagrams.\footnote{For detailed analysis of the tadpole diagrams, see ref.~\cite{EST}.}
Furthermore, aside from this problem, the theory has another intrinsic difficulty due to the existence of the tachyon:
quantization can be considered only in a formal manner.
By contrast, in superstring field theory, we expect that these difficulties will be absent.
Recently, analytic methods in classical bosonic string field theory, which developed mainly 
after Schnabl's analysis~\cite{Schnabl} of the tachyon vacuum, have been applied to superstring field theory.
Moreover, the work~\cite{KZ} by Kiermaier and Zwiebach opened up a vista for extending the methods at least to the one-loop level.
It seems that we are now at the stage for exploring earnestly the quantization of open superstring field theory.

The first covariant open superstring field theory was constructed by Witten~\cite{WittenSuper}.
It was a natural extension of the cubic bosonic string field theory~\cite{Witten}, with the action composed of the string fields in
the natural picture: a Neveu-Schwarz (NS) string field of picture number minus one and a Ramond string field of picture number minus a half.
However, it was pointed out by Wendt~\cite{Wendt} that the theory suffered divergences caused by the picture-changing operator~\cite{FMS1, FMS2}
inserted at the string midpoint.
To remedy this, various superstring field theories have been proposed. 
In one approach, the problem is circumvented within the framework of cubic superstring field theory, 
at the cost of the naturalness of the picture~\cite{AMZ, PTY}. 
This modified version of Witten's theory, as well as Witten's original one, is formulated in the small Hilbert space and 
still needs picture-changing operators.
By contrast, in another approach investigated in refs.~\cite{Berkovits-NS, Berkovits-R, Michishita},
two types of theory are constructed without using any picture-changing operators in virtue of 
the large Hilbert space formulation, although the resultant actions are non-polynomial.

In the sequence of papers~\cite{paper1, paper2}, 
the present author et al.\ have dealt with superstring field theory of the latter approach.\footnote{
See also refs.~\cite{Torii, Nathan, GS}.
While the authors were writing their papers, a work on the former approach appeared~\cite{KKK}.}
As a first step toward quantization, we have considered gauge fixing, concentrating on 
the NS-sector action~\cite{Berkovits-NS}, which is of the Wess-Zumino-Witten (WZW) type.
In particular, the first paper~\cite{paper1} is devoted to the analysis of the free theory.
There, we have gauge-fixed the free theory completely, and 
have found that it requires infinitely many ghost string fields $\Phi_{(-n,\hs m)}$ $(1\leq n,\ 0\leq m\leq n)$
and antighost string fields $\Phi_{(n+2,\hs -m)}$ $(1\leq m\leq n+1)$, as well as the original string field $\Phi_{(0,\hs 0)}$,
with a subscript $(g,p)$ on a string field indicating its world-sheet ghost number $g$ and picture number $p$.
Furthermore, we have also calculated propagators, including those in the ghost string field sector. 
However, validity of the gauges is not proved in the paper, nor is it transparent how the information about the gauge choices is 
incorporated into the propagators.
The main purpose of the present paper is to solve these problems in a manner such that the underlying structure is manifest.
We also aim at generalizing the results in ref.~\cite{paper1}.

Let us now give the outline of the present paper.
First, generalizing the gauge-fixing conditions considered in ref.~\cite{paper1},
we propose a larger class of conditions of the form
\begin{equation}
\bsymb{B_{n+2, n+1} \Phi_{-n} } =0\,,\quad 
\bsymb{\tB_{n+2, n+1} \Phi_{n+2}} =0\qquad \left( \forall n\geq 0\right).
\label{intro gfc}
\end{equation}
Here 
\begin{equation}
\bsymb{\Phi_{-n}} := 
\begin{bmatrix}
\Phi_{( -n ,\hs 0 )} \\[1ex]
\Phi_{( -n ,\hs 1 )} \\[.5ex]
\vdots \\
\Phi_{( -n ,\hs n )} 
\end{bmatrix}
,
\quad 
\bsymb{\Phi_{n+2}} := 
\begin{bmatrix}
\Phi_{( n+2 ,\hs -1 )} \\[1ex]
\Phi_{( n+2 ,\hs -2 )} \\[.5ex]
\vdots \\
\Phi_{( n+2 ,\hs -(n+1) )} 
\end{bmatrix}
\qquad \left( n\geq 0\right)
\end{equation}
are $(n+1)$-component vectors composed of string fields, whereas $\bsymb{B_{n+2, n+1}}$ and $\bsymb{\tB_{n+2, n+1}}$ are $(n+2)\times (n+1)$ matrices
characterized by zero modes of world-sheet oscillators.\footnote{For simplicity, we have omitted here some indices on the gauge-fixing matrices 
$\bsymb{B}$ and $\bsymb{\tB}$ in eq.~\eqref{gfc}.}

Second, we demonstrate the validity of the above gauge-fixing conditions.
The proof involves two steps: one has to show reachability and completeness of the conditions.
\begin{enumerate}
\item {\bf (Reachability)}\;
By using the gauge degrees of freedom, one can let string fields reach the configuration where the conditions hold.
\item {\bf (Completeness)}\;
If once the conditions are imposed, there remains no residual gauge symmetry that preserves the conditions. 
In other words, the conditions eliminate the gauge degrees of freedom completely.
\end{enumerate}

Third, with the aid of the byproducts of the proof, 
we elucidate how the gauge-fixing conditions are reflected in the structure of the propagators.
The result is similar to that obtained in refs.~\cite{b-gauges, gl-gauges} in the case of the bosonic string field theory~\cite{Witten}.
To be specific, suppose that we impose linear $b$-gauge conditions~\cite{b-gauges} on bosonic string fields:
\begin{equation}
\calB_{(g)} \Psi_{g} = 0 \quad \left( \forall g\in \mathbb{Z}\right).
\label{intro b}
\end{equation}
Here $\Psi_{g}$ is a bosonic string field of world-sheet ghost number $g$, and $\calB_{(g)}$ is a linear combination of 
$b$-ghost oscillators.\footnote{Strictly, in order to provide physically reasonable gauges, $\calB$'s have to satisfy some constraints.}
The subscript $(g)$ indicates that $\calB_{(g)}$ acts upon $\Psi_g$.
Under the conditions \eqref{intro b}, the ghost propagator $\Delta_{g}$ between $\Psi_g$ and $\Psi_{2-g}$ $(g\leq 0)$
is given by
\begin{equation}
\Delta_g = \frac{\calB_{(g)}}{\calL_{(g)}}\, Q\, \frac{\calB_{(g+1)}}{\calL_{(g+1)}} \quad
\text{with}\quad
\calL_{(g)} := \{ Q, \calB_{(g)}\}\,,
\label{intro prop}
\end{equation}
where $Q$ denotes the Becchi-Rouet-Stora-Tyutin (BRST) operator in the first-quantized theory.
We obtain a similar expression in the supersymmetric case. In the WZW-type superstring field theory,   
the ghost propagator $\bsymb{\Delta_{n+1,n}}$ between $\bsymb{\Phi_{-n}}$ and $\bsymb{\Phi_{n+1}}$ $(n\geq 1)$ essentially 
takes the form
\begin{equation}
\bsymb{\Delta_{n+1,n}} =
\bigl( \bsymb{P_{n+1,n+1}\hs B_{n+1,n}}\bigr)\hs \bsymb{Q_{n,n+1}}\hs \bigl(\bsymb{P'_{n+1,\hs n+1}\hs B'_{n+1,n}}\bigr),
\end{equation}
where all the symbols $\bsymb{P}$, $\bsymb{B}$, $\bsymb{Q}$, $\bsymb{P'}$, and $\bsymb{B'}$ represent matrices, whose sizes are indicated by
their subscripts (and thus $\bsymb{\Delta_{n+1,n}}$ is an $(n+1)\times n$ matrix).
In particular, $\bsymb{Q}$ is an extension of the BRST operator $Q$.
The matrices $\bsymb{P}$ and $\bsymb{P'}$ are the counterparts of $1/{\calL_{(g)}}$ and $1/{\calL_{(g+1)}}$ in eq.~\eqref{intro prop}, 
and $\bsymb{B}$ and $\bsymb{B'}$ are precisely the gauge-fixing matrices. 

It should be noted that the conditions \eqref{intro gfc} are analogous to the Siegel gauge condition~\cite{Siegel} in that they consist of
zero modes of world-sheet oscillators.
In order to develop our analysis further, 
lastly, we also consider the extension similar to linear $b$-gauges~\cite{b-gauges}.
We investigate those conditions which involve linear combinations of the world-sheet oscillators, including nonzero modes,
and find the corresponding propagators. 

The present paper is organized as follows.
In the next section, we review gauge fixing of the free superstring field theory, and explain how one can
obtain BRST-invariant actions under general gauge-fixing conditions.
We shall see that gauge-fixing matrices have to satisfy certain constraints in order for the gauges to be compatible with the BRST transformations.
After that, in section~\ref{gauge-fixing conditions}, we propose a class of gauge-fixing conditions which meet the compatibility constraints,
and prove its validity in section~\ref{validity} (and in appendix~\ref{reachability}).
In the process of showing the completeness of the gauges, we discover the key ingredients of propagators.
They play an essential role when we derive the propagators in section~\ref{propagators}.
Extending these results further, in section~\ref{ext}, we are going to consider a simple counterpart of linear $b$-gauges.
Finally, section~\ref{summary} is allocated for a summary and a discussion. 
Several appendices are provided to supply details and to help one's understanding.

\section{The free action and its gauge fixing}
\setcounter{equation}{0}
\indenths
The free action in the WZW-type open superstring field theory~\cite{Berkovits-NS} takes the form\footnote{
In our convention, the operator $Q\ez$ is anti-Hermitian, 
so that a factor of the imaginary unit is needed in order for the action to be real (see appendix~\ref{hc and reality}).
}
\begin{equation}
S_0 = -\frac{\iu}{2} \braket{ \Phi_{(0,\hs 0)}, Q\eta_0\hs\Phi_{(0,\hs 0)}} \,,
\label{free action} 
\end{equation}
where $\Phi_{(0,\hs 0)}$ is a Grassmann-even NS-sector string field of even parity under the Gliozzi-Scherk-Olive (GSO) projection.
Here and in the sequel, unless otherwise stated, a subscript $(g,p)$ on a string field indicates its world-sheet ghost number $g$
and picture number $p$.
In particular, $\Phi_{(0,\hs 0)}$ carries no world-sheet ghost number and no picture number.
The symbol $Q$ denotes the BRST operator in the first-quantized theory, and $\eta_0$
denotes the zero mode of $\eta$, which appears in the bosonization~\cite{FMS1, FMS2} of the superconformal ghosts:
$\beta = \e^{-\phi}\partial \xi,\ \gamma = \eta \e^\phi$.
We adopt the convention in which the quantum number $(g,p)$ of $Q$, of $\eta$, and of $\xi$ are $(1,0)$, $(1,-1)$, and $(-1,1)$, 
respectively.\footnote{We list $(g,p)$'s of relevant operators in appendix~\ref{algebraic relations}.
The SL$(2,\mathbb{R})$-invariant vacuum $\ket{0}$ is of $(g,p) = (0,0)$.}
Both $Q$ and $\ez$ act as the derivation upon string fields, satisfying
\begin{equation}
Q^2 = \ez^2 = \{ Q,\ez\} = 0\,.
\end{equation}

In eq.~\eqref{free action}, we have denoted by $\braket{\ \ ,\ \ }$ the Belavin-Polyakov-Zamolodchikov (BPZ) inner product~\cite{BPZ}.
Because the theory is formulated in the large Hilbert space, we need the insertion 
of $\xi$, of $c\p c\p^2 c$, and of $\e^{-2\phi}$ to obtain nonzero correlators. 
Owing to this anomaly, inner products of the form $\braket{\Phi_{(g_1,\hs p_1)}\hs,\hs\Phi_{(g_2,\hs p_2)}}$
vanish unless $(g_1 + g_2 ,\hs p_1 + p_2)=(2,-1)$.

The above action is invariant under the gauge transformation parameterized by the Grassmann-odd 
string fields $\Lambda_{(-1,\hs 0)}$ and $\Lambda_{(-1,\hs 1)}$:
\begin{equation}
\delta \Phi_{(0,\hs 0)} = Q\Lambda_{(-1,\hs 0)} +\eta_0 \Lambda_{(-1,\hs 1)}\,.
\label{gdof}
\end{equation}
In ref.~\cite{paper1}, we have fixed the gauge completely, and have shown that it requires infinitely many ghost string fields and 
antighost string fields of various $(g,p)$'s.
In the present section, we are going to review the result by the use of the BRST formalism,
in a manner which helps us to deal with a larger class of gauge-fixing conditions.

Throughout the present paper, we use the following notation.
Let $U$ and $V$ be vectors composed of string fields:
\begin{equation*}
U = 
\begin{bmatrix}
u_1 \\
\vdots \\
u_m \\
\end{bmatrix}
,\quad
V = 
\begin{bmatrix}
v_1 \\
\vdots \\
v_m \\
\end{bmatrix}
\quad 
\text{($u_i$ and $v_i$ $(1\leq i\leq m)$ are string fields).}
\end{equation*}
We define the inner product $\braket{U, V}$ as 
\begin{equation}
\braket{U, V} := \sum_{i=1}^m \braket{u_i, v_i} .
\end{equation}
Furthermore, we define BPZ conjugation of a matrix whose components are operators: the BPZ conjugate
$\bpz{{\bsymb M}} = \Bigl( \bigl( \bpz{{\bsymb M}} \bigr)_{ij} \Bigr) $ of a matrix 
${\bsymb M} = \left( M_{ij} \right)$ is defined as
\begin{equation}
\bigl( \bpz{{\bsymb M}} \bigr)_{ij}
:= \bpz{M_{ji}}. 
\label{bpzM}
\end{equation}
The $(i,\hs j)$-component of $\bpz{{\bsymb M}}$ is the BPZ conjugate of the $(j,\hs i)$-component of ${\bsymb M}$.
In other words, BPZ conjugation of a matrix is a combination of the conjugation of its components and transposition of the matrix.
For example, we have\footnote{For the BPZ conjugation of an operator, see eqs.~\eqref{bpz conj} and \eqref{bpzO}.
The operators $Q$ and $\ez$ are BPZ odd, whereas $\bz$ and $\xz$ are BPZ even.}
\begin{equation}
\bpz{
\bbm
Q & \ez
\ebm
}
=
-
\bbm
Q \\
\ez
\ebm
,\quad
\bpz{
\bbm
\bz \\
\xz
\ebm
}
=
\bbm
\bz & \xz
\ebm
.
\end{equation}
The definition \eqref{bpzM} is quite natural. To see this, let $V$, $W$, $\bsymb{ M_{l,m}}$, and $\bsymb{ N_{m,n}}$ 
be an $m$-vector, an $n$-vector, an $l\times m$ matrix, and an $m\times n$ matrix, respectively.
If all the components of $V$, $\bsymb{ M_{\hspace{.1em}l,m}}$, and $\bsymb{ N_{m,n}}$ have the same Grassmann parities 
$\gp{V}$, $\gp{{\bsymb M}}$, and $\gp{{\bsymb N}}$, respectively, then the following equations hold:
\bs \label{extension}
\begin{align}
\langle V,\hs \bsymb{ N_{m,n}} W \rangle &= (-1)^{\epsilon(V) \,\epsilon(\bsymb{N})} \langle \bpz{\bsymb{N_{m,n}}} V,\hs W \rangle \,,\\[1ex]
\bpz{\bsymb{ M_{l,m}}\hs \bsymb{ N_{m,n}}} &= (-1)^{\epsilon(\bsymb{M})\,\epsilon(\bsymb{N})} \bpz{\bsymb{N_{m,n}}} \bpz{\bsymb{M_{l,m}}}.
\end{align}
\es
These are the natural extensions of the properties of the ordinary BPZ inner product and BPZ conjugation, for which $l=m=n=1$.
Note that the sizes of the matrices $\bpz{\bsymb{ M_{l,m}}}$, $\bpz{\bsymb{ N_{m,n}}}$, and $\bpz{\bsymb{ M_{l,m}}\hs \bsymb{N_{m,n}}}$
are $m\times l$, $n\times m$, and $n\times l$, respectively.

\subsection{Gauge fixing with the BRST formalism}
\label{gf by BRST}
\indenths
Let us perform gauge fixing by the use of the BRST formalism.
All the string fields to appear in what follows have even parity under the GSO projection.
Because the Grassmann parity of a GSO-even basis state of the NS sector is congruent to its world-sheet ghost number, modulo two,
we have
\begin{equation}
\gp{\Phi_{(g,\hs p)}}\, \equiv\ g + \text{``Grassmann parity of the component fields of $\Phi_{(g,\hs p)}$''} \pmod{2}.
\label{parity}
\end{equation}
Here $\gp{\Phi_{(g,\hs p)}}$ denotes the Grassmann parity of a GSO-even string field $\Phi_{(g,\hs p)}$.
From eq.~\eqref{parity} it follows that the component fields of $\Phi_{(0,\hs 0)}$ are Grassmann even, as is expected.

Validity of the gauge-fixing conditions used below will be shown in section~\ref{validity} (and in appendix~\ref{reachability}).
\\[.5ex]
{\bf (I) Step 1: Gauge fixing of $\bsymb{S_0}$}\\*
\indenths
To eliminate the gauge symmetry \eqref{gdof} of the action \eqref{free action}, we impose on $\Phi_{(0,\hs 0)}$ the condition
\begin{equation}
\begin{bmatrix}
b_0 \\
\xi_0
\end{bmatrix}
\Phi_{(0,\hs 0)}
= 0\,.
\label{bxi p0}
\end{equation}
This can be achieved by introducing the gauge-fixing term
\begin{equation}
S^\GF_1 = 
- 
\left\langle
\begin{bmatrix}
N_{(3,\hs -1)} \\[.5ex]
N_{(3,\hs -2)}
\end{bmatrix}
,
\begin{bmatrix}
b_0 \\[.5ex]
\xi_0
\end{bmatrix}
\Phi_{(0,\hs 0)}
\right\rangle
= 
\left\langle
\bpz{
\begin{bmatrix}
b_0 \\[.5ex]
\xi_0
\end{bmatrix}
}
\begin{bmatrix}
N_{(3,\hs -1)} \\[.5ex]
N_{(3,\hs -2)}
\end{bmatrix}
,
\Phi_{(0,\hs 0)}
\right\rangle
,
\label{ns gf1}
\end{equation}
where $N_{(3,\hs -1)}$ and $N_{(3,\hs -2)}$ are auxiliary fields which obey a constraint to be mentioned.
Because the component fields of $\Phi_{(0,\hs 0)}$ are Grassmann even, those of $N_{(3,\hs -1)}$ and $N_{(3,\hs -2)}$, also, have to be
Grassmann even, in order that they may act as Lagrange multipliers for condition \eqref{bxi p0}.
Thus $N_{(3,\hs -1)}$ and $N_{(3,\hs -2)}$ are Grassmann odd (recall eq.~\eqref{parity}).

One obtains the (spacetime) BRST transformation of $\Phi_{(0,\hs 0)}$ 
merely by replacing the gauge parameters with corresponding ghost string fields:
\begin{equation}
\dB \Phi_{(0,\hs 0)} = 
\begin{bmatrix}
Q & \eta_0 \\
\end{bmatrix} 
\begin{bmatrix}
\Phi_{(-1,\hs 0)} \\[.5ex]
\Phi_{(-1,\hs 1)}
\end{bmatrix} .
\end{equation}
Because the parameters $\Lambda_{(-1,\hs 0)}$ and $\Lambda_{(-1,\hs 1)}$ are Grassmann odd,
the ghosts $\Phi_{(-1,\hs 0)}$ and $\Phi_{(-1,\hs 1)}$ are Grassmann even.
Nilpotency of the BRST transformation sets the following constraint on the ghosts:
\begin{equation}
0 = 
\delta_B \left( \delta_B \Phi_{(0,\hs 0)} \right) = 
- 
\begin{bmatrix}
Q & \eta_0 
\end{bmatrix} 
\left(
\delta_B
\begin{bmatrix}
\Phi_{( -1,\hs 0)} \\[.5ex]
\Phi_{( -1,\hs 1 )} 
\end{bmatrix} \right).
\label{const Phi-1}
\end{equation}
The BRST transformation of $\Phi_{(-1,\hs 0)}$ and $\Phi_{(-1,\hs 1)}$ will be determined later, consistently with this constraint.

The BRST-exact term below produces not only the gauge-fixing term $S^\GF_1$ but also the Faddeev-Popov (FP) term $S^\FP_1$:\footnote{In the present
paper, we use the convention in which FP terms have negative signs. This is different from the convention in ref.~\cite{paper1}, in which
FP terms have positive signs.}
\begin{equation}
S_1 
= 
\delta_B \braket{\Phi_{( 2,\hs -1 )}, \Phi_{( 0,\hs 0 )}}
= \underbrace{
\left\langle
\bpz{
\begin{bmatrix}
b_0 \\[.5ex]
\xi_0
\end{bmatrix}
}
\begin{bmatrix}
N_{(3,\hs -1 )} \\[.5ex]
N_{(3,\hs -2 )}
\end{bmatrix}
,
\Phi_{(0,\hs 0)}
\right\rangle}_{S^\GF_1}
\underbrace{
- \left\langle \Phi_{( 2,\hs -1)},
\begin{bmatrix}
Q & \eta_0 \\
\end{bmatrix} 
\begin{bmatrix}
\Phi_{( -1,\hs 0)} \\[.5ex]
\Phi_{( -1,\hs 1)}
\end{bmatrix}
\right\rangle}_{S^\FP_1} \,,
\end{equation}
where $\Phi_{( 2,\hs -1)}$ is an antighost string field, whose Grassmann parity is odd.
We have defined the BRST transformation of $\Phi_{( 2,\hs -1)}$, and of $N_{(3,\hs -1)}$ and $N_{(3,\hs -2)}$ as
\begin{equation}
\delta_B \Phi_{(2,\hs -1)}
=
\bpz{
\begin{bmatrix}
b_0 \\[.5ex]
\xi_0
\end{bmatrix}
}
\begin{bmatrix}
N_{(3,\hs -1)} \\[.5ex]
N_{(3,\hs -2)}
\end{bmatrix} 
=
\begin{bmatrix}
b_0 & \xi_0
\end{bmatrix}
\begin{bmatrix}
N_{(3,\hs -1)} \\[.5ex]
N_{(3,\hs -2)}
\end{bmatrix} ,
\quad
\delta_B
\begin{bmatrix}
N_{(3,\hs -1)} \\[.5ex]
N_{(3,\hs -2)}
\end{bmatrix}
= 0\,.
\label{BRST Phi2}
\end{equation}
The form of $S^\FP_1$ requires that we impose 
\begin{equation}
b_0 \xi_0 \Phi_{(2,\hs -1)} =0\,;
\label{bxi p2}
\end{equation}
otherwise, there remains redundant symmetry of the form\footnote{As we shall see in section~\ref{ex action},
this is part of the symmetry of the ``extended action''~\eqref{solution}.}
\begin{equation}
\Phi_{(2,\hs -1)} \ \longrightarrow \ 
\Phi_{(2,\hs -1)} + Q \eta_0\hs \Lambda_{(0,\hs 0)} \,.
\label{ns redundancy 1}
\end{equation}
Note that condition \eqref{bxi p2} is compatible with the BRST transformation \eqref{BRST Phi2}:
\begin{equation}
\bz\xz\, \dB \Phi_{(2,\hs -1)} = 0\quad
\left( \because\quad
\bz\xz\, \bpz{
\bbm
\bz \\
\xz
\ebm
}
= 0
\right).
\label{comp}
\end{equation}
Eq.~\eqref{comp} ensures that the BRST transform of $\Phi_{(2,\hs -1)}$ stays in the space where condition \eqref{bxi p2} holds.
Furthermore, because $N_{(3,\hs -1)}$ and $N_{(3,\hs -2)}$ appear in the form
\begin{equation*}
\bpz{
\begin{bmatrix}
b_0 \\[.5ex]
\xi_0
\end{bmatrix}
}
\begin{bmatrix}
N_{(3,\hs -1)} \\[.5ex]
N_{(3,\hs -2)}
\end{bmatrix} 
=
\begin{bmatrix}
b_0 & \xi_0
\end{bmatrix}
\begin{bmatrix}
N_{(3,\hs -1)} \\[.5ex]
N_{(3,\hs -2)}
\end{bmatrix} ,
\end{equation*}
we have to restrict also the space of $N_{(3,\hs -1)}$ and $N_{(3,\hs -2)}$ to remove redundancy of the form
\begin{equation}
\begin{bmatrix}
N_{(3,\hs -1)} \\[.5ex]
N_{(3,\hs -2)}
\end{bmatrix}
\ \longrightarrow \ 
\begin{bmatrix}
N_{(3,\hs -1)} \\[.5ex]
N_{(3,\hs -2)}
\end{bmatrix}
+
\bpz{
\begin{bmatrix}
b_0 & 0 \\
\xi_0 & b_0 \\
0 & \xi_0
\end{bmatrix}
}
\begin{bmatrix}
\ast \\
\ast \\
\ast 
\end{bmatrix}
=
\begin{bmatrix}
N_{(3,\hs -1)} \\[.5ex]
N_{(3,\hs -2)}
\end{bmatrix}
+
\begin{bmatrix}
b_0 & \xi_0 & 0 \\
0 & b_0 & \xi_0
\end{bmatrix}
\begin{bmatrix}
\ast \\
\ast \\
\ast 
\end{bmatrix}.
\end{equation} 
For example, we may adopt the following condition:
\begin{equation}
\begin{bmatrix}
c_0 & 0 \\
\eta_0 & c_0 \\
0 & \eta_0
\end{bmatrix}
\begin{bmatrix}
N_{(3,\hs -1)} \\[.5ex]
N_{(3,\hs -2)}
\end{bmatrix} 
= 0\,.
\label{ceta n3}
\end{equation}
Here and in the sequel, gauge-fixing conditions on the original field $\Phi_{(0,\hs 0)}$ and ghosts are realized indirectly through gauge-fixing terms, 
whereas constraints on antighosts and auxiliary fields are imposed directly.
In summary, we have finished eliminating the original gauge symmetry \eqref{gdof} by introducing the string fields listed below.
\begin{itemize}
\item The Grassmann-even ghost string fields $\Phi_{(-1,\hs 0)}$ and $\Phi_{(-1,\hs 1)}$ with
\begin{equation}
\delta_B \Phi_{(0,\hs 0)} = 
\begin{bmatrix}
Q & \eta_0 \\
\end{bmatrix} 
\begin{bmatrix}
\Phi_{(-1,\hs 0)} \\[.5ex]
\Phi_{(-1,\hs 1)}
\end{bmatrix} .
\end{equation}
\item The Grassmann-odd antighost string field $\Phi_{(2,\hs -1)}$ with
\begin{equation}
\bz\xz \Phi_{(2,\hs -1)} = 0\,,\quad
\delta_B \Phi_{(2,\hs -1)}
=
\bpz{
\begin{bmatrix}
b_0 \\[.5ex]
\xi_0
\end{bmatrix}
}
\begin{bmatrix}
N_{(3,\hs -1)} \\[.5ex]
N_{(3,\hs -2)}
\end{bmatrix} .
\end{equation}
\item The Grassmann-odd auxiliary string fields $N_{(3,\hs -1)}$ and $N_{(3,\hs -2)}$ with
\begin{equation}
\begin{bmatrix}
c_0 & 0 \\
\eta_0 & c_0 \\
0 & \eta_0
\end{bmatrix}
\begin{bmatrix}
N_{(3,\hs -1)} \\[.5ex]
N_{(3,\hs -2)}
\end{bmatrix} 
= 0\,,\quad
\delta_B
\begin{bmatrix}
N_{(3,\hs -1)} \\[.5ex]
N_{(3,\hs -2)}
\end{bmatrix}
= 0\,.
\end{equation}
\end{itemize}
The resultant action at this step is $S_0 + S_1$ with
\bs
\begin{align}
S_0 &= -\frac{\iu}{2} \braket{\Phi_{(0,\hs 0)}, Q \eta_0\hs \Phi_{(0,\hs 0)} },\\[1.8ex]
S_1 &= 
\underbrace{
- \left\langle \Phi_{( 2,\hs -1)},
\begin{bmatrix}
Q & \eta_0 \\
\end{bmatrix} 
\begin{bmatrix}
\Phi_{( -1,\hs 0)} \\[.5ex]
\Phi_{( -1,\hs 1)}
\end{bmatrix}
\right\rangle}_{S^\FP_1} 
+\underbrace{
\left\langle
\bpz{
\begin{bmatrix}
b_0 \\[.5ex]
\xi_0
\end{bmatrix}
}
\begin{bmatrix}
N_{(3,\hs -1 )} \\[.5ex]
N_{(3,\hs -2 )}
\end{bmatrix}
,
\Phi_{(0,\hs 0)}
\right\rangle}_{S^\GF_1}\,.
\end{align}
\es
\\[.5ex]
{\bf (II) Step 2: Gauge fixing of $\bsymb{S_0 + S_1}$}\\*
\indenths
The action $S_0 + S_1$ is invariant under the gauge transformation of the ghosts:
\begin{equation}
\begin{bmatrix}
\Phi_{( -1,\hs 0)} \\[.5ex]
\Phi_{( -1,\hs 1)} 
\end{bmatrix}
\ \longrightarrow \ 
\begin{bmatrix}
\Phi_{( -1,\hs 0)} \\[.5ex]
\Phi_{( -1,\hs 1)} 
\end{bmatrix}
+
\begin{bmatrix}
Q & \eta_0 & 0 \\[.5ex]
0 & Q      & \eta_0 
\end{bmatrix}
\begin{bmatrix}
\Lambda_{( -2,\hs 0 )} \\
\Lambda_{( -2,\hs 1 )} \\
\Lambda_{( -2,\hs 2 )} 
\end{bmatrix} .
\label{ghost gt}
\end{equation}
We can remove this symmetry by imposing the condition
\begin{equation}
\begin{bmatrix}
b_0 & 0 \\
\xi_0 & b_0 \\
0 & \xi_0
\end{bmatrix}
\begin{bmatrix}
\Phi_{( -1,\hs 0)} \\[.5ex]
\Phi_{( -1,\hs 1)}
\end{bmatrix}
=0\,.
\end{equation}
The BRST transformation of the ghosts, which originates from the symmetry \eqref{ghost gt}, is defined as
\begin{equation}
\delta_B
\begin{bmatrix}
\Phi_{\left( -1 ,\hs 0 \right)} \\[.5ex]
\Phi_{\left( -1 ,\hs 1 \right)} 
\end{bmatrix}
=
\begin{bmatrix}
Q & \eta_0 & 0 \\[.5ex]
0 & Q      & \eta_0 \\
\end{bmatrix}
\begin{bmatrix}
\Phi_{\left( -2 ,\hs 0 \right)} \\[.5ex]
\Phi_{\left( -2 ,\hs 1 \right)} \\[.5ex]
\Phi_{\left( -2 ,\hs 2 \right)} 
\end{bmatrix} ,
\end{equation}
where $\Phi_{(-2,\hs m)}$ $(0\leq m \leq 2)$ are Grassmann-even ghosts for ghosts.
We would like to remark that the above transformation is consistent with condition \eqref{const Phi-1}.
Nilpotency of the BRST transformation at this step requires that $\Phi_{(-2,\hs m)}$ be subject to the constraint
\begin{equation}
0 = 
\delta_B \left(
\delta_B
\begin{bmatrix}
\Phi_{\left( -1 ,\hs 0 \right)} \\[.5ex]
\Phi_{\left( -1 ,\hs 1 \right)} \\
\end{bmatrix}
\right)
= - 
\begin{bmatrix}
Q & \eta_0 & 0 \\[.5ex]
0 & Q      & \eta_0 
\end{bmatrix}
\left( \delta_B
\begin{bmatrix}
\Phi_{\left( -2 ,\hs 0 \right)} \\[.5ex]
\Phi_{\left( -2 ,\hs 1 \right)} \\[.5ex]
\Phi_{\left( -2 ,\hs 2 \right)} 
\end{bmatrix} 
\right) .
\end{equation}
The sum of the gauge-fixing term and the FP term is
\begin{align}
S_2
&= 
\delta_B \left\langle
\begin{bmatrix}
\Phi_{\left( 3 ,\hs -1 \right)} \\[.5ex]
\Phi_{\left( 3 ,\hs -2 \right)}
\end{bmatrix} , 
\begin{bmatrix}
\Phi_{\left( -1 ,\hs 0 \right)} \\[.5ex]
\Phi_{\left( -1 ,\hs 1 \right)} 
\end{bmatrix}
\right\rangle \nonumber \\*[2ex]
&=
\underbrace{
\left\langle
\bpz{
\begin{bmatrix}
b_0 & 0 \\[.5ex]
\xi_0 & b_0 \\[.5ex]
0 & \xi_0
\end{bmatrix}
}
\begin{bmatrix}
N_{\left( 4 ,\hs -1 \right)} \\[.5ex]
N_{\left( 4 ,\hs -2 \right)} \\[.5ex]
N_{\left( 4 ,\hs -3 \right)} 
\end{bmatrix} ,
\begin{bmatrix}
\Phi_{\left( -1 ,\hs 0 \right)} \\[.5ex]
\Phi_{\left( -1 ,\hs 1 \right)} 
\end{bmatrix}
\right\rangle}_{S_2^{\GF}}
\underbrace{
- \left\langle
\begin{bmatrix}
\Phi_{\left( 3 ,\hs -1 \right)} \\[.5ex]
\Phi_{\left( 3 ,\hs -2 \right)}
\end{bmatrix} , 
\begin{bmatrix}
Q & \eta_0 & 0 \\[.5ex]
0 & Q      & \eta_0 
\end{bmatrix}
\begin{bmatrix}
\Phi_{\left( -2 ,\hs 0 \right)} \\[.5ex]
\Phi_{\left( -2 ,\hs 1 \right)} \\[.5ex]
\Phi_{\left( -2 ,\hs 2 \right)} 
\end{bmatrix}
\right\rangle}_{S_2^{\FP}} \,.
\end{align}
Here $\Phi_{(3,\hs -m)}$ $(m=1, 2)$ and $N_{(4,\hs -m)}$ $(m=1, 2, 3)$,
both of which are Grassmann odd, are antighosts and auxiliary fields, respectively.
We have defined their BRST transformations as
\begin{equation}
\delta_B
\begin{bmatrix}
\Phi_{\left( 3 ,\hs -1 \right)} \\[.5ex]
\Phi_{\left( 3 ,\hs -2 \right)}
\end{bmatrix}
=
\bpz{
\begin{bmatrix}
b_0 & 0 \\[.5ex]
\xi_0 & b_0 \\[.5ex]
0 & \xi_0
\end{bmatrix}
}
\begin{bmatrix}
N_{\left( 4 ,\hs -1 \right)} \\[.5ex]
N_{\left( 4 ,\hs -2 \right)} \\[.5ex]
N_{\left( 4 ,\hs -3 \right)} 
\end{bmatrix}
,\quad
\delta_B
\begin{bmatrix}
N_{\left( 4 ,\hs -1 \right)} \\[.5ex]
N_{\left( 4 ,\hs -2 \right)} \\[.5ex]
N_{\left( 4 ,\hs -3 \right)} 
\end{bmatrix}
= 0\,.
\label{BRST Phi3}
\end{equation}
To get rid of the redundant symmetries\footnote{In section~\ref{ex action}, we shall see that the extended action \eqref{solution}
is invariant under the gauge transformation \eqref{rsp3}.}
\begin{align}
\begin{bmatrix}
\Phi_{\left( 3 ,\hs -1 \right)} \\[.5ex]
\Phi_{\left( 3 ,\hs -2 \right)}
\end{bmatrix}
&\ \longrightarrow \ 
\begin{bmatrix}
\Phi_{\left( 3 ,\hs -1 \right)} \\[.5ex]
\Phi_{\left( 3 ,\hs -2 \right)}
\end{bmatrix}
+
\begin{bmatrix}
Q \\[.5ex]
\eta_0
\end{bmatrix}
\Lambda_{\left( 2 ,\hs -1 \right)} \,,
\label{rsp3} \\[2ex]
\begin{bmatrix}
N_{\left( 4 ,\hs -1 \right)} \\[.5ex]
N_{\left( 4 ,\hs -2 \right)} \\[.5ex]
N_{\left( 4 ,\hs -3 \right)} 
\end{bmatrix}
&\ \longrightarrow \ 
\begin{bmatrix}
N_{\left( 4 ,\hs -1 \right)} \\[.5ex]
N_{\left( 4 ,\hs -2 \right)} \\[.5ex]
N_{\left( 4 ,\hs -3 \right)} 
\end{bmatrix}
+
\bpz{
\begin{bmatrix}
b_0 & 0 & 0 \\
\xi_0 & b_0 & 0 \\
0 & \xi_0 & b_0 \\
0 & 0 & \xi_0
\end{bmatrix}
}
\bbm
\ast \\
\ast \\
\ast \\
\ast
\ebm
,
\end{align}
we impose the following constraints directly:
\begin{equation}
\bbm
\bz &\xz
\ebm
\bbm
\Phi_{(3,\hs -1)} \\[.5ex]
\Phi_{(3,\hs -2)}
\ebm
= 0\,,\quad
\bbm
c_0 & 0 & 0 \\
\eta_0 & c_0 & 0 \\
0 & \eta_0 & c_0 \\
0 & 0 & \eta_0
\ebm
\begin{bmatrix}
N_{\left( 4 ,\hs -1 \right)} \\[.5ex]
N_{\left( 4 ,\hs -2 \right)} \\[.5ex]
N_{\left( 4 ,\hs -3 \right)} 
\end{bmatrix}
= 0\,.
\end{equation}
The constraint on $\Phi_{(3,\hs -m)}$ is compatible with the BRST transformation \eqref{BRST Phi3}:
\begin{equation}
\bbm
\bz &\xz
\ebm
\dB
\bbm
\Phi_{(3,\hs -1)} \\[.5ex]
\Phi_{(3,\hs -2)}
\ebm
= 0\quad
\left(
\because\quad
\bbm
\bz &\xz
\ebm
\bpz{
\begin{bmatrix}
b_0 & 0 \\[.5ex]
\xi_0 & b_0 \\[.5ex]
0 & \xi_0
\end{bmatrix}
}
=0
\right).
\end{equation}
To sum up, the gauge symmetry of $S_0 + S_1$ has been eliminated by introducing the following string fields.
\begin{itemize}
\item The Grassmann-even ghost string fields $\Phi_{(-2,\hs m)}$ $(0 \leq m \leq 2)$ with
\begin{equation}
\delta_B
\begin{bmatrix}
\Phi_{\left( -1 ,\hs 0 \right)} \\[.5ex]
\Phi_{\left( -1 ,\hs 1 \right)} 
\end{bmatrix}
=
\begin{bmatrix}
Q & \eta_0 & 0 \\[.5ex]
0 & Q      & \eta_0 
\end{bmatrix}
\begin{bmatrix}
\Phi_{\left( -2 ,\hs 0 \right)} \\[.5ex]
\Phi_{\left( -2 ,\hs 1 \right)} \\[.5ex]
\Phi_{\left( -2 ,\hs 2 \right)} 
\end{bmatrix} .
\end{equation}
\item The Grassmann-odd antighost string fields $\Phi_{(3,\hs -m)}$ $(m=1, 2)$ with
\begin{equation}
\bbm
\bz &\xz
\ebm
\bbm
\Phi_{(3,\hs -1)} \\[.5ex]
\Phi_{(3,\hs -2)}
\ebm
= 0\,,\quad
\delta_B
\begin{bmatrix}
\Phi_{\left( 3 ,\hs -1 \right)} \\[.5ex]
\Phi_{\left( 3 ,\hs -2 \right)}
\end{bmatrix}
=
\bpz{
\begin{bmatrix}
b_0 & 0 \\[.5ex]
\xi_0 & b_0 \\[.5ex]
0 & \xi_0
\end{bmatrix}
}
\begin{bmatrix}
N_{\left( 4 ,\hs -1 \right)} \\[.5ex]
N_{\left( 4 ,\hs -2 \right)} \\[.5ex]
N_{\left( 4 ,\hs -3 \right)} 
\end{bmatrix}
.
\end{equation}
\item The Grassmann-odd auxiliary string fields $N_{(4,\hs -m)}$ $(1\leq m \leq 3)$ with
\begin{align}
\bbm
c_0 & 0 & 0 \\
\eta_0 & c_0 & 0 \\
0 & \eta_0 & c_0 \\
0 & 0 & \eta_0
\ebm
\begin{bmatrix}
N_{\left( 4 ,\hs -1 \right)} \\[.5ex]
N_{\left( 4 ,\hs -2 \right)} \\[.5ex]
N_{\left( 4 ,\hs -3 \right)} 
\end{bmatrix}
= 0\,,\quad
\delta_B
\begin{bmatrix}
N_{\left( 4 ,\hs -1 \right)} \\[.5ex]
N_{\left( 4 ,\hs -2 \right)} \\[.5ex]
N_{\left( 4 ,\hs -3 \right)} 
\end{bmatrix}
= 0\,.
\end{align}
\end{itemize}
The resultant action is $S_0 + S_1 + S_2$ with
\bs
\begin{align}
S_0 &= -\frac{\iu}{2} \braket{\Phi_{(0,\hs 0)}, Q \eta_0\hs \Phi_{(0,\hs 0)} },\\[2.5ex]
S_1 &= 
\underbrace{
- \left\langle \Phi_{( 2,\hs -1)},
\begin{bmatrix}
Q & \eta_0 \\
\end{bmatrix} 
\begin{bmatrix}
\Phi_{( -1,\hs 0)} \\[.5ex]
\Phi_{( -1,\hs 1)}
\end{bmatrix}
\right\rangle}_{S^\FP_1} 
+\underbrace{
\left\langle
\bpz{
\begin{bmatrix}
b_0 \\[.5ex]
\xi_0
\end{bmatrix}
}
\begin{bmatrix}
N_{(3,\hs -1 )} \\[.5ex]
N_{(3,\hs -2 )}
\end{bmatrix}
,
\Phi_{(0,\hs 0)}
\right\rangle}_{S^\GF_1}\,,\\[1.5ex]
S_2 &=
\underbrace{
- \left\langle
\begin{bmatrix}
\Phi_{\left( 3 ,\hs -1 \right)} \\[.5ex]
\Phi_{\left( 3 ,\hs -2 \right)}
\end{bmatrix} , 
\begin{bmatrix}
Q & \eta_0 & 0 \\[.5ex]
0 & Q      & \eta_0 
\end{bmatrix}
\begin{bmatrix}
\Phi_{\left( -2 ,\hs 0 \right)} \\[.5ex]
\Phi_{\left( -2 ,\hs 1 \right)} \\[.5ex]
\Phi_{\left( -2 ,\hs 2 \right)} 
\end{bmatrix}
\right\rangle}_{S_2^{\FP}} 
+
\underbrace{
\left\langle
\bpz{
\begin{bmatrix}
b_0 & 0 \\[.5ex]
\xi_0 & b_0 \\[.5ex]
0 & \xi_0
\end{bmatrix}
}
\begin{bmatrix}
N_{\left( 4 ,\hs -1 \right)} \\[.5ex]
N_{\left( 4 ,\hs -2 \right)} \\[.5ex]
N_{\left( 4 ,\hs -3 \right)} 
\end{bmatrix} ,
\begin{bmatrix}
\Phi_{\left( -1 ,\hs 0 \right)} \\[.5ex]
\Phi_{\left( -1 ,\hs 1 \right)} 
\end{bmatrix}
\right\rangle}_{S_2^{\GF}}\,.
\end{align}
\es
\\[.5ex]
{\bf (III) The final result} \\*
\indenths
We can continue gauge fixing step by step in the above-mentioned manner.
After the $n$-th $(n\geq 1)$ step, we obtain the BRST-exact action
\begin{equation}
S_n = \dB \braket{\bsymb{\Phi_{n+1}}, \bsymb{\Phi_{-(n-1)}}}
=
\underbrace{ -\bigl\langle\bsymb{\Phi_{n+1}}, \bsymb{Q_{n,n+1} \Phi_{-n}}\bigr\rangle
}_{S^\FP_n}
+
\underbrace{ 
\bigl\langle\bpz{\bsymb{B_{n+1,n}}} \bsymb{N_{n+2}} \hs, \bsymb{\Phi_{-(n-1)}}\bigr\rangle
}_{S^\GF_n}
\quad  \left( n\geq 1\right),
\end{equation}
and the completely gauge-fixed action is given by
\begin{equation}
\sum^\infty_{n=0} S_n
= -\frac{\iu}{2} \braket{\bsymb{\Phi_0}, Q\ez \bsymb{\Phi_0}} 
-\sum^\infty_{n=1}\, \bigl\langle\bsymb{\Phi_{n+1}}, \bsymb{Q_{n,n+1} \Phi_{-n}}\bigr\rangle
+\sum^\infty_{n=1}\, \bigl\langle\bpz{\bsymb{B_{n+1,n}}} \bsymb{N_{n+2}} \hs, \bsymb{\Phi_{-(n-1)}}\bigr\rangle
\label{cgf action}
\end{equation}
with the constraints
\begin{align}
&\bz\xz \bsymb{\Phi_2} = 0\,,\quad \bsymb{\tB_{n-1,n}\hs \Phi_{n+1}} = 0\quad \left( n\geq 2\right),
\label{const ag} \\[1ex]
&\bsymb{C_{n+2,n+1}\hs N_{n+2}} = 0\quad \left( n\geq 1\right).
\label{restriction}
\end{align}
Here we have defined
\begin{align}
\bsymb{\Phi_{-n}} &:= 
\left.
\begin{bmatrix}
\Phi_{\left( -n ,\hs 0 \right)} \\
\vdots \\
\Phi_{\left( -n ,\hs n \right)} 
\end{bmatrix}
\right\} n+1
\quad \left( n\geq 0\right),\quad
\bsymb{\Phi_{n+1}} := 
\left.
\begin{bmatrix}
\Phi_{\left( n+1 ,\hs -1 \right)} \\
\vdots \\
\Phi_{\left( n+1 ,\hs -n \right)} 
\end{bmatrix}
\right\} n
\quad \left( n\geq 1\right) ,\\[3ex]
\bsymb{N_{n+2}} &:=
\left.
\bbm
N_{\left( n+2,\hs -1\right)} \\
\vdots \\
N_{\left( n+2,\hs -(n+1)\right)}
\ebm
\right\} n+1
\quad \left( n\geq 1\right),\\[1ex]
\bsymb{Q_{n, n+1}} &:= 
\underbrace{
\left.
\begin{bmatrix}
Q & \eta_0 & & \hsymbu{0} \\
  & \ddots & \ddots & \\
\hsymbl{0} & & Q & \eta_0 \\
\end{bmatrix}
\right\}
}_{n+1}
n
\quad \left( n\geq 1\right),\quad 
\bsymb{B_{n+1,n}} := 
\underbrace{
\left.
\begin{bmatrix}
b_0 & & \hsymbu{0} \\[-.5ex]
\xi_0 & \ddots & \\
 & \ddots & b_0 \\[.5ex]
\hsymbl{0} & & \xi_0 \\
\end{bmatrix}
\right\}
}_n
n+1
\quad \left( n\geq 1\right),
\label{QB} \\[1ex]
\bsymb{\tB_{n-1,n}} &:= 
\underbrace{
\left.
\begin{bmatrix}
b_0 & \xi_0 & & \hsymbu{0} \\
  & \ddots & \ddots & \\
\hsymbl{0} & & b_0 & \xi_0 \\
\end{bmatrix}
\right\}
}_n
n-1
\quad \left( n\geq 2\right),\quad
\bsymb{C_{n+2,n+1}} := 
\underbrace{
\left.
\begin{bmatrix}
c_0 & & \hsymbu{0} \\[-.5ex]
\eta_0 & \ddots & \\
 & \ddots & c_0 \\[.5ex]
\hsymbl{0} & & \eta_0 \\
\end{bmatrix}
\right\}
}_{n+1}
n+2
\quad \left( n\geq 1\right),
\label{tBC}
\end{align}
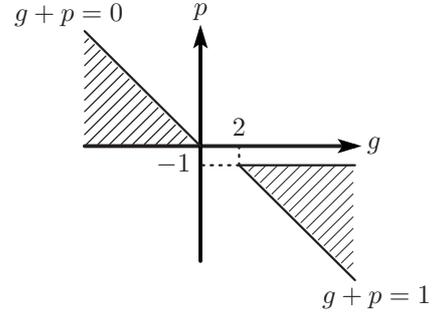
\begin{wrapfigure}[7]{r}[15pt]{8cm}
\unitlength 0.1in
\begin{picture}( 20.8000, 14.9500)( 13.2000,-27.2500)
%
{\color[named]{Black}{%
\special{pn 20}%
\special{pa 2000 1996}%
\special{pa 3400 1996}%
\special{fp}%
\special{sh 1}%
\special{pa 3400 1996}%
\special{pa 3334 1976}%
\special{pa 3348 1996}%
\special{pa 3334 2016}%
\special{pa 3400 1996}%
\special{fp}%
\special{pa 2600 2596}%
\special{pa 2600 1396}%
\special{fp}%
\special{sh 1}%
\special{pa 2600 1396}%
\special{pa 2580 1462}%
\special{pa 2600 1448}%
\special{pa 2620 1462}%
\special{pa 2600 1396}%
\special{fp}%
}}%
%
{\color[named]{Black}{%
\special{pn 13}%
\special{pa 2600 1996}%
\special{pa 2000 1396}%
\special{fp}%
}}%
%
{\color[named]{Black}{%
\special{pn 13}%
\special{pa 2800 2096}%
\special{pa 3400 2696}%
\special{fp}%
}}%
%
{\color[named]{Black}{%
\special{pn 8}%
\special{pa 2290 1706}%
\special{pa 2010 1986}%
\special{fp}%
\special{pa 2260 1676}%
\special{pa 2000 1936}%
\special{fp}%
\special{pa 2230 1646}%
\special{pa 2000 1876}%
\special{fp}%
\special{pa 2200 1616}%
\special{pa 2000 1816}%
\special{fp}%
\special{pa 2170 1586}%
\special{pa 2000 1756}%
\special{fp}%
\special{pa 2140 1556}%
\special{pa 2000 1696}%
\special{fp}%
\special{pa 2110 1526}%
\special{pa 2000 1636}%
\special{fp}%
\special{pa 2080 1496}%
\special{pa 2000 1576}%
\special{fp}%
\special{pa 2050 1466}%
\special{pa 2000 1516}%
\special{fp}%
\special{pa 2020 1436}%
\special{pa 2000 1456}%
\special{fp}%
\special{pa 2320 1736}%
\special{pa 2070 1986}%
\special{fp}%
\special{pa 2350 1766}%
\special{pa 2130 1986}%
\special{fp}%
\special{pa 2380 1796}%
\special{pa 2190 1986}%
\special{fp}%
\special{pa 2410 1826}%
\special{pa 2250 1986}%
\special{fp}%
\special{pa 2440 1856}%
\special{pa 2310 1986}%
\special{fp}%
\special{pa 2470 1886}%
\special{pa 2370 1986}%
\special{fp}%
\special{pa 2500 1916}%
\special{pa 2430 1986}%
\special{fp}%
\special{pa 2530 1946}%
\special{pa 2490 1986}%
\special{fp}%
}}%
%
{\color[named]{Black}{%
\special{pn 4}%
\special{pa 3390 2126}%
\special{pa 3120 2396}%
\special{fp}%
\special{pa 3390 2186}%
\special{pa 3150 2426}%
\special{fp}%
\special{pa 3390 2246}%
\special{pa 3180 2456}%
\special{fp}%
\special{pa 3390 2306}%
\special{pa 3210 2486}%
\special{fp}%
\special{pa 3390 2366}%
\special{pa 3240 2516}%
\special{fp}%
\special{pa 3390 2426}%
\special{pa 3270 2546}%
\special{fp}%
\special{pa 3390 2486}%
\special{pa 3300 2576}%
\special{fp}%
\special{pa 3390 2546}%
\special{pa 3330 2606}%
\special{fp}%
\special{pa 3390 2606}%
\special{pa 3360 2636}%
\special{fp}%
\special{pa 3360 2096}%
\special{pa 3090 2366}%
\special{fp}%
\special{pa 3300 2096}%
\special{pa 3060 2336}%
\special{fp}%
\special{pa 3240 2096}%
\special{pa 3030 2306}%
\special{fp}%
\special{pa 3180 2096}%
\special{pa 3000 2276}%
\special{fp}%
\special{pa 3120 2096}%
\special{pa 2970 2246}%
\special{fp}%
\special{pa 3060 2096}%
\special{pa 2940 2216}%
\special{fp}%
\special{pa 3000 2096}%
\special{pa 2910 2186}%
\special{fp}%
\special{pa 2940 2096}%
\special{pa 2880 2156}%
\special{fp}%
\special{pa 2880 2096}%
\special{pa 2850 2126}%
\special{fp}%
}}%
%
{\color[named]{Black}{%
\special{pn 13}%
\special{pa 3400 2096}%
\special{pa 2800 2096}%
\special{fp}%
}}%
\put(22.0000,-13.7000){\makebox(0,0)[rb]{{\small $g+p=0$}}}%
\put(32.3000,-27.2500){\makebox(0,0)[lt]{{\small $g+p=1$}}}%
\put(28.0000,-18.9500){\makebox(0,0){{\small $2$}}}%
%
{\color[named]{Black}{%
\special{pn 13}%
\special{pa 2800 1996}%
\special{pa 2800 2096}%
\special{dt 0.045}%
\special{pa 2800 2096}%
\special{pa 2600 2096}%
\special{dt 0.045}%
}}%
\put(24.6000,-20.9000){\makebox(0,0){{\small $-1$}}}%
\put(26.0000,-12.9500){\makebox(0,0){{\small $p$}}}%
\put(35.0000,-19.9500){\makebox(0,0){{\small $g$}}}%
\end{picture}%
\caption{Distribution of allowed $(g,p)$.}
\label{figure}
\end{wrapfigure}

\noindent
\begin{subequations}
\begin{align}
\delta_B \bsymb{\Phi_{-n}} &= \bsymb{Q_{n+1,n+2}}\hs \bsymb{\Phi_{-(n+1)}} \quad \left( n \geq 0 \right), 
\label{np brst g} \\*[1ex]
\delta_B \bsymb{\Phi_{n+1}} &= \bpz{\bsymb{B_{n+1,n}}} \bsymb{N_{n+2}} \quad \left( n \geq 1 \right), 
\label{np brst ag} \\*[1ex]
\delta_B \bsymb{N_{n+2}} &= 0 \quad \left( n \geq 1 \right).
\end{align}
\label{np brst}
\end{subequations}
\\ \\ \\[1ex]
The string fields $\Phi_{(-n,\hs m)}$ $( 1\leq n,\ 0 \leq m \leq n)$,
$\Phi_{(n+1,\hs -m)}$ $( 1 \leq m \leq n)$, and $N_{(n+2,\hs -m)}$ $( 1\leq n,\ 1 \leq m \leq n+1)$ are 
ghosts, antighosts, and auxiliary fields, respectively.
A string field $\Phi_{(g,\hs p)}$ is admissible only when the lattice point $(g,p)$ belongs to the region
shown in figure~\ref{figure}: the original field $\Phi_{(0,\hs 0)}$ and ghosts reside in the left region, 
and antighosts reside in the right region.
The subscripts on the vectors $\bsymb{\Phi}$ and $\bsymb{N}$ indicate their world-sheet ghost numbers,
whereas those on the matrices $\bsymb{Q}$, $\bsymb{B}$, $\bsymb{\tB}$, and $\bsymb{C}$ indicate the sizes of the matrices.
In what follows, unless otherwise stated, we use the same notation: the ghost number $g$ of a vector $\bsymb{V}$ 
and the size $m\times n$ of a matrix $\bsymb{M}$ are indicated by their subscripts as in $\bsymb{V_g}$ and $\bsymb{M_{m,n}}$.
Note that all the ghosts are Grassmann even, and all the antighosts and all the auxiliary fields are Grassmann odd, so that 
the vectors $\bsymb{\Phi_{-n}}$, $\bsymb{\Phi_{n+1}}$, and $\bsymb{N_{n+2}}$ have definite Grassmann parities:
\begin{equation}
\gp{\bsymb{\Phi_{-n}}} \equiv 0 \quad \left( n\geq 0\right),\quad
\gp{\bsymb{\Phi_{n+1}}} \equiv 1 \quad \left( n\geq 1\right),\quad
\gp{\bsymb{N_{n+2}}} \equiv 1 \quad \left( n\geq 1\right).
\end{equation}

By construction, the action \eqref{cgf action} is invariant under the BRST transformation \eqref{np brst},
whose nilpotency is guaranteed by
\begin{equation}
\bsymb{Q_{n+1,n+2} \,Q_{n+2,n+3}} = 0 \quad \left( n\geq 0 \right).
\label{Q^2=0}
\end{equation}
We would like to add that the constraints \eqref{const ag} are consistent with the transformations \eqref{np brst ag} because we have
\begin{equation}
\bz\xz \,\bpz{\bsymb{B_{2,1}}} = 0\,,\quad
\bsymb{\tB_{n-1,n}}\, \bpz{\bsymb{B_{n+1,n}}} = 0\quad \left( n\geq 2\right).
\end{equation}

\subsection{The extended action and its relation to gauge fixing}
\label{ex action}
\indenths
Let us decompose the completely gauge-fixed action \eqref{cgf action} into two parts: 
\begin{align}
\sum^\infty_{n=0} S_n &= S + S^\GF\,,\\
S &:= S_0 + \sum^\infty_{n=1} S^\FP_n = -\frac{\iu}{2} \braket{\bsymb{\Phi_0}, Q\ez \bsymb{\Phi_0}} 
-\sum^\infty_{n=1}\, \bigl\langle\bsymb{\Phi_{n+1}}, \bsymb{Q_{n,n+1} \Phi_{-n}}\bigr\rangle \,,
\label{solution} \\
S^\GF &:= \sum^\infty_{n=1} S^\GF_n = \sum^\infty_{n=1}\, \bigl\langle\bpz{\bsymb{B_{n+1,n}}} \bsymb{N_{n+2}} \hs, \bsymb{\Phi_{-(n-1)}}\bigr\rangle 
= -\sum^\infty_{n=1}\, \bigl\langle \bsymb{N_{n+2}}\hs, \bsymb{B_{n+1,n} {\Phi_{-(n-1)}}} \bigr\rangle \,.
\label{SGF}
\end{align}
The extended action \eqref{solution} is invariant under the gauge transformations (recall eq.~\eqref{Q^2=0})
\begin{equation}
\bsymb{\Phi_{-n}} \longrightarrow
\bsymb{\Phi_{-n}} + \bsymb{Q_{n+1,n+2}\hs \Lambda_{-(n+1)}} \,,
\quad
\bsymb{\Lambda_{-(n+1)}} :=
\left.
\bbm
\Lambda_{\left( -(n+1),\hs 0\right)} \\
\vdots \\
\Lambda_{\left( -(n+1),\hs n+1\right)}
\ebm
\right\} n+2
\quad \left( n\geq 0\right),
\label{g gt}
\end{equation}
from which the BRST transformations \eqref{np brst g} originate. 
The other action \eqref{SGF} plays the role of the gauge-fixing terms for these symmetries, imposing the conditions
\begin{equation}
\bsymb{B_{n+2,n+1}\hs \Phi_{-n}} = 0\quad \left( n\geq 0 \right).
\end{equation}
In order to investigate symmetries of $S$ further, 
suppose that antighosts are \emph{not} subject to the constraints \eqref{const ag}.
Then, we find that $S$ is also invariant under the gauge transformations of antighosts
\begin{equation}
\bsymb{\Phi_2} \longrightarrow
\bsymb{\Phi_2} + Q\eta_0\hs \bsymb{\Lambda_0} \,,
\quad
\bsymb{\Phi_{n+2}} \longrightarrow
\bsymb{\Phi_{n+2}} + \bsymb{Q_{n+1,n}\hs \Lambda_{n+1}}
\quad \left( n\geq 1\right)
\label{ag gt}
\end{equation}
with
\begin{align}
&\bsymb{\Lambda_0} := \Lambda_{(0,\hs 0)}\,,\quad
\bsymb{\Lambda_{n+1}} :=
\left.
\bbm
\Lambda_{(n+1,\hs -1)} \\
\vdots \\
\Lambda_{(n+1,\hs -n)}
\ebm
\right\} n
\quad
\left( n\geq 1\right), \\[1ex]
&\bsymb{Q_{n+1,n}} := 
\underbrace{
\left.
\begin{bmatrix}
Q & & \hsymbu{0} \\[-.5ex]
\eta_0 & \ddots & \\
 & \ddots & Q \\[.5ex]
\hsymbl{0} & & \eta_0 \\
\end{bmatrix}
\right\}
}_n
n+1 \quad
= -\bpz{\bsymb{Q_{n,n+1}}}
\quad 
\left( n \geq 1 \right).
\end{align}
The point is that \emph{the constraints \eqref{const ag} play exactly the role of the gauge-fixing conditions that eliminate these symmetries}. 
In fact, starting from the extended action \eqref{solution}, 
we can obtain a variety of completely gauge-fixed BRST-invariant actions in the following manner.\footnote{A similar approach 
is adopted in refs.~\cite{gl-gauges, a-gauge}.}
\begin{enumerate}
\item In order to eliminate the gauge symmetries of the extended action,
\bs
\begin{align}
&\bsymb{\Phi_{-n}} \longrightarrow
\bsymb{\Phi_{-n}} + \bsymb{Q_{n+1,n+2}\hs \Lambda_{-(n+1)}}
\qquad \left( n\geq 0\right),\\[1ex]
&\bsymb{\Phi_2} \longrightarrow
\bsymb{\Phi_2} + Q\eta_0\hs \bsymb{\Lambda_0} \,,
\quad
\bsymb{\Phi_{n+2}} \longrightarrow
\bsymb{\Phi_{n+2}} + \bsymb{Q_{n+1,n}\hs \Lambda_{n+1}}
\quad \left( n\geq 1\right),
\end{align}
\es
we impose gauge-fixing conditions on the original field $\bsymb{\Phi_0}$, the ghosts $\bsymb{\Phi_{-n}}$ $(n\geq 1)$,
and the antighosts $\bsymb{\Phi_{n+2}}$ $(n\geq 0)$. For example, we may use conditions of the form
\bs \label{1st and 2nd cond}
\begin{align}
\bsymb{B_{n+2,\hs n+1} \Phi_{-n}} = 0\quad \left( n\geq 0\right),
\label{1st cond} \\[1ex]
\bsymb{\tB_{m(n),\hs n+1} \Phi_{n+2}} = 0\quad \left( n\geq 0\right),
\label{2nd cond}
\end{align}
\es
where the number $m(n)$ depends on $n$.
(The matrices $\bsymb{B}$ and $\bsymb{\tB}$ do not have to be of the same form as those in eqs.~\eqref{QB} and \eqref{tBC}.)
Compatibility with the BRST transformation to be given in eq.~\eqref{BRST transf} requires that $\bsymb{B}$ and $\bsymb{\tB}$ satisfy
\begin{equation}
\bsymb{\tB_{m(n),\hs n+1}}\,\bpz{\bsymb{B_{n+2,\hs n+1}}} = 0\quad \left( n\geq 0\right).
\label{compatibility}
\end{equation}
\item The conditions on antighosts \eqref{2nd cond} are imposed directly, whereas those on the original field and ghosts \eqref{1st cond}
are realized indirectly by adding the gauge-fixing term
\begin{equation}
S^\GF = \sum^\infty_{n=1}\, \bigl\langle\bpz{\bsymb{B_{n+1,\hs n}}} \bsymb{N_{n+2}} \hs, \bsymb{\Phi_{-(n-1)}}\bigr\rangle 
= -\sum^\infty_{n=1}\, \bigl\langle \bsymb{N_{n+2}}\hs,\hs \bsymb{B_{n+1,\hs n} {\Phi_{-(n-1)}}} \bigr\rangle \,.
\end{equation}
The resultant action
\begin{align}
S + S^\GF
&= -\frac{\iu}{2} \braket{\bsymb{\Phi_0}, Q\ez \bsymb{\Phi_0}} 
-\sum^\infty_{n=1}\, \bigl\langle\bsymb{\Phi_{n+1}}, \bsymb{Q_{n,n+1} \Phi_{-n}}\bigr\rangle \nonumber \\*
&\quad +\sum^\infty_{n=1}\, \bigl\langle\bpz{\bsymb{B_{n+1,\hs n}}} \bsymb{N_{n+2}} \hs, \bsymb{\Phi_{-(n-1)}}\bigr\rangle
\end{align}
is invariant under the BRST transformation
\bs \label{BRST transf}
\begin{align}
\delta_B \bsymb{\Phi_{-n}} &= \bsymb{Q_{n+1,n+2}}\hs \bsymb{\Phi_{-(n+1)}} \quad \left( n \geq 0 \right), \\*[1ex]
\delta_B \bsymb{\Phi_{n+2}} &= \bpz{\bsymb{B_{n+2,\hs n+1}}} \bsymb{N_{n+3}} \quad \left( n \geq 0 \right), \\*[1ex]
\delta_B \bsymb{N_{n+3}} &= 0 \quad \left( n \geq 0 \right),
\end{align}
\es
because it is the sum of the original action $S_0$ and BRST-exact terms. 
\end{enumerate}
Note that in order to remove redundant symmetries associated with auxiliary fields, 
we have to restrict their space as well, as in eq.~\eqref{restriction}.
However, we do not have to specify the restrictions.
Suppose that we have integrated out all the auxiliary fields. Then, what remains is only the action $S$ with 
the constraints \eqref{1st and 2nd cond}, and it does not include auxiliary fields any longer.
In the rest of the present paper, we merely use this ``partially integrated'' version of the action.
Therefore, we do not need the specific forms of the restrictions on auxiliary fields.\footnote{Validity of  
condition \eqref{restriction} can be proved in the same manner as in section~\ref{validity} and appendix~\ref{reachability}.}

From the above point of view, the extended action $S$ is \emph{universal}, and various gauge-fixed actions can be
obtained \emph{merely by taking different gauge-fixing matrices}. This is a reflection of the fact that $S$ is essentially the
solution to the classical master equation, namely the classical part of the quantum master equation in the BV formalism~\cite{BV}.
For the relation between $S$ and the classical master equation, see ref.~\cite{paper1}.

Now that we have seen how to obtain BRST-invariant actions in general gauges, 
let us extend the conditions considered in section~\ref{gf by BRST}:
\bs
\begin{align}
\bz \Phi_{(-n,\hs 0)} \hspace{15.6ex}&=0 \quad \left( n\geq 0\right),
\label{ghost top} \\*[.3ex]
\xz \Phi_{(-n,\hs m)} + \bz \Phi_{(-n,\hs m+1)} &=0 \quad \left( 0\leq m\leq n-1\right),
\label{ghost middle} \\*[.3ex]
\xz \Phi_{(-n,\hs n)} \hspace{2.7ex}&=0 \quad \left( n\geq 0\right),
\label{ghost bottom}
\end{align}
\es
\bs
\begin{align}
\bz\xz \Phi_{(2,\hs -1)} &= 0\,,\\*[.3ex]
\bz \Phi_{(n+2,\hs -m)} + \xz \Phi_{(n+2,\hs -(m+1))} &= 0\quad \left( 1\leq m\leq n\right)
\label{antighost middle}.
\end{align}
\es
For this purpose, we introduce two real parameters $x$ and $y$ with $(x,y)\neq (0,0)$, and rescale $\bz$ and $\xz$ 
in eqs.~\eqref{ghost middle} and \eqref{antighost middle} as $\bz \to x\bz$, $\xz \to y\xz$.
This procedure provides a new set of gauge-fixing conditions, replacing 
eqs.~\eqref{ghost middle} and \eqref{antighost middle} with
\begin{align}
y\hs\xz \Phi_{(-n,\hs m)} + x\hs\bz \Phi_{(-n,\hs m+1)} &=0 \quad \left( 0\leq m\leq n-1\right),
\label{xy ghost middle} \\[.7ex]
x\hs\bz \Phi_{(n+2,\hs -m)} + y\hs\xz \Phi_{(n+2,\hs -(m+1))} &= 0\quad \left( 1\leq m\leq n\right).
\label{xy antighost middle}
\end{align}
In these gauges, however, cases in which $x=0$ or $y=0$ are exceptional:
if $x=0$ (resp.\ $y=0$), the antighosts $\Phi_{(n+2,\hs -1)}$ (resp.\ $\Phi_{\left( n+2,\hs -(n+1)\right)}$) $(n\geq 1)$ are not subject to any constraints.
To remedy this, we introduce the supplementary conditions
\bs
\begin{align}
\bz\xz\hs \Phi_{(n+2,\hs -1)} &= 0 \quad \left( n\geq 1\right),
\label{add 1} \\*
\bz\xz\hs \Phi_{\left( n+2,\hs -(n+1)\right)} &= 0 \quad \left( n\geq 1\right).
\label{add 2}
\end{align}
\es
Multiplying eq.~\eqref{xy antighost middle} by $\xz$ or $\bz$ from the left, we obtain
\bs
\begin{align}
0 &= \xz \bigl( x\hs \bz \Phi_{(n+2,\hs -1)} + y\hs \xz \Phi_{(n+2,\hs -2)}\bigr) = -x\hs \bz\xz\hs \Phi_{(n+2,\hs -1)} \,,\\*[.4ex]
0 &= \bz \bigl( x\hs \bz \Phi_{(n+2,\hs -n)} + y\hs \xz \Phi_{(n+2,\hs -(n+1))} \bigr) = y\hs \bz\xz\hs \Phi_{(n+2,\hs -(n+1))} \,.
\end{align}
\es
Thus, if $x\neq 0$ (resp.\ $y\neq 0$), eq.~\eqref{add 1} (resp.~\eqref{add 2}) follows from eq.~\eqref{xy antighost middle}.
However, if $x=0$ (resp.\ $y=0$), it supplies additional constraints.
New gauge-fixing matrices are given by
\begin{align}
\bsymb{B^{x,\hs y}_{n+2,\hs n+1}}
&:=
\underbrace{
\left.
\bbm
\bz &0 &\hdots &\hdots &0 \\[.5ex]
\hline \\[-2ex]
y\hs\xz &x\hs\bz &&&\hsymbu{0} \\[1ex]
&y\hs\xz &x\hs\bz && \\[1ex]
&&\ddots &\ddots && \\[1ex]
\hsymbl{0} &&& y\hs\xz &x\hs\bz \\[1ex]
\hline \\[-2ex]
0 &\hdots &\hdots &0& \xz
\ebm
\right\}
}_{n+1}
n+2
\qquad
\left( n\geq 0\right),
\label{Bxy} \\[2ex]
\bsymb{\tB^{x,\hs y}_{n+2,\hs n+1}}
&:=
\underbrace{
\left.
\bbm
\bz\xz &0 &\hdots &\hdots &0 \\[.5ex]
\hline \\[-2ex]
x\hs\bz &y\hs\xz &&&\hsymbu{0} \\[1ex]
&x\hs\bz &y\hs\xz && \\[1ex]
&&\ddots &\ddots && \\[1ex]
\hsymbl{0} &&& x\hs\bz &y\hs\xz \\[1ex]
\hline \\[-2ex]
0 &\hdots &\hdots &0& \bz\xz
\ebm
\right\}
}_{n+1}
n+2
\qquad
\left( n\geq 0\right).
\label{tBxy}
\end{align}
The two lines drawn in $\bsymb{B}$ (resp.\ $\bsymb{\tB}$) indicate the boundaries among the three submatrices 
associated with the three types of gauge-fixing conditions \eqref{ghost top},
\eqref{xy ghost middle}, and \eqref{ghost bottom} (resp.~\eqref{add 1}, \eqref{xy antighost middle}, and \eqref{add 2}).
We can readily confirm that the above matrices satisfy the compatibility conditions \eqref{compatibility}.

Thus far, we have been considering only those gauge-fixing matrices composed of $\bz$ and $\xz$.
However, it is also interesting to use, instead of $\xz$, 
the zero mode of the primary operator $d := \left[Q, b\xi\right]$\hs,
which has the same ghost number and picture number as $\xi$.
As a matter of fact, $\ez$ and $\dz$ are the counterparts of $Q$ and $\bz$ in the twisted $N=2$ superconformal algebra investigated by
Berkovits and Vafa~\cite{Berkovits-Vafa}.\footnote{The operators $Q$, $\ez$, $\bz$, and $\dz$ 
are identical to $G^+_0$, $\widetilde{G}^+_0$, $G^-_0$, and $\widetilde{G}^-_0$, respectively, 
in section 5 of ref.~\cite{Berkovits-Vafa}.
Useful algebraic relations are summarized in appendix~\ref{algebraic relations}.}
The crucial point is that algebraic relations are symmetric in $\bz$ and $\dz$:
\begin{equation}
\bz^2 = \dz^2 = \{ \bz ,\dz \} = \{Q,\dz\} = \{\ez,\bz\} = 0\,,\quad \{Q,\bz\} = \{\ez,\dz\} =L_0\,.
\label{N=2}
\end{equation}
As we shall see, in virtue of these properties, the propagators in $\bz$-$\dz$ gauges are much
simpler than those in $\bz$-$\xz$ gauges.

Here is another remark on the forms of the matrices \eqref{Bxy} and \eqref{tBxy}:
one does not have to use the same $(x,y)$ for every $n$.
In the next section, a larger class of gauge-fixing matrices will be proposed.

\section{Gauge-fixing conditions}
\label{gauge-fixing conditions}
\setcounter{equation}{0}
\indenths
We consider a class of gauge-fixing conditions of the form
\bs \label{gfc}
\begin{align}
\bsymb{B^{\zz^{(n)};\hs x_n,\hs y_n}_{n+2,\hs n+1} \Phi_{-n} } &=0\quad \left( n\geq 0\right),
\label{gfc B} \\*[1ex]
\bsymb{\tB^{\zz^{(n)};\hs x_n,\hs y_n}_{n+2,\hs n+1} \Phi_{n+2}} &=0\quad \left( n\geq 0\right),
\label{gfc tB}
\end{align}
\es
where
\begin{align}
\bsymb{B^{\zz^{(n)};\hs x_n,\hs y_n}_{n+2,\hs n+1}}
&:=
\underbrace{
\left.
\bbm
\bz &0 &\hdots &\hdots &0 \\[.5ex]
\hline \\[-2ex]
y_n\zz^{(n)} &x_n\bz &&&\hsymbu{0} \\[1ex]
&y_n\zz^{(n)} &x_n\bz && \\[1ex]
&&\ddots &\ddots && \\[1ex]
\hsymbl{0} &&& y_n\zz^{(n)} &x_n\bz \\[1ex]
\hline \\[-2ex]
0 &\hdots &\hdots &0& \zz^{(n)}
\ebm
\right\}
}_{n+1}
n+2
\qquad
\left( n\geq 0\right),
\label{B} \\[2ex]
\bsymb{\tB^{\zz^{(n)};\hs x_n,\hs y_n}_{n+2,\hs n+1}}
&:=
\underbrace{
\left.
\bbm
\bz\zz^{(n)} &0 &\hdots &\hdots &0 \\[.5ex]
\hline \\[-2ex]
x_{n}\bz &y_{n}\zz^{(n)} &&&\hsymbu{0} \\[1ex]
&x_{n}\bz &y_{n}\zz^{(n)} && \\[1ex]
&&\ddots &\ddots && \\[1ex]
\hsymbl{0} &&& x_{n}\bz &y_{n}\zz^{(n)} \\[1ex]
\hline \\[-2ex]
0 &\hdots &\hdots &0& \bz\zz^{(n)}
\ebm
\right\}
}_{n+1}
n+2
\qquad
\left( n\geq 0\right).
\label{tB}
\end{align}
Here $x_n$ and $y_n$ $\left( n\geq 0\right)$ are real parameters satisfying $(x_n,\hs y_n) \neq (0,\hs 0)$,
and $\zz^{(n)}$ $\left( n\geq 0\right)$ are either $\xz$ or $\dz$.
The quantum number $(g,p)$ of $\zz^{(n)}$ is $(-1,1)$, and we have
\begin{equation}
\bz^2 = \bigl( \zz^{(n)}\bigr)^2 = \bigl\{ \bz,\hs\zz^{(n)}\bigr\} = 0\,.
\end{equation} 
As will be mentioned in section~\ref{on-shell}, when $\zz^{(n)} = \dz$, we have to require the condition $x_n + y_n \neq 0$ as well.
It should be mentioned that the gauges \eqref{gfc} depend on $x_n$ and $y_n$ only through the \emph{ratio} of $x_n$ to $y_n$.
In order to avoid misunderstanding, we write down the forms of small-size matrices explicitly.
(As one can see from the equations below,
the matrices $\bsymb{B^{\zz^{(0)};\hs x_0,\hs y_0}_{2,\hs 1}}$ and
$\bsymb{\tB^{\zz^{(0)};\hs x_0,\hs y_0}_{2,\hs 1}}$ are independent of $x_0$ and $y_0$.)
\begin{align}
\bsymb{B^{\zz^{(0)};\hs x_0,\hs y_0}_{2,\hs 1}}
&= 
\bbm
\bz \\[1ex]
\zz^{(0)}
\ebm
,\quad
\bsymb{B^{\zz^{(1)};\hs x_1,\hs y_1}_{3,\hs 2}}
= 
\bbm
\bz &0 \\[.5ex]
\hline\\[-2ex]
y_1\zz^{(1)} &x_1\bz \\[1ex]
\hline\\[-2ex]
0 &\zz^{(1)}
\ebm
,\quad
\bsymb{B^{\zz^{(2)};\hs x_2,\hs y_2}_{4,\hs 3}}
=
\bbm
\bz &0 &0 \\[.5ex]
\hline\\[-2ex]
y_2\zz^{(2)} &x_2\bz &0 \\[1ex]
0 &y_2\zz^{(2)} &x_2\bz \\[1ex]
\hline\\[-2ex]
0 &0 &\zz^{(2)}
\ebm
,\\[1ex]
\bsymb{\tB^{\zz^{(0)};\hs x_0,\hs y_0}_{2,\hs 1}}
&= 
\bbm
\bz\zz^{(0)} \\[1ex]
\bz\zz^{(0)}
\ebm
,\quad
\bsymb{\tB^{\zz^{(1)};\hs x_1,\hs y_1}_{3,\hs 2}}
= 
\bbm
\bz\zz^{(1)} &0 \\[1ex]
\hline\\[-2ex]
x_1\bz &y_1\zz^{(1)} \\[1ex]
\hline\\[-2ex]
0 &\bz\zz^{(1)}
\ebm
,\quad
\bsymb{\tB^{\zz^{(2)};\hs x_2,\hs y_2}_{4,\hs 3}}
=
\bbm
\bz\zz^{(2)} &0 &0 \\[1ex]
\hline\\[-2ex]
x_2\bz &y_2\zz^{(2)} &0 \\[1ex]
0 &x_2\bz &y_2\zz^{(2)} \\[1ex]
\hline\\[-2ex]
0 &0 &\bz\zz^{(2)}
\ebm
.
\end{align}
Obviously, eq.~\eqref{gfc tB} can be decomposed as follows:
\bs \label{xy}
\begin{align}
&\bz\zz^{(n)}\Phi_{(n+2,\hs -1)} =0
\qquad \left( n\geq 0\right),
\label{xy1} \\[2ex]
n
&\underbrace{
\left\{
\bbm
x_n\bz &y_n\zz^{(n)} &&& \hsymbu{0} \\[1ex]
&x_n\bz &y_n\zz^{(n)} && \\[1ex]
&&\ddots &\ddots & \\[1ex]
\hsymbl{0} &&&x_n\bz &y_n\zz^{(n)}
\ebm
\right.
}_{n+1}
\bbm
\Phi_{(n+2,\hs -1)} \\[1ex]
\vdots \\[1ex]
\Phi_{(n+2,\hs -(n+1))}
\ebm
= 0
\qquad \left( n\geq 1 \right),
\label{xy2} \\[1ex]
&\bz\zz^{(n)} \Phi_{(n+2,\hs -(n+1))} =0
\qquad \left( n\geq 0\right).
\label{xy3}
\end{align}
\es
Note that if $x_n\neq 0$ $\left( n\geq 1\right)$ (resp.\ $y_n\neq 0$ $\left( n\geq 1\right)$),
condition \eqref{xy1} (resp.~\eqref{xy3}) follows from \eqref{xy2}, and thus we do not need that condition,
but if not, it is independent of \eqref{xy2}.
Note also that when $n=0$, condition \eqref{gfc tB} is reduced to $\bz\zz^{(0)} \bsymb{\Phi_2} = 0$.

The above class of gauge-fixing conditions contains two types of gauges for \emph{each} $n\geq 0$.
One is $\bz$-$\xz$ gauges, in which $\zz^{(n)} = \xz$, and the other is $\bz$-$\dz$ gauges, in which $\zz^{(n)} = \dz$.
Both of these are consistent with the BRST symmetry, because
\begin{equation}
\bsymb{\tB^{\zz^{(n)};\hs x_n,\hs y_n}_{n+2,\hs n+1}} \bpz{\bsymb{B^{\zz^{(n)};\hs x_n,\hs y_n}_{n+2,\hs n+1}}} = 0 
\quad \left( n\geq 0\right).
\label{tBB}
\end{equation}
It can be readily confirmed that we also have
\begin{equation}
\bsymb{B^{\zz^{(n)};\hs x_n,\hs y_n}_{n+2,\hs n+1}} \bpz{\bsymb{\tB^{\zz^{(n)};\hs x_n,\hs y_n}_{n+2,\hs n+1}}} = 0 
\quad \left( n\geq 0\right).
\label{BtB}
\end{equation}
We would like to emphasize that one does not have to adopt the same type of gauge-fixing conditions for \emph{every} $n\geq 0$.
However, the simplest choice would be
\begin{equation} 
\zz^{(0)}=\zz^{(1)}=\zz^{(2)}=\cdots\,,\quad x_0 = x_1 = x_2 =\cdots\,,\quad y_0 = y_1 = y_2 = \cdots\,.
\label{simplest choice}
\end{equation}
In the next section (and in appendix~\ref{reachability}), we investigate the validity of the gauge-fixing conditions \eqref{gfc} in detail.

\section{Validity of the gauge-fixing conditions}
\label{validity}
\setcounter{equation}{0}
\subsection{On-shell values of \texorpdfstring{$L_0$}{L0}}
\label{on-shell}
\indenths
In general, on-shell gauge structure is a little different from the off-shell one.
Before considering the validity of the conditions \eqref{gfc}, let us see what values $L_0$ can take
on the mass shell.
We first consider the equation of motion of $\Phi_{(0,\hs 0)}$
\begin{equation}
Q\eta_0 \Phi_{(0,\hs 0)} = 0
\label{eom 0 0}
\end{equation}
under the gauge-fixing condition
\begin{equation}
\bsymb{B^{\zz ;\hs x,\hs y}_{2,\hs 1} \Phi_0} = 0\,,\quad
\mrm{i.e.,}\quad
\bz \Phi_{(0,\hs 0)} = \zz \Phi_{(0,\hs 0)} = 0\,.
\label{gfc 0 0}
\end{equation}
Here and in the sequel, $\bsymb{B^{\zz ;\hs x,\hs y}_{n+2,\hs n+1}}$ should be understood as 
$\bigl[\bsymb{B^{\zz^{(n)} ;\hs x_n,\hs y_n}_{n+2,\hs n+1}} \bigr]_{\zz^{(n)}=\zz,\,x_n =x,\,y_n =y}$\,, with
$\zeta$ denoting either $\xi$ or $d$
(and thus $\zz$ denoting either $\xz$ or $\dz$). 
From eqs.~\eqref{eom 0 0} and \eqref{gfc 0 0} we obtain
\begin{equation}
0= \zz \bz Q \eta_0 \Phi_{(0,\hs 0)} = \zz \bigl( L_0 - Q\bz \bigr) \eta_0 \Phi_{(0,\hs 0)}
= L_0 \zz\eta_0 \Phi_{(0,\hs 0)} = L_0 \bigl( \Lbar - \eta_0 \zz \bigr) \Phi_{(0,\hs 0)} = L_0 \Lbar \Phi_{(0,\hs 0)}
\label{on-shell 0 0}
\end{equation}
with
\begin{align}
\Lbar := \{ \eta_0 , \zz^{(n)} \} =
\left\{
\begin{aligned}
&1 \quad \bigl( \text{for } \zz^{(n)} = \xz \bigr),\\
&L_0 \quad \bigl( \text{for } \zz^{(n)} = \dz \bigr).
\end{aligned}
\right.
\end{align}
Eq.~\eqref{on-shell 0 0} tells us that on the mass shell of $\Phi_{(0,\hs 0)}$, we have $L_0 = 0$.

As a next example, let us consider the equation of motion
\begin{equation}
Q\Phi_{(-1,\hs 0)} + \eta_0 \Phi_{(-1,\hs 1)} = 0
\end{equation}
originating from the ghost action
\begin{equation}
- \left\langle \Phi_{( 2,\hs -1)}\,,
\begin{bmatrix}
Q & \eta_0 \\
\end{bmatrix} 
\begin{bmatrix}
\Phi_{( -1,\hs 0)} \\[.5ex]
\Phi_{( -1,\hs 1)}
\end{bmatrix}
\right\rangle
=
-\bigl\langle\Phi_{(2,\hs -1)}\, , Q\Phi_{(-1,\hs 0)} + \eta_0 \Phi_{(-1,\hs 1)} \bigr\rangle\,.
\end{equation}
If we impose the condition
\begin{equation}
\bsymb{B^{\zz;\hs x,\hs y}_{3,\hs 2} \Phi_{-1}} = 0\,,\quad
\mrm{i.e.,}\quad
\left\{
\begin{aligned}
&\bz \Phi_{(-1,\hs 0)} = 0\,,\\
& y\hs \zz \Phi_{(-1,\hs 0)} + x\hs \bz \Phi_{(-1,\hs 1)} = 0\,,\\
&\zz \Phi_{(-1,\hs 1)} =0\,,
\end{aligned}
\right.
\end{equation}
then we obtain
\begin{align}
0 &= x\hs \zz\bz \bigl( Q\Phi_{(-1,\hs 0)} + \eta_0 \Phi_{(-1,\hs 1)} \bigr) 
= x\hs \zz( L_0 - Q\bz) \Phi_{(-1,\hs 0)} - \zz\eta_0 \bigl( x\hs \bz \Phi_{(-1,\hs 1)} \bigr) \nonumber \\*[.5ex]
&= x\hs \zz L_0 \Phi_{(-1,\hs 0)} + \zz\eta_0 \bigl( y\hs \zz \Phi_{(-1,\hs 0)} \bigr)
= \zz L_{x,y} \Phi_{(-1,\hs 0)} \,,
\label{on-shell -1 0}
\end{align}
with
\begin{equation}
\Lxy := x L_0 + y\Lbar =
\left\{
\begin{aligned}
&x L_0 + y \quad \bigl( \text{for } \zz^{(n)} = \xz \bigr),\\
&\left(x+y\right) L_0 \quad \bigl( \text{for } \zz^{(n)} = \dz \bigr).
\end{aligned}
\right.
\label{Lxy}
\end{equation}
Eq.~\eqref{on-shell -1 0} suggests that $L_{x,y}$ will be zero on the mass shell.

As we shall see in section~\ref{propagators}, propagators indeed have poles at $L_0 =0$ and $L_{x_n,y_n} = 0$.
We would like to remark that when $\zz^{(n)} = \dz$, we have to require $x_n + y_n \neq 0$ in order to exclude the case in which 
the operator $L_{x_n,y_n} = \left( x_n + y_n \right) L_0$ is trivially zero.
The ``off-shell conditions'' $L_0 \neq 0$ and $L_{x_n ,y_n} \neq 0$, which are reduced to $L_0\neq 0$ and $x_n + y_n \neq 0$ for $\zz^{(n)}=\dz$, 
will be assumed when we prove the validity of the gauge-fixing conditions.

\subsection{Validity of the gauge-fixing conditions}
\indenths
In the present subsection (and in appendix~\ref{reachability}), we prove that \emph{off the mass shell} the symmetries
\begin{equation}
\bsymb{\Phi_{-n}} \longrightarrow
\bsymb{\Phi_{-n}} + \bsymb{Q_{n+1,n+2}\hs \Lambda_{-(n+1)}}
\qquad \left( n\geq 0\right)
\label{gt of ghosts}
\end{equation}
and
\begin{equation}
\bsymb{\Phi_2} \longrightarrow
\bsymb{\Phi_2} + Q\eta_0\hs \bsymb{\Lambda_0} \,,
\quad
\bsymb{\Phi_{n+2}} \longrightarrow
\bsymb{\Phi_{n+2}} + \bsymb{Q_{n+1,n}\hs \Lambda_{n+1}}
\quad \left( n\geq 1\right)
\label{gt of antighosts}
\end{equation}
can be eliminated respectively by imposing the conditions
\begin{equation}
\bsymb{B^{\zz^{(n)};\hs x_n,\hs y_n}_{n+2,\hs n+1} \Phi_{-n} } =0\quad \left( n\geq 0\right)
\label{gfc for ghosts}
\end{equation}
and
\begin{equation}
\bsymb{\tB^{\zz^{(n)};\hs x_n,\hs y_n}_{n+2,\hs n+1} \Phi_{n+2}} =0\quad \left( n\geq 0\right).
\label{gfc for antighosts}
\end{equation}
This involves two steps.
We have to show reachability and completeness of the conditions.
\begin{enumerate}
\item
{\bf (Reachability)}\;
By the use of the transformation \eqref{gt of ghosts} (resp.~\eqref{gt of antighosts}), 
one can let the string fields $\Phi_{(-n,\hs m)}$ $\left( 0 \leq m \leq n \right)$ (resp.\ $\Phi_{(n+2,\hs -m)}$ $\left( 1\leq m\leq n+1\right)$)
reach the configuration where condition \eqref{gfc for ghosts} (resp.~\eqref{gfc for antighosts}) holds.
\item
{\bf (Completeness)}\;
If once condition \eqref{gfc for ghosts} (or \eqref{gfc for antighosts}) is imposed, there remains no 
residual gauge symmetry that preserves the condition. In other words, the condition eliminates the gauge degrees of freedom completely.
\end{enumerate}
We first consider the reachability and the completeness of condition \eqref{gfc for ghosts},
and then consider those of \eqref{gfc for antighosts}.
The way itself to prove the completeness is significant: byproducts of the proofs shall play 
an essential role when we derive propagators in section~\ref{propagators}.
In what follows, we assume the ``off-shell conditions'' $L_0 \neq 0$ and $L_{x_n,y_n} \neq 0$.
\vspace{2ex}

The validity of condition \eqref{gfc for ghosts} is ensured by the following two propositions.\footnote{
\label{footnote:validity}
In propositions~\ref{reachability bxi} and \ref{completeness bxi}, we intensionally use the symbols
$x$, $y$, and $\zz$ rather than $x_n$, $y_n$, and $\zz^{(n)}$, in order to avoid notational complexity in their proofs.
One may think that in these propositions, only those gauge-fixing conditions satisfying eq.~\eqref{simplest choice}
are considered, but it is not the case.
The point is that the propositions hold for \emph{each} $n\geq 0$, without the assumption that $x$, $y$, and $\zz$,
which are arbitrary as long as they satisfy $(x,y)\neq (0,0)$ and $\zz\in \{\xz, \dz\}$, 
take the same values for \emph{every} $n$.
Hence it goes without saying that they hold even if $x$, $y$, and $\zz$ in eqs.~\eqref{bxi gfc} and \eqref{bxi gf cond} are replaced with
$x_n$, $y_n$, and $\zz^{(n)}$. 
}
(For the proof of proposition~\ref{reachability bxi}, see appendix~\ref{reachability}.)

\begin{Prop}[Reachability of \eqref{gfc for ghosts}]
\label{reachability bxi}
For any set of string fields of the form $\{ \Phi_{(-n,\hs m)} \mid 0 \leq m \leq n \}$ ($n\geq 0$ is fixed),
there exists a set of string fields 
$\{ \Lambda_{(-(n+1),\hs m)} \mid 0 \leq m \leq n+1 \}$ such that
\begin{equation}
\bsymb{B^{\zz;\hs x,\hs y}_{n+2,\hs n+1}} \Bigl( \bsymb{\Phi_{-n}} + \bsymb{Q_{n+1,n+2}\hs \Lambda_{-(n+1)}} \Bigr) =0
\quad \left( n\geq 0\right).
\label{bxi gfc}
\end{equation}
\end{Prop}
\vspace{1ex}

\begin{Prop}[Completeness of \eqref{gfc for ghosts}]
\label{completeness bxi}
Let $\{ \Phi_{( -n ,\hs m )} \mid  0 \leq m \leq n \}$ be a set of string fields
($n\geq 0$ is fixed) satisfying
\begin{equation}
\bsymb{B^{\zz;\hs x,\hs y}_{n+2,\hs n+1}} \bsymb{\Phi_{-n}} =0
\quad \left( n\geq 0\right).
\label{bxi gf cond}
\end{equation}
If the transformation
\begin{equation}
\bsymb{\Phi_{-n}} \longrightarrow
\bsymb{\Phi'_{-n}} := 
\bsymb{\Phi_{-n}} + \bsymb{Q_{n+1,n+2}\hs \Lambda_{-(n+1)}}
\quad \left( n\geq 0\right)
\label{transfPhi}
\end{equation}
preserves condition \eqref{bxi gf cond}, then one obtains
\begin{equation}
\bsymb{Q_{n+1,n+2}\hs \Lambda_{-(n+1)}} = 0\,,
\quad \text{i.e.,}\quad
\bsymb{\Phi'_{-n}} = \bsymb{\Phi_{-n}}
\quad \left( n\geq 0\right).
\label{QL}
\end{equation}
(This proposition ensures that there exists no residual gauge symmetry of the form \eqref{transfPhi} that preserves
condition \eqref{bxi gf cond}.)
\end{Prop}
{\bf \underline{Proof.}}\quad Because the ``transformation'' parameterized by $\bsymb{\Theta}$
\begin{equation}
\bsymb{\Lambda_{-(n+1)}} \longrightarrow
\bsymb{\Lambda_{-(n+1)}} + \bsymb{Q_{n+2,n+3}\hs \Theta_{-(n+2)}}\,,
\quad
\bsymb{\Theta_{-(n+2)}} :=
\bbm
\Theta_{\left( -(n+2),\hs 0\right)} \\
\vdots \\
\Theta_{\left( -(n+2),\hs n+2\right)}
\ebm
\end{equation}
does not change the value of $\bsymb{Q_{n+1,n+2}\hs \Lambda_{-(n+1)}}$,
we may assume
\begin{equation}
\bsymb{B^{\zz;\hs x,\hs y}_{n+3,\hs n+2}}\hs \bsymb{\Lambda_{-(n+1)}} =0
\label{b xi Lambda}
\end{equation}
without loss of generality (see proposition~\ref{reachability bxi}).\footnote{
If one replaces the symbols $x$, $y$, and $\zz$ in eq.~\eqref{bxi gf cond} with $x_n$, $y_n$, and $\zz^{(n)}$, then 
those in eq.~\eqref{b xi Lambda}, also, have to be replaced with $x_n$, $y_n$, and $\zz^{(n)}$, \emph{not} with
$x_{n+1}$, $y_{n+1}$, and $\zz^{(n+1)}$. This does not mean, however, that we assume condition~\eqref{simplest choice}.
We emphasize again that proposition~\ref{reachability bxi} holds for \emph{each} $n\geq 0$, with $x$, $y$, and $\zz$ arbitrary.
}
Hence, for the purpose of proving the proposition, it is sufficient to show that under condition \eqref{b xi Lambda}
we obtain
\begin{equation}
\bsymb{B^{\zz;\hs x,\hs y}_{n+2,\hs n+1} Q_{n+1,n+2}\hs \Lambda_{-(n+1)}} = 0 \ 
\Longrightarrow \ 
\bsymb{\Lambda_{-(n+1)}} = 0\,.
\label{BQL}
\end{equation}
Indeed, if eq.~\eqref{BQL} holds, then we find 
\begin{align}
&\ \bsymb{B^{\zz;\hs x,\hs y}_{n+2,\hs n+1}} \bsymb{\Phi_{-n}} = \bsymb{B^{\zz;\hs x,\hs y}_{n+2,\hs n+1}} \bsymb{\Phi'_{-n}} = 0 \nonumber \\*[1ex]
\Longrightarrow &\  
\bsymb{B^{\zz;\hs x,\hs y}_{n+2,\hs n+1} Q_{n+1,n+2}\hs \Lambda_{-(n+1)}} = 0
\ \Longrightarrow \ 
\text{\eqref{QL}}\,,
\end{align}
which completes the proof.
Eq.~\eqref{BQL} is ensured by showing the existence of an $(n+2) \times (n+2)$ matrix $\bsymb{P^{\zz;\hs x,\hs y}_{n+2,\hs n+2}}$ such that 
\begin{equation}
\bsymb{P^{\zz;\hs x,\hs y}_{n+2,\hs n+2}\hs B^{\zz;\hs x,\hs y}_{n+2,\hs n+1}\hs Q_{n+1,n+2}}
=
\bsymb{1_{n+2}}
+
\bsymb{M^{\zz;\hs x\hs y}_{n+2,\hs n+3}\hs B^{\zz;\hs x,\hs y}_{n+3,\hs n+2} }\,.
\label{P}
\end{equation}
Here $\bsymb{1_{n+2}}$ is the unit matrix of size $n+2$, and $\bsymb{M^{\zz;\hs x,\hs y}_{n+2,\hs n+3}}$ is some $(n+2) \times (n+3)$
matrix. Given eq.~\eqref{P}, we indeed obtain \eqref{BQL}:
\begin{align}
&\bsymb{B^{\zz;\hs x,\hs y}_{n+2,\hs n+1} Q_{n+1,n+2}\hs \Lambda_{-(n+1)}} = 0 \nonumber \\*[1ex]
\Longrightarrow \ &
0 = \Bigl(\bsymb{P^{\zz;\hs x,\hs y}_{n+2,\hs n+2}\hs B^{\zz;\hs x,\hs y}_{n+2,\hs n+1}\hs Q_{n+1,n+2}}\Bigr) \bsymb{\Lambda_{-(n+1)}}
=
\Bigl( \bsymb{1_{n+2}}
+
\bsymb{M^{\zz;\hs x,\hs y}_{n+2,\hs n+3}\hs B^{\zz;\hs x,\hs y}_{n+3,\hs n+2} } \Bigr) \bsymb{\Lambda_{-(n+1)}} \nonumber \\*[1ex]
\Longrightarrow \ & \bsymb{\Lambda_{-(n+1)}} = 0\quad \left(\because \text{\eqref{b xi Lambda}}\right).
\end{align}
As a matter of fact, we can solve eq.~\eqref{P} explicitly.
The matrices $\bsymb{P^{\zz;\hs x,\hs y}_{n+2,\hs n+2}}$ and $\bsymb{M^{\zz;\hs x,\hs y}_{n+2,\hs n+3}}$ which solve eq.~\eqref{P}
are listed in appendix~\ref{solutions to PtP}.
\qquad $\Box$
\\[3ex]
\indenths
Now that we have confirmed the validity of condition \eqref{gfc for ghosts},
let us see that of \eqref{gfc for antighosts}. 
The proof goes along the same lines. (For the proof of proposition~\ref{reachability bxi anti}, see appendix~\ref{reachability}.)

\begin{Prop}[Reachability of \eqref{gfc for antighosts}]
\label{reachability bxi anti}
\ \\
{\bf {\upshape (A)}}
For any string field $\Phi_{(2,\hs -1)}$ of indicated quantum number,
there exists a string field $\Lambda_{(0,\hs 0)}$ such that
\begin{equation}
\bz\zz \left( \Phi_{(2,\hs -1)} + Q\eta_0\hs \Lambda_{(0,\hs 0)} \right) =0 \,.
\label{bxi gfc 2,-1}
\end{equation}
{\bf {\upshape (B)}}
Let $n\geq 1$ be fixed.
For any set of string fields of the form $\{ \Phi_{(n+2,\hs -m)} \mid 1 \leq m \leq n+1 \}$,
there exists a set of string fields $\{ \Lambda_{(n+1,\hs  -m)} \mid 1 \leq m \leq n \}$ such that
\begin{equation}
\bsymb{\tB^{\zz;\hs x,\hs y}_{n+2,\hs n+1}} \Bigl( \bsymb{\Phi_{n+2}} + \bsymb{Q_{n+1,n}\hs \Lambda_{n+1}} \Bigr) =0
\quad \left( n\geq 1\right).
\label{bxi gfc anti}
\end{equation}
\end{Prop}
\vspace{1ex}

\begin{Prop}[Completeness of \eqref{gfc for antighosts}]
\label{completeness bxi anti}
Let $\{ \Phi_{(n+2,\hs -m )} \mid  1 \leq m \leq n+1 \}$ be a set of string fields
($n\geq 0$ is fixed) satisfying
\begin{equation}
\bsymb{\tB^{\zz;\hs x,\hs y}_{n+2,\hs n+1}} \bsymb{\Phi_{n+2}} =0
\quad \left( n\geq 0\right).
\label{bxi gf cond anti}
\end{equation}
If the transformation
\bs \label{transfPhi anti}
\begin{align}
\bsymb{\Phi_2} &\longrightarrow
\bsymb{\Phi_2} + Q\eta_0\hs \bsymb{\Lambda_0} 
\quad \left( \mrm{for\ }n= 0 \right)\,,\\*[1ex]
\bsymb{\Phi_{n+2}} &\longrightarrow
\bsymb{\Phi_{n+2}} + \bsymb{Q_{n+1,n}\hs \Lambda_{n+1}}
\quad \left( \mrm{for\ }n\geq 1 \right)
\end{align}
\es
preserves condition \eqref{bxi gf cond anti}, then one obtains
\bs
\begin{align}
Q\eta_0\hs \bsymb{\Lambda_0} &= 0 \quad \left( \mrm{for\ }n= 0 \right)\,,\\*[1ex]
\bsymb{Q_{n+1,n}\hs \Lambda_{n+1}} &= 0 \quad \left( \mrm{for\ }n\geq 1 \right).
\end{align}
\es
\end{Prop}
{\bf \underline{Proof.}}\quad The ``transformation'' parameterized by $\bsymb{\Theta}$
\bs
\begin{align}
\bsymb{\Lambda_0} &\longrightarrow \bsymb{\Lambda_0} + \bsymb{Q_{1,2}\hs \Theta_{-1}}\,, 
\quad
\bsymb{\Theta_{-1}} :=
\bbm
\Theta_{\left( -1,\hs 0\right)} \\[.5ex]
\Theta_{\left( -1,\hs 1\right)}
\ebm
\quad \left( \mrm{for\ }n= 0 \right)\,,
\\[1ex]
\bsymb{\Lambda_2} &\longrightarrow \bsymb{\Lambda_2} + Q\eta_0\hs \bsymb{\Theta_0}\,, 
\quad
\bsymb{\Theta_0} := \Theta_{(0,\hs 0)}
\quad \left( \mrm{for\ }n= 1 \right)\,, 
\\[1ex]
\bsymb{\Lambda_{n+1}} &\longrightarrow \bsymb{\Lambda_{n+1}} + \bsymb{Q_{n,n-1}\hs \Theta_n}\,, 
\quad
\bsymb{\Theta_n} :=
\bbm
\Theta_{\left( n,\hs -1\right)} \\
\vdots \\
\Theta_{\left( n,\hs -(n-1)\right)}
\ebm
\quad \left( \mrm{for\ }n\geq 2 \right)
\end{align}
\es
does not affect the value of
\bs
\begin{align}
Q\eta_0 \bsymb{\Lambda_0} \quad \left( \mrm{for\ }n= 0 \right)\,,\\[.5ex]
\bsymb{Q_{n+1,n}\hs \Lambda_{n+1}} \quad \left( \mrm{for\ }n\geq 1 \right)\,.
\end{align}
\es
Therefore, we may assume
\bs \label{b xi Lambda anti}
\begin{align}
\bsymb{B^{\zz;\hs x,\hs y}_{2,\hs 1} \Lambda_0} &=0 \quad \left( \mrm{for\ }n= 0 \right)\,,\\[1ex]
\bsymb{\tB^{\zz;\hs x,\hs y}_{n+1,\hs n}}\hs \bsymb{\Lambda_{n+1}} &=0 \quad \left( \mrm{for\ }n\geq 1 \right)
\end{align}
\es
without loss of generality (see propositions~\ref{reachability bxi} and~\ref{reachability bxi anti}).
Hence, for the purpose of proving the proposition, it is sufficient to show that under condition \eqref{b xi Lambda anti}
we gain
\bs \label{tB Q L}
\begin{align}
\bz\zz Q\eta_0 \bsymb{\Lambda_0} = 0 \ &\Longrightarrow \ \bsymb{\Lambda_0} = 0\quad \left( \mrm{for\ }n= 0 \right)\,,\\[1ex]
\bsymb{\tB^{\zz;\hs x,\hs y}_{n+2,\hs n+1} Q_{n+1,n}\hs \Lambda_{n+1}} = 0 \ 
&\Longrightarrow \ \bsymb{\Lambda_{n+1}} = 0\quad \left( \mrm{for\ }n\geq 1 \right)\,.
\end{align}
\es
Eq.~\eqref{tB Q L} is ensured by the existence of an operator $\bsymb{P^{\zz;\hs x,\hs y}_{1,\hs 1}}$ (for the $n=0$ case)
and of an $n \times (n+2)$ matrix $\bsymb{\widetilde{P}^{\zz;\hs x,\hs y}_{n,\hs n+2}}$ (for the $n\geq 1$ case) such that 
\bs \label{P anti}
\begin{align}
\bsymb{P^{\zz;\hs x,\hs y}_{1,\hs 1} B^{\zz;\hs x,\hs y}_{1,\hs 0} Q_{0,1}} 
&= 1 + \bsymb{M^{\zz;\hs x,\hs y}_{1,\hs 2} B^{\zz ;\hs x,\hs y}_{2,\hs 1}} \quad \left( \mrm{for\ }n= 0 \right)\,,
\label{1001}\\[1ex]
\bsymb{\widetilde{P}^{\zz;\hs x,\hs y}_{n,\hs n+2}\hs \tB^{\zz;\hs x,\hs y}_{n+2,\hs n+1} Q_{n+1,n}}
&= \bsymb{1_n} + \bsymb{\tM^{\zz;\hs x,\hs y}_{n,\hs n+1} \tB^{\zz;\hs x,\hs y}_{n+1,\hs n} } \quad \left( \mrm{for\ }n\geq 1 \right)\,.
\end{align}
\es
Here $\bsymb{1_n}$ is the unit matrix of size $n$, $\bsymb{M^{x,\hs y}_{1,\hs 2}}$ is some $1\times 2$ matrix, 
and $\bsymb{\tM^{x,\hs y}_{n,\hs n+1}}$ is some $n \times (n+1)$ matrix.
We have defined $\bsymb{B^{\zz;\hs x,\hs y}_{1,\hs 0}}$ and $\bsymb{Q_{0,1}}$ as
\begin{equation}
\bsymb{B^{\zz;\hs x,\hs y}_{1,\hs 0}} := \bz\zz \,,\quad
\bsymb{Q_{0,1}} := \iu\, Q\ez\,.
\end{equation}
(Compare eq.~\eqref{1001} with eq.~\eqref{P}.)
The solutions to eq.~\eqref{P anti} are listed in appendix~\ref{solutions to PtP}.
\qquad $\Box$

\subsection{Byproducts of the proofs}
\indenths
In the proofs of propositions~\ref{completeness bxi} and~\ref{completeness bxi anti},
we have obtained interesting equations \eqref{P} and \eqref{P anti},
which connect the gauge-fixing matrices $\bsymb{B}$'s and $\bsymb{\tB}$'s with the kinetic-term matrices $\bsymb{Q}$'s
in the extended action.
In fact, the matrices
\begin{equation}
\bsymb{K^{\zz;\hs x,\hs y}_{n+1,\hs n}} 
:= \bsymb{P^{\zz;\hs x,\hs y}_{n+1,\hs n+1}\hs B^{\zz;\hs x,\hs y}_{n+1,\hs n}} \quad \left( n\geq 0\right),\quad
\bsymb{\tK^{\zz;\hs x,\hs y}_{n,\hs n+1}} 
:= \bsymb{\widetilde{P}^{\zz;\hs x,\hs y}_{n,\hs n+2}\, \tB^{\zz;\hs x,\hs y}_{n+2,\hs n+1}} \quad \left( n\geq 1\right),
\label{K and tK}
\end{equation}
have a close relation to propagators.
In the next section, we shall see that propagators are expressed as the products of
$\bsymb{K}$'s (or $\mrm{bpz}\bigl(\bsymb{\tK}\bigr)$'s) and $\bsymb{Q}$.
The concrete forms of $\bsymb{K}$ and $\bsymb{\tK}$, which can be immediately obtained from 
$\bsymb{P}$ and $\bsymb{\widetilde{P}}$ in appendix~\ref{solutions to PtP},
are listed in appendix~\ref{concrete forms}.

We would like to mention some important properties of $\bsymb{K}$ and
$\bsymb{\tK}$ needed in the next section, in the form of a proposition. 
Let us define $\bsymb{\tK^{\zz;\hs x,\hs y}_{0,\hs 1}}$ and $\bsymb{Q_{1,0}}$ as
\begin{align}
&\bsymb{\tK^{\zz;\hs x,\hs y}_{0,\hs 1}} := \bpz{\bsymb{K^{\zz;\hs x,\hs y}_{1,\hs 0}}}\,,\\[1ex]
&\bsymb{Q_{1,0}} := +\bpz{\bsymb{Q_{0,1}}} = \bsymb{Q_{0,1}} = \iu\, Q\ez\,.
\end{align}
\begin{Prop}[Properties of $\bsymb{K}$ and $\bsymb{\tK}$]
\label{properties of K and tK} 
The following equations hold.
\begin{equation}
\bsymb{K^{\zz;\hs x,\hs y}_{n+1,\hs n}}\, \bpz{\bsymb{\tB^{\zz;\hs x,\hs y}_{n+1,\hs n}}} =0 \quad \left( n\geq 1\right),\quad
\bsymb{\tK^{\zz;\hs x,\hs y}_{n,\hs n+1}}\, \bpz{\bsymb{B^{\zz;\hs x,\hs y}_{n+2,\hs n+1}}} =0\quad \left( n\geq 0\right),
\label{KB}
\end{equation}
\begin{align}
\bsymb{K^{\zz;\hs x,\hs y}_{n+1,\hs n}\, Q_{n,n+1}}
&=
\bsymb{1_{n+1}}
+
\bsymb{M^{\zz;\hs x,\hs y}_{n+1,\hs n+2}\hs B^{\zz;\hs x,\hs y}_{n+2,\hs n+1} } \quad \left( n\geq 0\right),
\label{KQ} \\[1ex]
\bsymb{\tK^{\zz;\hs x,\hs y}_{n,\hs n+1}\, Q_{n+1,n}}
&= \bsymb{1_n} + \bsymb{\tM^{\zz;\hs x,\hs y}_{n,\hs n+1}\hs \tB^{\zz;\hs x,\hs y}_{n+1,\hs n} } \quad \left( n\geq 1 \right)\,,
\label{tKQ}
\end{align}
\\[-5ex]
\begin{equation}
\bsymb{\tK^{\zz;\hs x,\hs y}_{n,\hs n+1}} = \bpz{\bsymb{K^{\zz;\hs x,\hs y}_{n+1,\hs n}}}\,,\quad
\bsymb{K^{\zz;\hs x,\hs y}_{n+1,\hs n}} = \bpz{\bsymb{\tK^{\zz;\hs x,\hs y}_{n,\hs n+1}}}\quad \left( n\geq 0\right),
\label{bpz KtK}
\end{equation}
\begin{align}
&\bsymb{K^{\zz;\hs x,\hs y}_{n,\hs n-1}\, Q_{n-1,n}} + \bsymb{Q_{n,n+1}\hs K^{\zz;\hs x,\hs y}_{n+1,\hs n}} = \bsymb{1_n} \quad \left( n\geq 1\right),
\label{KQ+QK} \\[1ex] 
&\bsymb{\tK^{\zz;\hs x,\hs y}_{n,\hs n+1}\, Q_{n+1,n}} + \bsymb{Q_{n,n-1}\hs \tK^{\zz;\hs x,\hs y}_{n-1,\hs n}} = \bsymb{1_n} \quad \left( n\geq 1\right).
\label{tKQ+QtK}
\end{align}
\end{Prop}

Eq.~\eqref{KB} immediately follows from the definition \eqref{K and tK} and eqs.~\eqref{tBB} and \eqref{BtB}.
Eqs.~\eqref{KQ} and \eqref{tKQ} are identically the relations \eqref{P} and \eqref{P anti}.
Eqs.~\eqref{bpz KtK}, \eqref{KQ+QK}, and \eqref{tKQ+QtK} can be confirmed with straightforward calculation.
(Note that because eqs.~\eqref{KQ+QK} and \eqref{tKQ+QtK} are the BPZ conjugates of each other,
one does not have to confirm both.) We would like to stress that proposition~\ref{properties of K and tK} holds for 
\emph{each} $n$, with $x$, $y$, and $\zz$ arbitrary (see also footnote~\ref{footnote:validity}). 

\section{Propagators}
\label{propagators}
\setcounter{equation}{0}
\indenths
In order to find the propagators under the gauge-fixing conditions \eqref{gfc},
we rewrite the action \eqref{solution} as\footnote{When we would like to explicitly show the size $m\times n$ of a zero matrix,
we will append the subscript ``$m,n$'' and will denote the zero matrix by $\bsymb{0_{m,n}}$\,.}
\begin{equation}
S= -\frac{1}{2} \bigl\langle \bsymb{\Phi_0} , \bsymb{Q_{0,1}} \bsymb{\Phi_0} \bigr\rangle
- \frac{1}{2} \sum^\infty_{n=1} \left\langle 
\bbm
\bsymb{\Phi_{-n}} \\[1ex]
\bsymb{\Phi_{n+1}}
\ebm
,
\bbm
\bsymb{0_{n+1,n+1}} & \bsymb{Q_{n+1,n}} \\[1ex]
\bsymb{Q_{n,n+1}} & \bsymb{0_{n,n}}
\ebm
\bbm
\bsymb{\Phi_{-n}} \\[1ex]
\bsymb{\Phi_{n+1}}
\ebm
\right\rangle
,
\end{equation}
and consider the inverses of $\bsymb{Q_{0,1}} = \iu\,Q\eta_0$ and 
$\bbm
\bsymb{0_{n+1,n+1}} & \bsymb{Q_{n+1,n}} \\[1ex]
\bsymb{Q_{n,n+1}} & \bsymb{0_{n,n}}
\ebm$
in the restricted subspaces
\bs
\begin{align}
\RS^\zeta_0 &:= \Set{ \bsymb{\Phi_0} | \bsymb{B^{\zz^{(0)};\hs x_0,\hs y_0}_{2,\hs 1} \Phi_0} = 0 } \,,\\*[2ex]
\RS^\zeta_n &:= \Set{
\bbm
\bsymb{\Phi_{-n}} \\
\bsymb{\Phi_{n+1}}
\ebm
| \bsymb{B^{\zz^{(n)};\hs x_n,\hs y_n}_{n+2,\hs n+1} \Phi_{-n} } =0\,,\quad
\bsymb{\tB^{\zz^{(n-1)};\hs x_{n-1},\hs y_{n-1}}_{n+1,\hs n} \Phi_{n+1}} =0 } \quad \left( n\geq 1\right).
\end{align}
\es

\subsection{\texorpdfstring{$\bsymb{\Phi_0}$}{Phi0}-propagator}
\indenths
We denote the $\bsymb{\Phi_0}$-propagator by $\bsymb{\Delta_{1,0}}$.
It is the inverse of $\bsymb{Q_{0,1}} = \iu\,Q\eta_0$ in the subspace $\RS^\zeta_0$, where the equation
\begin{equation}
\bsymb{B^{\zz^{(0)};\hs x_0,\hs y_0}_{2,\hs 1} \Phi_0} = 0
\quad \left(\iff \bpz{\bsymb{\Phi_0}} \mrm{bpz}\Bigl(\bsymb{B^{\zz^{(0)};\hs x_0,\hs y_0}_{2,\hs 1}}\Bigr) = 0\right)
\label{B 2 1}
\end{equation}
holds. Thus, it has to satisfy the following relations:
\bs \label{D10}
\begin{align}
\bsymb{\Delta_{1,0}\hs Q_{0,1}}
&= 1 + \bsymb{\calM_{1,2}\hs B^{\zz^{(0)};\hs x_0,\hs y_0}_{2,\hs 1}}\,,
\label{D10 Q} \\[1ex]
\bsymb{Q_{0,1}\hs \Delta_{1,0}} 
&= 1 + \mrm{bpz}\Bigl(\bsymb{B^{\zz^{(0)};\hs x_0,\hs y_0}_{2,\hs 1}}\Bigr) \bsymb{\calM_{2,1}}
\label{Q01 D}
\end{align}
\es
with $\bsymb{\calM_{1,2}}$ and $\bsymb{\calM_{2,1}}$
being some $1\times 2$ matrix and some $2\times 1$ matrix, respectively.
Eq.~\eqref{D10 Q}/\eqref{Q01 D} ensures that $\bsymb{\Delta_{1,0}}$ is the left/right inverse of $\bsymb{Q_{0,1}}$
in $\RS^\zeta_0$:
the operator \eqref{D10 Q}/\eqref{Q01 D} acts as the identity operator upon $\bsymb{\Phi_0}$/$\bpz{\bsymb{\Phi_0}}$ from the left/right 
under condition \eqref{B 2 1}.  
It should be noted that if the propagator $\bsymb{\Delta_{1,0}}$ is BPZ-even, 
eq.~\eqref{Q01 D} immediately follows from the BPZ conjugate of \eqref{D10 Q}.\footnote{
In this case, the matrices $\bsymb{\calM_{1,2}}$ and $\bsymb{\calM_{2,1}}$ shall be related to each other.}
As a matter of fact, we have already known the BPZ-even $\bsymb{\Delta_{1,0}}$ satisfying eq.~\eqref{D10 Q}. From eqs.~\eqref{KQ}, \eqref{small K^xi},
and \eqref{small K^d}, we find that $\bsymb{\Delta_{1,0}}$ is precisely $\bsymb{K^{\zz^{(0)};\hs x_0,\hs y_0}_{1,\hs 0}}$:
\begin{equation}
\bsymb{\Delta_{1,0}} = \bsymb{K^{\zz^{(0)};\hs x_0,\hs y_0}_{1,\hs 0}} 
= \bsymb{P^{\zz^{(0)};\hs x_0,\hs y_0}_{1,\hs 1}\hs B^{\zz^{(0)};\hs x_0,\hs y_0}_{1,\hs 0}} \,.
\end{equation}

\subsection{Ghost propagators}
\indenths
Let us calculate the propagator $\bsymb{\Delta_{n+1,n}}$ between $\bsymb{\Phi_{-n}}$ and $\bsymb{\Phi_{n+1}}$ $(n\geq 1)$
under the conditions \eqref{gfc}.
For this purpose, we consider the \emph{BPZ-even} inverse of the matrix
\begin{equation}
\bbm
\bsymb{0_{n+1,n+1}} & \bsymb{Q_{n+1,n}} \\[1ex]
\bsymb{Q_{n,n+1}} & \bsymb{0_{n,n}}
\ebm
=
\bbm
\bsymb{0_{n+1,n+1}} & \bsymb{Q_{n+1,n}} \\[1ex]
-\bpz{\bsymb{Q_{n+1,n}}} & \bsymb{0_{n,n}}
\ebm
.
\label{Q matrix}
\end{equation}
We may suppose that it takes the form\footnote{The propagator matrices in ref.~\cite{paper1}
correspond to not $\bsymb{\Delta_{n+1,n}}$ but $\bsymb{\Delta_{n,n+1}} := \bpz{\bsymb{\Delta_{n+1,n}}}$.}
\begin{equation}
\bbm
\bsymb{0_{n+1,n+1}} & \bsymb{\Delta_{n+1,n}} \\[1ex]
\bpz{\bsymb{\Delta_{n+1,n}}} & \bsymb{0_{n,n}}
\ebm
\,.
\label{prop matrix}
\end{equation}
Because the matrix \eqref{prop matrix} is BPZ-even, if it is the left inverse of \eqref{Q matrix}, it is automatically the right inverse.
The propagator $\bsymb{\Delta_{n+1,n}}$ is obtained by solving the equation
\begin{align}
&
\bbm
\bsymb{0_{n+1,n+1}} & \bsymb{\Delta_{n+1,n}} \\[1ex]
\bpz{\bsymb{\Delta_{n+1,n}}} & \bsymb{0_{n,n}}
\ebm
\bbm
\bsymb{0_{n+1,n+1}} & \bsymb{Q_{n+1,n}} \\[1ex]
\bsymb{Q_{n,n+1}} & \bsymb{0_{n,n}}
\ebm
\nonumber \\*[1.5ex]
=&\ \bsymb{1_{2n+1}} + 
\bbm
\bsymb{\calM_{n+1,n+2}\hs B^{\zz^{(n)};\hs x_n,\hs y_n}_{n+2,\hs n+1}} & \bsymb{0_{n+1,n}} \\[1ex]
\bsymb{0_{n,n+1}} & \bsymb{\tcalM_{n,n+1}}\hs \bsymb{\tB^{\zz^{(n-1)};\hs x_{n-1},\hs y_{n-1}}_{n+1,\hs n}}
\ebm
\qquad (n\geq 1)
\,.
\label{prop eq}
\end{align}
Here $\bsymb{\calM_{n+1,n+2}}$ and $\bsymb{\tcalM_{n,n+1}}$ are some $(n+1)\times (n+2)$ matrix and some $n\times (n+1)$ matrix, respectively.
Note that the right-hand side of eq.~\eqref{prop eq} acts as the identity $\bsymb{1_{2n+1}}$ in $\RS^\zeta_n$.
The following proposition tells us the form of the propagator.
\begin{Prop}[Ghost propagators]
\label{ghost propagators}
Eq.~\eqref{prop eq} is solved by
\begin{align}
\bsymb{\Delta_{n+1,n}} &=
\bsymb{K^{\zz^{(n)};\hs x_n,\hs y_n}_{n+1,\hs n}\hs Q_{n,n+1}\hs K^{\zz^{(n-1)};\hs x_{n-1},\hs y_{n-1}}_{n+1,\hs n}} \nonumber \\*[1ex]
&= \bigl( \bsymb{P^{\zz^{(n)};\hs x_n,\hs y_n}_{n+1,\hs n+1}\hs B^{\zz^{(n)};\hs x_n,\hs y_n}_{n+1,\hs n}} \bigr)
\hs \bsymb{Q_{n,n+1}}\hs
\bigl( \bsymb{P^{\zz^{(n-1)};\hs x_{n-1},\hs y_{n-1}}_{n+1,\hs n+1}\hs B^{\zz^{(n-1)};\hs x_{n-1},\hs y_{n-1}}_{n+1,\hs n}} \bigr)\,.
\label{Delta}
\end{align}
\end{Prop}
\vspace{1ex}
{\bf \underline{Proof.}}\quad Eq.~\eqref{prop eq} is equivalent to the following set of equations:
\bs \label{Deq}
\begin{align}
&\bsymb{\Delta_{n+1,n}}\hs \bsymb{Q_{n,n+1}} = \bsymb{1_{n+1}} + \bsymb{\calM_{n+1,n+2}\hs B^{\zz^{(n)};\hs x_n,\hs y_n}_{n+2,n+1}}\,,\\[1ex]
&\bpz{\bsymb{\Delta_{n+1,n}}} \bsymb{Q_{n+1,n}} = \bsymb{1_n} + \bsymb{\tcalM_{n,n+1}}\hs \bsymb{\tB^{\zz^{(n-1)};\hs x_{n-1},\hs y_{n-1}}_{n+1,\hs n}}\,.
\end{align}
\es
Let us prove that \eqref{Delta} is the solution to these.
For this purpose, we use proposition~\ref{properties of K and tK}.
From eqs.~\eqref{KQ+QK} and \eqref{tKQ+QtK}, and
\begin{equation}
\bsymb{Q_{n-1,n}\, Q_{n,n+1}} = 0\,,\quad
\bsymb{Q_{n+1,n}\, Q_{n,n-1}} = 0\quad \left( n\geq 1\right),
\end{equation}
it follows that
\begin{align}
\bsymb{Q_{n,n+1}\hs K^{\zz;\hs x,\hs y}_{n+1,\hs n}\, Q_{n,n+1}} = \bsymb{Q_{n,n+1}}\quad \left( n\geq 1\right),
\label{QKQ} \\[1ex]
\bsymb{Q_{n+1,n}\hs \tK^{\zz;\hs x,\hs y}_{n,\hs n+1}\, Q_{n+1,n}} = \bsymb{Q_{n+1,n}}\quad \left( n\geq 1\right).
\label{QtKQ}
\end{align}
Thus we gain the equations below:
\begin{align}
&\ \Bigl( \bsymb{K^{\zz^{(n)};\hs x_n,\hs y_n}_{n+1,\hs n}\hs Q_{n,n+1}\hs K^{\zz^{(n-1)};\hs x_{n-1},\hs y_{n-1}}_{n+1,\hs n}} \Bigr) \bsymb{Q_{n,n+1}} 
\nonumber \\*[1ex]
=&\ \bsymb{K^{\zz^{(n)};\hs x_n,\hs y_n}_{n+1,\hs n}} \Bigl( \bsymb{Q_{n,n+1}\hs K^{\zz^{(n-1)};\hs x_{n-1},\hs y_{n-1}}_{n+1,\hs n}\hs Q_{n,n+1}} \Bigr) 
\nonumber \\*[1ex]
=&\ \bsymb{K^{\zz^{(n)};\hs x_n,\hs y_n}_{n+1,\hs n}}\hs \bsymb{Q_{n,n+1}} 
= \bsymb{1_{n+1}} + \bsymb{M^{\zz^{(n)};\hs x_n,\hs y_n}_{n+1,\hs n+2}\hs B^{\zz^{(n)};\hs x_n,\hs y_n}_{n+2,\hs n+1} } \qquad \left( n\geq 1\right),
\label{bpzD Q} \\[3ex]
&\ \mrm{bpz}
\Bigl( \bsymb{K^{\zz^{(n)};\hs x_n,\hs y_n}_{n+1,\hs n}\hs Q_{n,n+1}\hs K^{\zz^{(n-1)};\hs x_{n-1},\hs y_{n-1}}_{n+1,\hs n}} \Bigr)\hs \bsymb{Q_{n+1,n}}
\nonumber \\*[1ex]
=&\ \Bigl(\bsymb{\tK^{\zz^{(n-1)};\hs x_{n-1},\hs y_{n-1}}_{n,\hs n+1}\hs Q_{n+1,n}\hs \tK^{\zz^{(n)};\hs x_n,\hs y_n}_{n,\hs n+1}}\Bigr) \bsymb{Q_{n+1,n}}
\nonumber \\*[1ex]
=&\ \bsymb{\tK^{\zz^{(n-1)};\hs x_{n-1},\hs y_{n-1}}_{n,\hs n+1}}
\Bigl(\bsymb{Q_{n+1,n}\hs \tK^{\zz^{(n)};\hs x_n,\hs y_n}_{n,\hs n+1}}\hs \bsymb{Q_{n+1,n}} \Bigr)
\nonumber \\*[1ex]
=&\ \bsymb{\tK^{\zz^{(n-1)};\hs x_{n-1},\hs y_{n-1}}_{n,\hs n+1}}\hs \bsymb{Q_{n+1,n}}
= \bsymb{1_n} 
+ \bsymb{\tM^{\zz^{(n-1)};\hs x_{n-1},\hs y_{n-1}}_{n,\hs n+1} \tB^{\zz^{(n-1)};\hs x_{n-1},\hs y_{n-1}}_{n+1,\hs n} } \qquad \left( n\geq 1 \right).
\label{D Q}
\end{align}
In the second and the last equality of \eqref{bpzD Q}, we have used eqs.~\eqref{QKQ} and \eqref{KQ}, respectively,
and in the first, the third, and the last equality of \eqref{D Q}, we have used eqs.~\eqref{bpz KtK}, \eqref{QtKQ}, and \eqref{tKQ}, respectively.
Eqs.~\eqref{bpzD Q} and \eqref{D Q} ensure that eq.~\eqref{Deq} is solved by \eqref{Delta}.
\qquad $\Box$\\

As was mentioned in section~\ref{on-shell}, the propagator \eqref{Delta} has poles at the points where 
$L_0 =0\hs$, $L_{x_{n-1},\hs y_{n-1}} = 0$, or $L_{x_n ,\hs y_n}= 0\hs$ because $\bsymb{K}$'s have. 
We would like to remark that 
\emph{all the information about gauge-fixing conditions is incorporated into the propagator 
through the matrix $\bsymb{K} = \mrm{bpz}\bigl(\bsymb{\tK}\bigr)$}
(recall the definition \eqref{K and tK}).
As is shown in the corollary below, the expression \eqref{Delta} is simplified in the special case in which
\begin{equation}
\zz^{(n-1)} = \zz^{(n)} \left( = \zz\right),\quad
x_{n-1} = x_n \left( = x\right),\quad
y_{n-1} = y_n \left( = y\right).
\label{zzxy}
\end{equation}

\begin{Cor}
\label{corollary}
Under condition \eqref{zzxy}, the propagator \eqref{Delta} is simplified to
\begin{equation}
\bsymb{\Delta_{n+1,n}} = \bsymb{K^{\zz;\hs x,\hs y}_{n+1,\hs n} }
= \bsymb{P^{\zz;\hs x,\hs y}_{n+1,\hs n+1}\hs B^{\zz;\hs x,\hs y}_{n+1,\hs n}} \,.
\label{simple D}
\end{equation}
\end{Cor}
\vspace{1ex}
{\bf \underline{Proof.}}\quad From eqs.~\eqref{bpz KtK}, \eqref{K and tK}, and \eqref{KB}, we obtain
\begin{align}
&\ \bsymb{K^{\zz;\hs x,\hs y}_{n+2,\hs n+1}\hs K^{\zz;\hs x,\hs y}_{n+1,\hs n}}
= \bsymb{K^{\zz;\hs x,\hs y}_{n+2,\hs n+1}}\, \bpz{\bsymb{\tK^{\zz;\hs x,\hs y}_{n,\hs n+1}}}
= \bsymb{K^{\zz;\hs x,\hs y}_{n+2,\hs n+1}}\, \bpz{\bsymb{\widetilde{P}^{\zz;\hs x,\hs y}_{n,\hs n+2}\hs \tB^{\zz;\hs x,\hs y}_{n+2,\hs n+1}}} 
\nonumber \\*[1ex]
=&\ \bsymb{K^{\zz;\hs x,\hs y}_{n+2,\hs n+1}}\, \bpz{\bsymb{\tB^{\zz;\hs x,\hs y}_{n+2,\hs n+1}}} \bpz{\bsymb{\widetilde{P}^{\zz;\hs x,\hs y}_{n,\hs n+2}}}
=0\qquad \left( n\geq 1\right).
\label{tKtK}
\end{align}
Thus, using eqs.~\eqref{tKtK} and \eqref{KQ+QK}, we find that
\begin{align}
\bsymb{\Delta_{n+1,n}} &= \bsymb{K^{\zz;\hs x,\hs y}_{n+1,\hs n}\, Q_{n,n+1}\hs K^{\zz;\hs x,\hs y}_{n+1,\hs n}}
\nonumber \\*[1ex]
&= \Bigl( \bsymb{K^{\zz;\hs x,\hs y}_{n+1,\hs n}\, Q_{n,n+1}} + \bsymb{Q_{n+1,n+2}\, K^{\zz;\hs x,\hs y}_{n+2,\hs n+1}}\Bigr)
\bsymb{K^{\zz;\hs x,\hs y}_{n+1,\hs n}}
= \bsymb{K^{\zz;\hs x,\hs y}_{n+1,\hs n} }\,.
\end{align}
$\Box$

\subsection{Some examples}
\label{some examples}
\indenths
In the present subsection, let us see some examples in which the propagators become simple.
We focus on gauge-fixing conditions such that 
\begin{equation}
\zz^{(n)} = \zz \,,\quad 
x_n = x\,,\quad
y_n = y\qquad \left(\text{for every $n$}\right).
\end{equation}
In this case, according to corollary~\ref{corollary}, propagators are given by
\begin{equation}
\bsymb{\Delta_{n+1,n}} = \bsymb{K^{\zz;\hs x,\hs y}_{n+1,\hs n}}\quad \left( \forall n\geq 0\right).
\end{equation}
For the concrete forms of the matrices $\bsymb{K^{\zz;\hs x,\hs y}_{n+1,\hs n}}$, see appendix~\ref{concrete forms}.\\[1ex]
{\bf (I) \boldmath{$\zz = \xz$} case} \\*
\indenths
For $\zz = \xz$, propagators take complicated forms because $\{ Q, \xz\}$ is non-vanishing.
However, they are dramatically simplified when $x=0$ or $y=0$.\\
{\bf (i) \boldmath{$x=0$} (and \boldmath{$y\neq 0$}) case}\\*
\indenths
When $x$ is zero, operators composing the matrices $\bsymb{K^{\xz;\hs x,\hs y}_{n+1,\hs n}}$ listed in appendix~\ref{concrete forms} are reduced to
\begin{equation}
P_\xi = \xz\ez\,,\quad
P_X = -X_0\,,\quad
\frac{K_\xi}{\Lxy} = \xz\,,\quad
\KXb = \xz\Bigl( 1-\frac{X_0 \ez\bz}{L_0}\Bigr),
\end{equation}
where $X_0 := \{Q,\xz\}$ is the zero mode of the picture-changing operator $X(z) = \{Q, \xi(z)\}$.
Thus we obtain
\bs
\begin{align}
&\bsymb{\Delta_{1,0}} = \iu\,\frac{\bz\xz}{L_0}\,,\quad
\bsymb{\Delta_{2,1}} =
\bbm
\frac{\xz\ez\bz}{L_0} \\[2ex]
\xz\bigl( 1 - \frac{X_0 \ez\bz}{L_0}\bigr)
\ebm
=:
\bbm
E \\[2ex]
F
\ebm
,
\\[2ex]
&\bsymb{\Delta_{3,2}} = 
\bbm
E &0 \\[.5ex]
F &0 \\[.5ex]
(-X_0) F &\xz 
\ebm
,
\quad
\bsymb{\Delta_{4,3}} = 
\bbm
E &0&0 \\[.5ex]
F &0&0 \\[.5ex]
(-X_0) F &\xz &0 \\[1ex]
(-X_0)^2 F &(-X_0)\xz &\xz 
\ebm
,
\end{align}
\es
\vspace{1ex}
\begin{equation}
\bsymb{\Delta_{n+1,n}} =
\bbm
E &0&0&0&\cdots &0&0 \\[.5ex]
F &0&0&0&\cdots &0&0 \\[.5ex]
(-X_0) F &\xz &0 &0&\cdots &0&0 \\[1ex]
(-X_0)^2 F &(-X_0)\xz &\xz &0&\cdots &0&0 \\[1ex]
(-X_0)^3 F &(-X_0)^2\xz &(-X_0)\xz &\xz &\cdots &0&0 \\[.5ex] 
\vdots &\vdots &\vdots &\vdots &\vdots &\vdots &\vdots \\[.5ex]
(-X_0)^{n-2} F &(-X_0)^{n-3}\xz &(-X_0)^{n-4}\xz &(-X_0)^{n-5}\xz &\cdots &\xz &0 \\[1ex]
(-X_0)^{n-1} F &(-X_0)^{n-2}\xz &(-X_0)^{n-3}\xz &(-X_0)^{n-4}\xz &\cdots &(-X_0)\xz &\xz
\ebm
\quad \left( n\geq 1\right).
\end{equation}
It is noteworthy that the propagators include $X_0$.
We would like to remark that the propagator matrices in ref.~\cite{paper1}
correspond to the BPZ conjugates $\bsymb{\Delta_{n,n+1}} := \bpz{\bsymb{\Delta_{n+1,n}}}$
rather than $\bsymb{\Delta_{n+1,n}}$ themselves.
In order to help ones's understanding, we list $\bsymb{\Delta_{n,n+1}}$ as well.
\bs
\begin{align}
&\bsymb{\Delta_{0,1}} = \iu\,\frac{\bz\xz}{L_0}\,,\quad
\bsymb{\Delta_{1,2}} =
\bbm
\frac{\ez\xz\bz}{L_0} &\bigl( 1 + \frac{X_0 \ez\bz}{L_0}\bigr)\xz
\ebm
=
\bbm
E^\star & F^\star
\ebm
,
\\[2ex]
&\bsymb{\Delta_{2,3}} = 
\bbm
E^\star &F^\star &(-X_0) F^\star \\[.5ex]
\hline \\[-2ex]
0 &0 &\xz
\ebm
,
\quad
\bsymb{\Delta_{3,4}} = 
\bbm
E^\star &F^\star &(-X_0) F^\star &(-X_0)^2 F^\star \\[.5ex]
\hline \\[-2ex]
0 &0 &\xz &(-X_0)\xz \\[1ex]
0 &0 &0 &\xz
\ebm
,
\end{align}
\es
\vspace{.5ex}
\begin{equation}
\bsymb{\Delta_{n,n+1}} =
\bbm
E^\star &F^\star &(-X_0) F^\star &(-X_0)^2 F^\star &\cdots &(-X_0)^{n-2} F^\star &(-X_0)^{n-1} F^\star \\[1ex]
\hline \\[-2ex]
0 &0 &\xz &(-X_0)\xz &\cdots &(-X_0)^{n-3} \xz &(-X_0)^{n-2} \xz \\[1ex]
0 &0 &0 &\xz &\cdots &(-X_0)^{n-4} \xz &(-X_0)^{n-3} \xz \\[.5ex]
\vdots &\vdots &\vdots &\vdots &\cdots &\vdots &\vdots \\[.5ex]
0 &0 &0 &0 &\cdots &\xz &(-X_0)\xz \\[1ex]
0 &0 &0 &0 &\cdots &0 &\xz
\ebm
\quad \left( n\geq 1\right),
\end{equation}
where $\star$ denotes the BPZ conjugation.
\\[1ex]
{\bf (ii) \boldmath{$y=0$} (and \boldmath{$x\neq 0$}) case}\\*
\indenths
When $y$ is zero, we have
\begin{equation}
P_\xi =1 \,,\quad
P_X = \frac{K_\xi}{\Lxy} = \KXb = 0\,,
\end{equation}
so that propagators are given by
\begin{align}
&\bsymb{\Delta_{1,0}} = \iu\,\frac{\bz\xz}{L_0}\,,\quad
\bsymb{\Delta_{2,1}} =
\bbm
\frac{\bz}{L_0} \\[1.5ex] 
\frac{\xz\bz Q}{L_0}
\ebm
,\quad
\bsymb{\Delta_{3,2}} = 
\bbm
\frac{\bz}{L_0} &0 \\[1ex]
0 &\frac{\bz}{L_0} \\[1.5ex]
0 &\frac{\xz\bz Q}{L_0}
\ebm
,\\[2ex]
&\bsymb{\Delta_{n+1,n}} = 
\bbm
\frac{\bz}{L_0} &&&&\hsymbu{0} \\[1ex]
&\frac{\bz}{L_0} &&& \\[1ex]
&& \ddots && \\[1ex]
&&&\frac{\bz}{L_0} & \\[1ex]
&&&&\frac{\bz}{L_0} \\[1.5ex]
\hsymbl{0} &&&&\frac{\xz\bz Q}{L_0}
\ebm
\quad \left( n\geq 1 \right).
\end{align}
\ \\
{\bf (II) \boldmath{$\zz =\dz$} case (with general \boldmath{$x$} and \boldmath{$y$})} \\*
\indenths
In the $\zz = \dz$ case, unlike in the $\zz = \xz$ case, propagators take simple forms for general $x$ and $y$
(with $(x,y)\neq (0,0)$ and $x+y\neq 0$). 
This is the reflection of the algebraic relations \eqref{N=2} symmetric in $\bz$ and $\dz$.
It is crucial that the anticommutator $\{Q,\zz\}$ vanishes for $\zz = \dz$.
Small-size propagators are given by
\begin{equation}
\bsymb{\Delta_{1,0}} = \iu\,\frac{\bz\dz}{L_0^2}\,,\quad
\bsymb{\Delta_{2,1}} =
\bbm
K_b \\[1ex]
K_d
\ebm
,\quad
\bsymb{\Delta_{3,2}} = 
\bbm
K_b &0 \\[1ex]
\frac{y\dz}{\Lxy} &\frac{x\bz}{\Lxy} \\[1ex]
0 &K_d
\ebm
,
\end{equation}
with
\begin{equation}
K_b := \frac{\bz}{L_0} \Bigl( 1-y\frac{\ez\dz}{\Lxy} \Bigr),\quad
K_d := \frac{\dz}{L_0} \Bigl( 1-x\frac{Q\bz}{\Lxy} \Bigr).
\end{equation}
The symbol $\Lxy$ is defined in eq.~\eqref{Lxy}.
For an arbitrary $n \left( \geq 1\right)$, we have
\begin{equation}
\bsymb{\Delta_{n+1,n}} =
\bbm
K_b &&&&\hsymbu{0} \\[1.5ex]
\frac{y\dz}{\Lxy} &\frac{x\bz}{\Lxy} &&&\\[1.5ex]
&\frac{y\dz}{\Lxy} &\frac{x\bz}{\Lxy} &&\\[1.5ex]
&&\ddots &\ddots &\\[1.5ex]
&&&\frac{y\dz}{\Lxy} &\frac{x\bz}{\Lxy}\\[1.5ex]
\hsymbl{0} &&&&K_d
\ebm
\quad \left( n\geq 1\right).
\end{equation}
It is worth while mentioning that in $\bz$-$\dz$ gauges, propagators involve the inverse of $L^2_0$
(see, for example, the form of $\bsymb{\Delta_{1,0}}$).
Since $L_0$ corresponds to $p^2 + m^2$ in field theory of particles, the existence of $1/{L^2_0}$ seems to imply 
the appearance of $(p^2 + m^2)^{-2}$ in the propagators.
One may think that this is problematic, but in ref.~\cite{amp}, it is shown that the correct on-shell four-point tree amplitude
is reproduced by using the propagator $\bsymb{\Delta_{1,0}}$.
We will discuss more about $\bz$-$\dz$ gauges in section~\ref{summary}.

\section{Extension similar to linear \texorpdfstring{\boldmath{$b$}}{b}-gauges}
\label{ext}
\setcounter{equation}{0}
\indenths
Thus far, we have been dealing with a simple class of gauge-fixing conditions.
Gauge-fixing matrices were composed of only the zero modes $\bz$ and $\zz$.
In this sense, condition \eqref{gfc} is analogous to the Siegel gauge~\cite{Siegel} in cubic string field theory~\cite{Witten, WittenSuper},
under which string fields are annihilated by $\bz$. 
In fact, as is shown in ref.~\cite{paper1}, the gauge-fixed action in the $\bz$-$\xz$ gauge with $x=0$ corresponds to
that in the Siegel gauge in Witten's superstring field theory~\cite{WittenSuper}.\footnote{Note that 
as long as one restricts oneself within the free theory, no problem exists in Witten's superstring field theory.}
The form of the propagator \eqref{Delta} tells us that the gauges are suitable to see the relation to the world-sheet theory
on the upper half-plane: on-shell poles can be specified in terms of $L_0$.

However, recent analytic methods in open string field theory have been developing 
in another conformal frame, the sliver frame~\cite{sliver}, in which computations involving Witten's star product~\cite{Witten} 
are facilitated.
This series of developments was triggered by Schnabl's analysis~\cite{Schnabl} of the tachyon vacuum in bosonic string field theory~\cite{Witten}.
He utilized a gauge naturally fitting into the above-mentioned frame, which was crucial to the classical solution describing the tachyon condensation. 
Although the Schnabl gauge itself turned out to be singular~\cite{off-shell_amp}, Kiermaier, Sen, and Zwiebach have shown that 
it can be realized as a limit of a one-parameter family of regular gauges~\cite{b-gauges}.
Their gauge-fixing conditions were constructed from linear combinations of $b$-ghost oscillators, 
after which they were named ``linear $b$-gauges.''

In the present section, we consider the extension of the conditions \eqref{gfc} in this direction.
First, in section~\ref{linear-b}, we review linear $b$-gauges in bosonic string field theory.
We explain the subject in detail because our analysis in the case of the superstring field theory is based on 
the same argument.
Second, in section~\ref{extended gfc}, we propose extended gauge-fixing conditions in which $\bz$'s and $\zz$'s in condition \eqref{gfc}
are replaced with linear combinations of $b$-modes and of $\zeta$-modes, respectively.
Lastly, in section~\ref{ext propagators}, we consider the propagators under the extended gauge-fixing conditions.

\subsection{Linear \texorpdfstring{$b$}{b}-gauges in bosonic string field theory}
\label{linear-b}
\subsubsection{The extended action and gauge-fixing conditions}
\indenths
The free bosonic string field theory action~\cite{Witten} takes the form
\begin{equation}
S_0 = -\frac{1}{2} \langle \Psi_1 , Q\Psi_1 \rangle \,,
\label{free bosonic}
\end{equation}
where $\Psi_1$ is a Grassmann-odd string field of world-sheet ghost number one.
In what follows, the world-sheet ghost number $g$ of a bosonic string field is indicated by a subscript
as in $\Psi_g$.
Owing to the ghost number anomaly, BPZ inner products of the form $\langle \Psi_{g_1} , \Psi_{g_2} \rangle$ vanish unless $g_1 + g_2 = 3$.
Note that in eq.~\eqref{free bosonic}, unlike in eq.~\eqref{free action}, $Q$ denotes the BRST operator for bosonic strings.

As is well-known, the extended action $S$ in this theory is given by~\cite{Baulieu, Bochicchio1, Bochicchio2, Thorn, BS}
\begin{equation}
S = -\frac{1}{2} \langle \Psi , Q\Psi \rangle 
= -\-\frac{1}{2} \langle \Psi_1 , Q\Psi_1 \rangle - \sum_{g\leq 0}\,\langle \Psi_{2-g} , Q\Psi_g \rangle \,.
\label{S bosonic}
\end{equation}
Here the symbol
\begin{equation}
\Psi := \sum^\infty_{g=-\infty} \Psi_g
\label{Psi}
\end{equation}
is the sum of the original field $\Psi_1$, all the ghosts $\Psi_g$ $(g\leq 0)$, and all the antighosts $\Psi_g$ $(g\geq 2)$.
Every string field on the right-hand side of eq.~\eqref{Psi} is Grassmann odd.
The action \eqref{S bosonic} is invariant under the following gauge transformations:
\begin{equation}
\Psi_g \ \longrightarrow \ \Psi_g + Q \Lambda_{g-1} \quad \left( g\in \mathbb{Z}\right).
\label{bosonic gt}
\end{equation}
In order to fix the gauges, we impose on each string field a condition of the form
\begin{equation}
\calB_{(g)} \Psi_g = 0 \quad \left( g\in \mathbb{Z} \right),
\label{BPsi}
\end{equation}
where $\calB_{(g)}$ is some Grassmann-odd nilpotent operator of world-sheet ghost number minus one.
We have specified by the subscript $(g)$ that $\calB_{(g)}$ acts upon $\Psi_g$.
BRST-invariant gauge-fixed actions can be constructed in the same manner that was explained in section~\ref{ex action}.
We impose the conditions on antighosts $\Psi_g$ $(g\geq 2)$ directly, whereas
those on $\Psi_1$ and ghosts $\Psi_g$ $(g\leq 0)$ are achieved indirectly by introducing the gauge-fixing term
\begin{equation}
S^\GF = - \sum_{g\leq 1}\, \langle N_{4-g} , \calB_{(g)} \Psi_g \rangle 
= \sum_{g\leq 1}\, \langle \calB^\star_{(g)} N_{4-g} , \Psi_g \rangle \,.
\end{equation}
Here $N_{4-g}$ are Grassmann-odd auxiliary fields, and $\star$ denotes the BPZ conjugation.
Needless to say, in order to remove the redundant symmetries ($\Theta$'s are parameters)
\begin{equation}
N_{4-g} \ \longrightarrow \ N_{4-g} + \calB^\star_{(g)} \Theta_{5-g} \quad \left( g\leq 1\right),
\end{equation}
one has to restrict the space of $N$'s; however, we do not specify the restrictions.
The completely gauge-fixed action
\begin{equation}
S + S^\GF = -\frac{1}{2} \langle \Psi_1 , Q\Psi_1 \rangle - \sum_{g\leq 0}\,\langle \Psi_{2-g} , Q\Psi_g \rangle 
+ \sum_{g\leq 1}\, \langle \calB^\star_{(g)} N_{4-g} , \Psi_g \rangle
\label{bosonic gf action}
\end{equation}
is invariant under the BRST transformation of the form
\bs
\begin{align}
\dB \Psi_g &= Q\Psi_{g-1} \quad \left( g\leq 1\right),
\label{ghost brst} \\[.5ex]
\dB \Psi_g &= \calB^\star_{(3-g)} N_{g+1} \quad \left( g\geq 2\right),\\[.5ex]
\dB N_g &= 0 \quad \left( g\geq 3\right).
\end{align}
\es
We would like to remark that the expression \eqref{bosonic gf action} can be rewritten as
\begin{equation}
S+S^\GF = -\frac{1}{2} \langle \Psi_1 , Q\Psi_1 \rangle 
+ \dB \sum_{g\leq 0}\, \braket{\Psi_{2-g} , \Psi_{g+1}} \,.
\end{equation}
This time, compatibility between the constraints on and the BRST transformations of antighosts requires
\begin{equation}
\calB_{(g)}\hs \calB^\star_{(3-g)} = 0\quad \left( g\geq 2\right),
\label{BBstar}
\end{equation}
which implies
\begin{equation}
\calB_{(g)} = \calB^\star_{(3-g)} \quad \left( g\geq 2\right).
\label{B constraint}
\end{equation}
Thus, if $\calB_{(g)}$ with $g\leq 1$ are given, then the other $\calB$'s are automatically determined.

In ref.~\cite{b-gauges}, a class of gauges fulfilling the requirements \eqref{B constraint} is explored.
There, $\calB$'s are composed of linear combinations of $b$-modes, weighted by vector fields $v^{(g)}(z)$:
\begin{equation}
\calB_{(g)} = \sum^\infty_{k=-\infty} v^{(g)}_k\, b_k = \oint_\Cu \frac{\diff z}{2\pi \mrm{i}}\, v^{(g)}(z)\hs b(z)\quad\left( g\leq 1\right)
\quad\text{with}\quad v^{(g)}(z) = \sum^\infty_{k=-\infty} v^{(g)}_k z^{k+1}\,.
\label{linear b-gauges}
\end{equation}
Here $\Cu$ denotes the counterclockwise unit circle centered at the origin. We have assumed that
the Laurent expansions of $v^{(g)}(z)$ are valid on $\Cu$.
In order for these linear $b$-gauges to be physically reasonable,
$v^{(g)}$ have to obey some restraints explained below.\\[1ex]
{\bf (i) Compatibility with the reality condition}\\*[.5ex]
\indenths
In order to guarantee the reality of the open string field theory action, one has to impose a certain condition 
on each string field~\cite{tensor const}.
It requires that string fields $\Psi_g$ be invariant under the composition of Hermitian conjugation (hc) and
the inverse BPZ conjugation ($\mrm{bpz}^{-1}$):\footnote{For pedagogical explanation of the reality condition, 
see appendix C of ref.~\cite{a-gauge}.}
\begin{equation}
\Psi^\ddagger_g :=  \mrm{bpz}^{-1} \circ \mrm{hc} \left(\Psi_g\right) = \Psi_g \quad \left(g\in\mathbb{Z}\right).
\end{equation}
Under these constraints, eq.~\eqref{BPsi} implies 
\begin{equation}
\bigl( \calB_{(g)} \Psi_g \bigr)^\ddagger = 0\,, \quad \text{i.e.,}\quad \calB^\ddagger_{(g)} \Psi_g = 0 \quad \left( g\in\mathbb{Z}\right).
\end{equation}
Thus, in order for the gauges \eqref{BPsi} to be compatible with the reality condition, $\calB$'s have to satisfy
\begin{equation}
\calB^\ddagger_{(g)} = \alpha_{(g)}\, \calB_{(g)}\ \iff \ 
\sum^\infty_{k=-\infty} \overline{v^{(g)}_k}\, (-1)^k\hs b_k \,= \,\alpha_{(g)} \hspace{-.5em} \sum^\infty_{k=-\infty} v^{(g)}_k b_k \ \iff \ 
(-1)^k\, \overline{v^{(g)}_k} = \alpha_{(g)}\hs v^{(g)}_k \;\, \left( \forall k\in\mathbb{Z} \right)
\label{propto}
\end{equation}
for some \,$\alpha_{(g)}\in \mathbb{C} \setminus \{0\}$.
Here we have used
\begin{equation}
\mrm{hc}\hs (b_k) = b_{-k}\,,\quad \bpz{b_k} = (-1)^k\hs b_{-k} \,,
\end{equation}
and have denoted by $\overline{v^{(g)}_k}$ the complex conjugate of $v^{(g)}_k$.
The last equation in \eqref{propto} tells us that the absolute values of $\alpha_{(g)}$ are equal to one:
\begin{equation}
(-1)^k\, \overline{v^{(g)}_k} = \alpha_{(g)}\hs v^{(g)}_k \;\, \left( \forall k\in\mathbb{Z} \right) \ \Longrightarrow \ 
| v^{(g)}_k | = |\alpha_{(g)} |\hs | v^{(g)}_k | \quad \left( \forall k\in\mathbb{Z} \right) \ \iff \ 
|\alpha_{(g)}| = 1\,.
\end{equation}
{\bf (ii) Validity of the Riemann surface interpretation and the Schwinger representation}\\*[.5ex]
\indenths
The equations of motion derived from the action \eqref{S bosonic} are of the form
\begin{equation}
Q \Psi_g = 0\quad \left( g\in\mathbb{Z} \right).
\label{eom bosonic}
\end{equation}
From eqs.~\eqref{BPsi} and \eqref{eom bosonic} it follows that
\begin{equation}
\calL_{(g)} \Psi_g = \{Q , \calB_{(g)}\}\hs \Psi_g = 0 \quad \text{with}\quad
\calL_{(g)} := \{ Q, \calB_{(g)} \} \quad\left( g\in \mathbb{Z}\right).
\end{equation}
Consequently, on the mass shell of $\Psi_g$, we have $\calL_{(g)} = 0$, which implies that the propagator involves the inverse of $\calL_{(g)}$.\\
\indenths
In the case of the Siegel gauge, in which $\calB_{(g)} = \bz$ $(\forall g\in\mathbb{Z})$ and thus $\calL_{(g)} = L_0$,
we may utilize the Schwinger representation
\begin{equation}
\frac{1}{L_0} = \lim_{\Lambda \to \infty} \int^\Lambda_0 \diff s\, \e^{-sL_0} 
\end{equation}
because for states of positive conformal weight, we have
\begin{equation}
L_0 \int^\Lambda_0 \diff s\, \e^{-s L_0} = 1-\e^{-\Lambda L_0} \to 1 \quad \left( \Lambda \to \infty\right).
\end{equation}
The merit of the above representation is that Feynman diagrams in string field theory can be naturally interpreted 
as correlation functions on Riemann surfaces, with insertions of the operators $\e^{-sL_0}$ providing parts of the surfaces.
We would like to apply this interpretation also to the case of general linear $b$-gauges,
in which we shall consider the operators
\begin{equation}
\calL_{(g)} = \{ Q, \calB_{(g)} \} = \oint_\Cu \frac{\diff z}{2\pi \iu}\, v^{(g)}(z)\, T(z)
= \int_{C_s} \biggl( \frac{\diff z}{2\pi \iu}\, v^{(g)}(z)\, T(z) - \frac{\diff \zbar}{2\pi \iu}\, v^{(g)}(\zbar)\, T(\zbar) \biggr).
\end{equation}
Here $C_s$ denotes the origin-centered counterclockwise unit semicircle on the upper half-plane, and 
$T$ denotes the energy-momentum tensor satisfying
\begin{equation}
\overline{T(z)} = T(\zbar) 
\label{T}
\end{equation}
with the bars representing complex conjugation.

First, in order that $\calL_{(g)}$ may generate conformal transformations and thus insertions of the form $\e^{-s\calL_{(g)}}$ 
may fit into the Riemann surface interpretation, the vector fields $v^{(g)}(z)$ have to satisfy the conditions corresponding to \eqref{T}:
\begin{equation}
\overline{v^{(g)}(z)} = v^{(g)}(\overline{z}) \,.
\label{doubling}
\end{equation}
As a necessary result, all the coefficients $v^{(g)}_k$ are real, and we may assume 
$v^{(g)}_0 \geq 0$ \hs $(\forall g\leq 1)$ without loss of generality.
(If necessary, we can multiply some of the equations \eqref{BPsi} by minus one.)

Second, in order for the Schwinger representation of ${1}/{\calL_{(g)}}$ to be valid,
the operator $\e^{-\Lambda_{(g)}\calL_{(g)}}$ in
\begin{equation}
\calL_{(g)} \int^{\Lambda_{(g)}}_0 \diff s\, \e^{-s\calL_{(g)}} = 1 - \e^{-\Lambda_{(g)}\calL_{(g)}}
\end{equation}
has to vanish in the $\Lambda_{(g)}\to \infty$ limit. 
It is shown in ref.~\cite{b-gauges} that if the vector field $v^{(g)}$ obeying the constraints \eqref{propto} and \eqref{doubling} 
is analytic in some neighborhood of the unit circle $|z| =1$ and satisfies
\begin{equation}
v^{(g)}_\perp (z) := \Re\bigl[\overline{z}\,v^{(g)}(z)\bigr] > 0\quad \text{for}\quad |z| =1\,,
\label{properness}
\end{equation}
then the term $\e^{-\Lambda_{(g)}\calL_{(g)}}$ vanishes in the $\Lambda_{(g)}\to \infty$ limit, and the representation
\begin{equation}
\frac{1}{\calL_{(g)}} = \lim_{\Lambda_{(g)}\to \infty} \int^{\Lambda_{(g)}}_0 \diff s\, \e^{-s\calL_{(g)}} 
\end{equation}
provides a proper definition of ${1}/{\calL_{(g)}}$.\footnote{Condition \eqref{properness} is \emph{sufficient}
but \emph{may not be necessary}.}
Substituting $z=\e^{\mrm{i}\theta}$ $(\theta \in \mathbb{R})$ into eq.~\eqref{properness}, we find that $v^{(g)}_0$ should be positive:
\begin{equation}
0 < \frac{1}{2\pi} \int^{2\pi}_0 \diff \theta\ v^{(g)}_\perp (\e^{\mrm{i}\theta}) = 
\int^{2\pi}_0 \diff\theta\, \Bigl( v^{(g)}_0 + \sum_{k\neq 0} v^{(g)}_{k} \cos(k\theta) \Bigr) = v^{(g)}_0 \,.
\label{v_0}
\end{equation}
Note that eq.~\eqref{propto} and the positivity of $v^{(g)}_0 \in \mathbb{R}\,$ entail $\alpha_{(g)} = 1$, so that 
taking account of the reality of $v^{(g)}_k$, we obtain
\begin{equation}
v^{(g)} (z) = \sum^\infty_{k=-\infty} v^{(g)}_{2k} z^{2k+1}\,,\quad \text{i.e.,}\quad
v^{(g)}(-z) = -v^{(g)}(z)\,.
\label{v_2k}
\end{equation}
\ \\

In summary, we require that $v^{(g)}$ be analytic in some neighborhood of the unit circle $|z|=1$ and satisfy
conditions \eqref{doubling}, \eqref{properness}, and \eqref{v_2k}:
\bs \label{v}
\begin{align}
\overline{v^{(g)}(z)} &= v^{(g)}(\overline{z}) \,,\quad v^{(g)}(-z) = -v^{(g)}(z)\,,
\label{over v} \\[.5ex]
v^{(g)}_\perp (z) &:= \Re\bigl[\overline{z}\,v^{(g)}(z)\bigr] > 0\quad \text{for}\quad |z| =1\,.
\label{Rev}
\end{align}
\es
These requirements are consistent with the constraints \eqref{B constraint}.
To see this, we use the expression
\begin{equation}
\calB^\star_{(g)} = \oint_\Cu \frac{\diff z}{2\pi \mrm{i}}\, \bigl( v^{(g)}\bigr)^\star (z)\hs b(z)
\quad \text{with}\quad \bigl( v^{(g)}\bigr)^\star (z) := -z^2 v^{(g)}(-1/z)\,.
\end{equation}
It can be readily confirmed that $\bigl( v^{(g)}\bigr)^\star (z)$ fulfills condition \eqref{over v} if and only if $v^{(g)}(z)$ does.
Furthermore, under condition \eqref{over v}, we have
\begin{equation}
\zbar\, \bigl( v^{(g)}\bigr)^\star (z) = -z\hs v^{(g)}(-1/z) = z\hs v^{(g)} (1/z) = z\hs v^{(g)}(\zbar) = \overline{\hs\zbar\, v^{(g)}(z)}
\quad \text{for}\quad |z| =1\,.
\label{vstar}
\end{equation}
This equation, together with \eqref{Rev}, ensures that $\bigl( v^{(g)}\bigr)^\star (z)$ indeed obeys the condition corresponding to \eqref{Rev}:
\begin{equation}
\Re\bigl[\zbar\,\bigl( v^{(g)}\bigr)^\star (z)\bigr] = \Re\bigl[\zbar\,v^{(g)}(z)\bigr] > 0\quad
\text{for}\quad |z| =1\,.
\end{equation}

Here is an additional remark. 
The analyticity of $v^{(g)}$ in some neighborhood of the unit circle guarantees that 
each vector field $v^{(g)}(z)$ is analytic in some ring domain
\begin{equation}
R(r_{(g)},r'_{(g)}) := \Set{z\in \mathbb{C}| 0< r_{(g)} < |z| < r'_{(g)}}\quad \bigl( r_{(g)} < 1 < r'_{(g)}\bigr),
\end{equation}
and thus the Laurent expansion 
\begin{equation}
v^{(g)}(z) = \sum^\infty_{k=-\infty} v^{(g)}_k z^{k+1}
\end{equation}
is valid there.
However, the same Laurent expansion cannot necessarily be utilized for $0 < |z| < r_{(g)}$: the vector field may have some
singularities.

In ref.~\cite{b-gauges}, gauge-fixing conditions meeting the requirements \eqref{v} are said to be regular. 
The Siegel gauge, under which $v^{(g)}(z) = z$ $(\forall g\in\mathbb{Z})$, is obviously regular.
However, as we shall see, the Schnabl gauge is not.

\subsubsection{Regularization of the Schnabl gauge}
\label{regularization}
\indenths
The Schnabl gauge is characterized by the zero mode $\hb_0$ of the $b$-ghost in the sliver frame~\cite{sliver}:
\begin{equation}
\hb_0 := \oint_\Cu \frac{\diff \hz}{2\pi\mrm{i}}\, \hz\, \hb(\hz) \,,\quad 
\hz := f(z) = \frac{2}{\pi} \arctan (z) \,,
\label{hb_0}
\end{equation}
where
\begin{equation}
\hb(\hz) = \left( \frac{\diff \hz}{\diff z} \right)^{-2} b(z)
\label{hb}
\end{equation}
is the conformal transform of $b$ by the function $f$.
Expressing the integral \eqref{hb_0} in terms of $z$, we obtain
\begin{equation}
\hb_0 = \oint_\Cu \frac{\diff z}{2\pi\mrm{i}}\, \frac{f(z)}{f'(z)}\, b(z) = \oint_\Cu \frac{\diff z}{2\pi\mrm{i}}\,(1+z^2) \arctan (z)\, b(z) \,.
\end{equation}
Therefore, if we choose $\hb_0$ as $\calB_{(1)}$, then the corresponding vector field $v^{(1)} = \hv$ takes the form
\begin{equation}
\hv (z) = \frac{f(z)}{f'(z)} = (1+z^2) \arctan (z) \,,
\end{equation}
whose non-vanishing coefficients are
\begin{equation}
\hv_0 = 1 \,,\quad
\hv_{2k} = \frac{2(-1)^{k+1}}{4k^2-1}\quad \left( k = 1,2,...\right).
\end{equation}
One should note that this gauge is \emph{not} regular because it does not satisfy condition \eqref{Rev}
for $z = \pm\hs\iu\hs$:
\begin{equation}
\hv (\pm\hs\iu) = \pm\hs\iu\Bigl( \hv_0 + \sum^\infty_{k=1} \hv_{2k} (-1)^k \Bigr)
=\pm\hs\iu\Bigl( 1- \sum^\infty_{k=1}\frac{2}{4k^2 -1} \Bigr) = 0 \,.
\label{hv iu}
\end{equation}
However, there exists a one-parameter family of regular gauges $\calB_{(1)} = \hbl_0$ with $0<\lambda<\infty$
which realizes the Schnable gauge in the $\lambda \to 0$ limit:
\begin{equation}
\hbl_0 := \e^{\lambda L_0}\, \hb_0\, \e^{-\lambda L_0} \quad \left( 0<\lambda<\infty\right).
\label{hbl_0}
\end{equation}
Moreover, it approaches the Siegel gauge as $\lambda$ tends to infinity.
Indeed, because $\hb_0$ contains only the non-negative modes $b_{2k}$ $(k\geq 0)$,
the relation
\begin{equation}
\e^{\lambda L_0}\,b_{2k}\,\e^{-\lambda L_0} = \e^{-2k\lambda} b_{2k}
\label{b2k}
\end{equation}
ensures the existence of the limit, which is exactly $\bz$:
\begin{equation}
\lim_{\lambda\to\infty} \hbl_0 = \bz\,.
\end{equation}
From eqs.~\eqref{hbl_0} and \eqref{b2k}, we find that the vector fields $\hvl$ associated with $\hbl_0$
are related to $\hv$ by
\begin{equation}
\hvl_{2k} = \e^{-2k\lambda}\, \hv_{2k} \quad \left( k\geq 0\right).
\end{equation}
Thus we have
\begin{equation}
\hvl (z) = \sum^\infty_{k=-\infty} \hv_{2k}\, \e^{-2k\lambda} z^{2k+1} = \e^\lambda\,\hv (\e^{-\lambda}z)
= \frac{f^\lambda (z)}{\bigl( f^\lambda\bigr)' (z)} 
\label{hvl}
\end{equation}
with
\begin{equation}
f^\lambda (z) := f(\e^{-\lambda}z) = \frac{2}{\pi}\,\arctan(\e^{-\lambda}z) .
\label{f lambda}
\end{equation}
Consequently, the operator $\hbl_0$ can be regarded as the zero mode in the conformal frame defined by the map
$z \mapsto f^\lambda (z)$.
Regularity of the gauges follows from the inequality
\begin{equation}
|\hvl_\perp (z)| = \left| \Re\bigl[ \overline{z}\,\hvl (z)\bigr]\right|
\geq 1 - \sum^{\infty}_{k=1} |\hs\hvl_{2k}| > 1 - \sum^\infty_{k=1} |\hs \hv_{2k} | = 0\quad \left(\,|z|=1;\ \forall \lambda > 0\right).
\end{equation}

\subsubsection{Propagators in linear \texorpdfstring{$b$}{b}-gauges}
\indenths
Let us calculate propagators under conditions \eqref{BPsi} and \eqref{B constraint}.
For this purpose, we rewrite the action \eqref{S bosonic} as
\begin{equation}
S = -\frac{1}{2}\braket{\Psi_1 , Q\Psi_1}
-\frac{1}{2}\sum_{g\leq 0}\, 
\left\langle 
\bbm
\Psi_g \\[.1ex]
\Psi_{2-g}
\ebm
,
\bbm
0 & Q \\[.1ex]
Q & 0
\ebm
\bbm
\Psi_g \\[.1ex]
\Psi_{2-g}
\ebm
\right\rangle
,
\end{equation}
and seek the BPZ-even left inverses of $Q$ and 
$
\bbm
0 & Q \\[.1ex]
Q & 0
\ebm
$
in the restricted subspaces
\begin{align}
\RS_1 &:= \Set{\Psi_1 | \calB_{(1)}\Psi_1 = 0 }\,,\\[1.7ex]
\RS_g &:= \Set{ 
\bbm
\Psi_g \\[.1ex]
\Psi_{2-g}
\ebm
|
\calB_{(g)} \Psi_g =0\,,\quad \calB_{(g+1)}^\star \Psi_{2-g} = 0 } \quad \left( g\leq 0\right).
\end{align}
We would like to remark that the BPZ-even left inverses are automatically the right inverses. 

First, we consider the $\Psi_1$-propagator $\Delta_1$, which is the inverse of $Q$ in the subspace $\RS_1$.
Using the identity
\begin{equation}
\biggl( \frac{\calB_{(1)}}{\calL_{(1)}}\, Q\, \frac{\calB_{(1)}^\star}{\calL_{(1)}^\star}\biggr) Q
= \frac{\calB_{(1)}}{\calL_{(1)}}\biggl( Q\, \frac{\calB_{(1)}^\star}{\calL_{(1)}^\star}\,Q\biggr)
= \frac{\calB_{(1)}}{\calL_{(1)}}\,Q 
= 1- Q\,\frac{\calB_{(1)}}{\calL_{(1)}}\,,
\end{equation}
we find with ease that
\begin{equation}
\Delta_1 = \frac{\calB_{(1)}}{\calL_{(1)}}\, Q\, \frac{\calB_{(1)}^\star}{\calL_{(1)}^\star}\,.
\label{Delta_1}
\end{equation}
Next, in order to obtain the ghost propagator $\Delta_g$ between $\Psi_g$ and $\Psi_{2-g}$ $\left(g\leq 0\right)$,
we consider the BPZ-even left inverse of the matrix 
\begin{equation}
\bbm
0 & Q \\[.1ex]
Q & 0
\ebm
=
\bbm
0 & Q \\[.1ex]
-\bpz{Q} & 0
\ebm
.
\end{equation}
We may suppose that it takes the form
\begin{equation}
\bbm
0 & \Delta_g \\[.1ex]
\bpz{\Delta_g} & 0
\ebm
.
\label{Delta_g}
\end{equation}
The matrix \eqref{Delta_g} is determined by the equation
\begin{equation}
\bbm
0 & \Delta_g \\[1ex]
\bpz{\Delta_g} & 0
\ebm
\bbm
0 & Q \\[1ex]
Q & 0
\ebm
=
\bbm
1 & 0 \\[1ex]
0 & 1 
\ebm
+
\bbm
\calM_{(g)} \calB_{(g)} & 0\\[1ex]
0 & \calM_{(2-g)}\calB_{(g+1)}^\star
\ebm
,
\end{equation}
the right-hand side of which acts as the unit matrix in $\RS_g$.
Here $\calM_{(g)}$ and $\calM_{(2-g)}$ are some operators.
The following two identities
\bs
\begin{align}
\biggl( \frac{\calB_{(g)}}{\calL_{(g)}}\, Q\, \frac{\calB_{(g+1)}}{\calL_{(g+1)}}\biggr) Q
&= \frac{\calB_{(g)}}{\calL_{(g)}}\biggl( Q\, \frac{\calB_{(g+1)}}{\calL_{(g+1)}}\,Q\biggr)
= \frac{\calB_{(g)}}{\calL_{(g)}}\,Q 
= 1- Q\,\frac{\calB_{(g)}}{\calL_{(g)}}\,, \\[1.8ex]
\mrm{bpz}\biggl( \frac{\calB_{(g)}}{\calL_{(g)}}\, Q\, \frac{\calB_{(g+1)}}{\calL_{(g+1)}}\biggr) Q
&= \frac{\calB_{(g+1)}^\star}{\calL_{(g+1)}^\star}\biggl( Q\, \frac{\calB_{(g)}^\star}{\calL_{(g)}^\star}\,Q\biggr)
= \frac{\calB_{(g+1)}^\star}{\calL_{(g+1)}^\star}\,Q 
= 1- Q\,\frac{\calB_{(g+1)}^\star}{\calL_{(g+1)}^\star}\,,
\end{align}
\es
tell us that 
\begin{equation}
\Delta_g = \frac{\calB_{(g)}}{\calL_{(g)}}\, Q\, \frac{\calB_{(g+1)}}{\calL_{(g+1)}} \quad \left( g\leq 0\right).
\label{bosonic Delta}
\end{equation}
In particular, if $\calB_{(g)}=\calB_{(g+1)}$, then we obtain
\begin{equation}
\Delta_g = \frac{\calB_{(g)}}{\calL_{(g)}} \quad \left(\text{for $\calB_{(g)}=\calB_{(g+1)}$}\right).
\label{reduced Delta}
\end{equation}

It should be noted that the structures of the propagators \eqref{bosonic Delta} and \eqref{reduced Delta} are similar to 
those of \eqref{Delta} and \eqref{simple D} in the superstring field theory.
In fact, the matrices $\bsymb{Q}$, $\bsymb{B}$, and $\bsymb{P}$ in eqs.~\eqref{Delta} and \eqref{simple D}
correspond respectively to $Q$, $\calB$, and $1/\calL$ in eqs.~\eqref{bosonic Delta} and \eqref{reduced Delta}.
As we shall see in the following subsections, we can readily extend the results in the previous sections,
simply by replacing $\bz$'s and $\zz$'s in gauge-fixing matrices with
linear combinations of $b$-modes and of $\zeta$-modes.

\subsection{Extended gauge-fixing conditions}
\label{extended gfc}
\indenths
Replacing $\bz$'s and $\zz$'s in the matrices \eqref{B} and \eqref{tB}
with linear combinations of $b$-modes and of $\zeta$-modes,
we introduce the following new gauge-fixing matrices:
\begin{align}
\bsymb{B^{\calZ_{(n)};\hs x_n,\hs y_n}_{n+2,\hs n+1}}
&:=
\underbrace{
\left.
\bbm
\calB_{(n)} &0 &\hdots &\hdots &0 \\[.5ex]
\hline \\[-2ex]
y_n\calZ_{(n)} &x_n\calB_{(n)} &&&\hsymbu{0} \\[1ex]
&y_n\calZ_{(n)} &x_n\calB_{(n)} && \\[1ex]
&&\ddots &\ddots && \\[1ex]
\hsymbl{0} &&& y_n\calZ_{(n)} &x_n\calB_{(n)} \\[1ex]
\hline \\[-2ex]
0 &\hdots &\hdots &0& \calZ_{(n)}
\ebm
\right\}
}_{n+1}
n+2
\qquad
\left( n\geq 0\right),
\label{ext B} \\[2ex]
\bsymb{\tB^{\calZ_{(n)};\hs x_n,\hs y_n}_{n+2,\hs n+1}}
&:=
\underbrace{
\left.
\bbm
\calB_{(n)}^\star \calZ_{(n)}^\star &0 &\hdots &\hdots &0 \\[.8ex]
\hline \\[-2ex]
x_{n}\calB_{(n)}^\star &y_{n}\calZ_{(n)}^\star &&&\hsymbu{0} \\[1ex]
&x_{n}\calB_{(n)}^\star &y_{n}\calZ_{(n)}^\star && \\[1ex]
&&\ddots &\ddots && \\[1ex]
\hsymbl{0} &&& x_{n}\calB_{(n)}^\star &y_{n}\calZ_{(n)}^\star \\[1ex]
\hline \\[-2ex]
0 &\hdots &\hdots &0& \calB_{(n)}^\star \calZ_{(n)}^\star
\ebm
\right\}
}_{n+1}
n+2
\qquad
\left( n\geq 0\right).
\label{ext tB}
\end{align}
Here $x_n$ and $y_n$ are real parameters satisfying $(x_n,\hs y_n) \neq (0,\hs 0)$,
and $\calB_{(n)}$ are defined as
\begin{equation}
\calB_{(n)} := \oint_\Cu \frac{\diff z}{2\pi \mrm{i}}\, v^{(n)}(z)\, b(z)
= \sum^\infty_{k=-\infty} v^{(n)}_k\hs b_k 
\quad\text{with}\quad v^{(n)}(z) = \sum^\infty_{k=-\infty} v^{(n)}_k z^{k+1} \,. 
\label{calB}
\end{equation}
The other symbols $\calZ_{(n)}$ are utilized for denoting either $\Xi_{(n)}$ or $\calD_{(n)}$ given by
\begin{align}
\Xi_{(n)} &:= \oint_\Cu \frac{\diff z}{2\pi \mrm{i}}\, w^{(n)}(z)\, \xi(z)
= \sum^\infty_{k=-\infty} w^{(n)}_k\hs \xi_k 
\quad\text{with}\quad w^{(n)}(z) := \frac{\calN_{(n)}}{v^{(n)}(z)} = \sum^\infty_{k=-\infty} w^{(n)}_k z^{k-1} \,,
\label{Xi} \\[.5ex]
\calD_{(n)} &:= \oint_\Cu \frac{\diff z}{2\pi \mrm{i}}\, v^{(n)}(z)\, d(z)
= \sum^\infty_{k=-\infty} v^{(n)}_k\hs d_k \,.
\label{calD}
\end{align} 
Note that the identical vector fields $v^{(n)}(z)$ are used in the above definitions \eqref{calB}-\eqref{calD},
and $w^{(n)}(z)$ are essentially ${1}/{v^{(n)}(z)}$, whose normalization constants $\calN_{(n)} \in \mathbb{R}\hs $ are determined later.
Although one does not necessarily have to use the identical vector fields, we consider only the above forms of $\calB_{(n)}$, $\Xi_{(n)}$, 
and $\calD_{(n)}$ for simplicity.
In order to guarantee the reasonableness of the gauges, we impose the same restrictions on $v^{(n)}$ as in \eqref{v}:
$v^{(n)}$ should be analytic in some neighborhood of the unit circle $|z|=1$ and should satisfy
\bs \label{ext v}
\begin{align}
\overline{v^{(n)}(z)} &= v^{(n)}(\overline{z}) \,,\quad v^{(n)}(-z) = -v^{(n)}(z)
\quad \Bigl( \iff v^{(n)}(z) = \sum^\infty_{k=-\infty} v^{(n)}_{2k} z^{2k+1}\,,\quad v^{(n)}_{2k} \in \mathbb{R}\Bigr),
\label{ext conj v} \\*
v^{(n)}_\perp (z) &:= \Re\bigl[\overline{z}\,v^{(n)}(z)\bigr] > 0\quad \text{for}\quad |z| =1\,.
\label{ext Rev}
\end{align}
\es
These restrictions ensure the compatibility with the reality conditions, and the validity of the Riemann surface interpretation and
the Schwinger representation\footnote{The proof of the validity of the Schwinger representation given in ref.~\cite{b-gauges} 
can be applied also to the present case.} 
of ${1}/{\calL_{(n)}}$ with
\begin{equation}
\calL_{(n)} = \{ Q, \calB_{(n)}\} = \{\ez, \calD_{(n)} \} = \sum^\infty_{k=-\infty} v^{(n)}_k\hs L_k \,.
\end{equation}
We would like to remark that the reality conditions in the present case take the form\footnote{
In particular, the reality condition of $\Phi_{(0,\hs 0)}$ is investigated in detail in appendix~\ref{hc and reality}.}
\begin{equation}
\bsymb{\Phi^\ddagger_{-n}} := 
\bbm
\Phi_{(-n,\hs 0)}^\ddagger \\
\vdots \\
\Phi_{(-n,\hs n)}^\ddagger
\ebm
\propto
\bsymb{\Phi_{-n}}
\,,\quad
\bsymb{\Phi^\ddagger_{n+2}} := 
\bbm
\Phi_{(n+2,\hs -1)}^\ddagger \\
\vdots \\
\Phi_{(n+2,\hs -(n+1))}^\ddagger
\ebm
\propto
\bsymb{\Phi_{n+2}}
\qquad \left( n\geq 0\right).
\end{equation}

Here is another important consequence of the analyticity requirement on $v^{(n)}$ and condition \eqref{ext Rev}:
They imply that $v^{(n)}$ is nonzero in some neighborhood of the unit circle.
Thus the function $w^{(n)} \propto \frac{1}{v^{(n)}}$, also, is analytic in this neighborhood, which ensures the existence of 
its Laurent series valid on $\Cu$.

The gauge-fixing conditions characterized by the operators \eqref{calB}-\eqref{calD},
\bs \label{gfc ext}
\begin{align}
\bsymb{B^{\calZ_{(n)};\hs x_n,\hs y_n}_{n+2,\hs n+1} \Phi_{-n} } &=0\quad \left( n\geq 0\right),
\label{gfc ext B} \\*[1.5ex]
\bsymb{\tB^{\calZ_{(n)};\hs x_n,\hs y_n}_{n+2,\hs n+1} \Phi_{n+2}} &=0\quad \left( n\geq 0\right),
\label{gfc ext tB}
\end{align}
\es
satisfy the consistency condition corresponding to eqs.~\eqref{tBB} and \eqref{BtB}:
\begin{equation}
\bsymb{\tB^{\calZ_{(n)};\hs x_n,\hs y_n}_{n+2,\hs n+1}}\hs \bpz{\bsymb{B^{\calZ_{(n)};\hs x_n,\hs y_n}_{n+2,\hs n+1}}} = 0 
\quad
\left( \iff
\bsymb{B^{\calZ_{(n)};\hs x_n,\hs y_n}_{n+2,\hs n+1}}\hs \bpz{\bsymb{\tB^{\calZ_{(n)};\hs x_n,\hs y_n}_{n+2,\hs n+1}}} = 0 \right)
\quad \left( n\geq 0\right).
\label{ext tBB}
\end{equation}
This immediately follows from the equations below
\begin{equation}
\calB_{(n)}^2 = 0\,,\quad \calZ_{(n)}^2 = 0\,,\quad \{\calB_{(n)},\calZ_{(n)}\} = 0\,.
\label{BD}
\end{equation}
To obtain the last equality for $\calZ = \calD$, we have used eq.~\eqref{db}:
\begin{align}
&\ \{\calB_{(n)},\calD_{(n)}\} 
=\  \oint_\Cu \frac{\diff z_1}{2\pi \iu} \oint_\Cu \frac{\diff z_2}{2\pi \iu}\, v^{(n)}(z_1) v^{(n)}(z_2)
\Bigl( \frac{2}{z^2_{12}}\hs \xi b (z_2) + \frac{1}{z_{12}}\hs \p (\xi b) (z_2) \Bigr) \nonumber \\*[1ex]
=&\ \oint_\Cu \frac{\diff z}{2\pi \iu} \Bigl( 2\hs \p v^{(n)} v^{(n)}\, \xi b + v^{(n)} v^{(n)} \,\p(\xi b) \Bigr) (z)
= \oint_\Cu \frac{\diff z}{2\pi \iu}\, \p\Bigl( v^{(n)} v^{(n)}\, \xi b \Bigr)(z) = 0\,.
\end{align}

One should note that the gauge-fixing conditions \eqref{gfc ext} contain two types of gauges for each $n\geq 0$:
$\calB_{(n)}$-$\Xi_{(n)}$ gauges and $\calB_{(n)}$-$\calD_{(n)}$ gauges. 
As was emphasized at the end of section~\ref{gauge-fixing conditions}, one does not have to adopt the same type of gauge-fixing conditions
for every $n\geq 0$.

\subsubsection{\texorpdfstring{$\calB_{(n)}$-$\Xi_{(n)}$}{B(n)-Xi(n)} gauges}
\label{BXi gauges}
\indenths
For a reason to be mentioned later, we choose the normalization constant $\calN_{(n)}$ in eq.~\eqref{Xi} such that $w^{(n)}_0$ is set equal to one.
In order for this to be possible, in $\calB_{(n)}$-$\Xi_{(n)}$ gauges, 
we require\footnote{In $\calB_{(n)}$-$\calD_{(n)}$ gauges, we do not impose this constraint on $v^{(n)}$.}
\begin{equation}
\oint_\Cu \frac{\diff z}{2\pi \iu}\,\frac{1}{v^{(n)}(z)} \neq 0\,.
\label{nonzero Res}
\end{equation}
In this case, the constant $\calN_{(n)}$ is given by
\begin{equation}
\calN_{(n)} = \biggl( \oint_\Cu \frac{\diff z}{2\pi \iu}\,\frac{1}{v^{(n)}(z)} \biggr)^{-1},
\end{equation}
realizing the normalization
\begin{equation}
w^{(n)}_0 = \oint_\Cu \frac{\diff z}{2\pi \iu}\,w^{(n)}(z) = \oint_\Cu \frac{\diff z}{2\pi \iu}\,\frac{\calN_{(n)}}{v^{(n)}(z)} = 1\,.
\label{w normalization}
\end{equation}
Using this $w^{(n)}$, we define $\calC_{(n)}$ as
\begin{equation}
\calC_{(n)} := \oint_\Cu \frac{\diff z}{2\pi \iu}\,\frac{w^{(n)}(z)}{v^{(n)}(z)} \,c(z)
=\oint_\Cu \frac{\diff z}{2\pi \iu}\,\frac{\calN_{(n)}}{\bigl( v^{(n)}(z)\bigr)^2} \,c(z)\,.
\end{equation}
Under the above normalization \eqref{w normalization}, 
the operators $Q$, $\ez$, $\calB_{(n)}$, $\calC_{(n)}$, and $\Xi_{(n)}$ satisfy the same algebraic relations 
as $Q$, $\ez$, $\bz$, $\cz$, and $\xz$:
\bs \label{calB calC Xi}
\begin{align}
Q^2 &= \ez^2 = \calB_{(n)}^2 = \calC_{(n)}^2 = \Xi_{(n)}^2 = 0\,,\\[.5ex]
\{ Q,\ez\} &= \{\ez, \calB_{(n)}\} = \{\ez,\calC_{(n)}\} = \{\calB_{(n)},\Xi_{(n)}\} = \{\calC_{(n)}, \Xi_{(n)}\} = 0\,,\\[.5ex]
\{\calB_{(n)}, \calC_{(n)}\} &= \{\ez, \Xi_{(n)}\} = 1\,,\\[.5ex]
\{ Q, \calB_{(n)}\} &= \calL_{(n)}\,,\quad \{Q, \Xi_{(n)} \} =: \calX_{(n)} \,,\\[.5ex]
[\calL_{(n)}, Q] &= [\calL_{(n)}, \ez] = [\calL_{(n)}, \calB_{(n)}] = [\calL_{(n)}, \calC_{(n)}] = [\calL_{(n)}, \Xi_{(n)}] = 0\,,\\[.5ex]
[\calX_{(n)}, Q] &= [\calX_{(n)}, \ez] = [\calX_{(n)}, \calB_{(n)}] = [\calX_{(n)}, \Xi_{(n)}] = 0\,,
\end{align}
\es
which can be shown in the manner explained below eq.~\eqref{BD}. 
The reason why we have adjusted $w^{(n)}_0$ to one is to achieve these relations.
When we proved the validity of conditions \eqref{gfc for ghosts} and \eqref{gfc for antighosts} for $\zz = \xz$,
we used only the algebraic relations among $Q$, $\ez$, $\bz$, $\cz$, and $\xz$. 
Hence, for $\calZ_{(n)} = \Xi_{(n)}$, the corresponding relations \eqref{calB calC Xi} and their BPZ conjugates suffice to ensure 
the validity of condition \eqref{gfc ext B} and that of \eqref{gfc ext tB}, respectively, \emph{if} $x_n = 0$ or $y_n = 0$.
However, if $x_n$ and $y_n$ are nonzero, there may exist a problem
related to the inverse of the operator $x_n \calL_{(n)} + y_n$, which is the counterpart of $x_n L_0 + y_n$ in $\bz$-$\xz$ gauges:
we may not be able to take the well-defined inverse of $x_n \calL_{(n)} + y_n$.
In order to avoid this potential complexity, in $\calB_{(n)}$-$\Xi_{(n)}$ gauges,
we will restrict ourselves within the cases in which $x_n = 0$ or $y_n = 0$. 
 
\subsubsection{\texorpdfstring{$\calB_{(n)}$-$\calD_{(n)}$}{B(n)-D(n)} gauges}
\indenths
In $\calB_{(n)}$-$\calD_{(n)}$ gauges, we have the following algebraic relations
among $Q$, $\ez$, $\calB_{(n)}$, and $\calD_{(n)}$, without requiring condition \eqref{nonzero Res}.
\bs \label{calB calD}
\begin{align}
Q^2 &= \ez^2 = \calB_{(n)}^2 = \calD_{(n)}^2 = 0\,,\\[.5ex]
\{ Q,\ez\} &= \{Q, \calD_{(n)}\} = \{\ez, \calB_{(n)}\} = \{\calB_{(n)} ,\calD_{(n)}\} = 0\,,\\[.5ex]
\{ Q, \calB_{(n)}\} &= \{\ez ,\calD_{(n)}\} = \calL_{(n)}\,,\\[.5ex]
[\calL_{(n)}, Q] &= [\calL_{(n)}, \ez] = [\calL_{(n)}, \calB_{(n)}] = [\calL_{(n)}, \calD_{(n)}] = 0\,.
\end{align}
\es
These relations and their BPZ conjugates are the same as the relations among $Q$, $\ez$, $\bz$, and $\dz$;
hence the validity of conditions \eqref{gfc ext B} and \eqref{gfc ext tB} for $\calZ_{(n)} = \calD_{(n)}$ is guaranteed.
In $\calB_{(n)}$-$\calD_{(n)}$ gauges, for the same reason explained in section~\ref{on-shell},
the parameters $x_n$ and $y_n$ have to satisfy $x_n + y_n \neq 0$. 

\subsubsection{A one-parameter family of regular gauges}
\indenths
Let us consider the one-parameter family of vector fields \eqref{hvl} in the case of $\calB_{(n)}$-$\calZ_{(n)}$ gauges.
The vector field $\hvl$ $\left( 0<\lambda <\infty\right)$ satisfies
\begin{equation}
\oint_\Cu \frac{\diff z}{2\pi \iu}\,\frac{1}{\hvl (z)} = 1\,.
\end{equation}
Therefore, if we set $v^{(n)} = \hvl$, we have $\calN_{(n)}=1$ and find that the operators
\begin{align}
\calB_{(n)} &= \hbl_0 = \oint_\Cu \frac{\diff z}{2\pi \mrm{i}}\, \hvl (z)\, b(z) \,,\\[1ex]
\Xi_{(n)} &= \hxil_0 := \oint_\Cu \frac{\diff z}{2\pi \mrm{i}}\, \frac{1}{\hvl (z)}\, \xi(z) \,,\\[1ex]
\calD_{(n)} &= \hdl_0 := \oint_\Cu \frac{\diff z}{2\pi \mrm{i}}\, \hvl (z)\, d(z)
\end{align}
are precisely the zero modes in the conformal frame associated with the function $f^\lambda$ in eq.~\eqref{f lambda}.
We have already investigated $\hbl_0$ in detail in section~\ref{regularization}: it interpolates between the canonical zero mode $\bz$ 
and the zero mode in the sliver frame $\hb_0$,
providing a regularization of the latter.
It can be confirmed that the same applies to $\hxil_0$ and $\hdl_0$, by noting that
\begin{equation}
\hvl (z) \to \left\{
\begin{aligned}
&z\quad \left( \lambda \to \infty\right) ,\\
&\hv (z) \quad \left( \lambda \to 0\right).
\end{aligned}
\right.
\end{equation}
However, the limit 
\begin{equation}
\lim_{\lambda\to 0} \hxil_0 = \lim_{\lambda\to 0} \oint_\Cu \frac{\diff z}{2\pi \mrm{i}}\, \frac{1}{\hvl (z)}\, \xi(z) 
\end{equation}
needs careful treatment.
To see this, we consider the naive zero modes in the sliver frame
\begin{equation}
\hb_0 = \oint_\Cu \frac{\diff z}{2\pi \mrm{i}}\, \hv (z)\, b(z)\,,\quad 
\hxi_0 = \oint_\Cu \frac{\diff z}{2\pi \mrm{i}}\, \frac{1}{\hv (z)}\, \xi(z) \,,\quad 
\hd_0 = \oint_\Cu \frac{\diff z}{2\pi \mrm{i}}\, \hv (z)\, d(z)\,.
\end{equation}
The point is that unlike $\hb_0$ and $\hd_0$, the operator $\hxi_0$ itself is singular:
the function $1/{\hv}$ diverges at $z = \pm\hs \iu$ (see eq.~\eqref{hv iu}), and thus there does not exist a Laurent expansion 
valid on the unit circle $\Cu$.
Hence one must not perform the interchange of the integral and the limit:
\begin{equation}
\lim_{\lambda\to 0} \oint_\Cu \frac{\diff z}{2\pi \mrm{i}}\, \frac{1}{\hvl (z)}\, \xi(z)
\ \neq\ \oint_\Cu \frac{\diff z}{2\pi \mrm{i}}\, \lim_{\lambda\to 0}\frac{1}{\hvl (z)}\, \xi(z)\,.
\end{equation}
By contrast, in the $\lambda\to\infty$ limit, there exists no such complexity.
The parameterized gauges are reduced exactly to the $\bz$-$\zz$ gauges \eqref{gfc} investigated in 
sections~\ref{gauge-fixing conditions}-\ref{propagators}.

\subsection{Extended propagators}
\label{ext propagators}
\indenths
Propagators in $\calB_{(n)}$-$\calZ_{(n)}$ gauges can be readily derived in the manner explained in section~\ref{propagators}.
For $\calB_{(n)}$-$\Xi_{(n)}$ gauges, we implicitly restrict ourselves within the cases in which $x_n = 0$ or $y_n = 0$.
(See the last two sentences in section~\ref{BXi gauges}.)

\subsubsection{Extended \texorpdfstring{$\bsymb{\Phi_0}$}{Phi0}-propagator}
\indenths
First, let us determine the $\bsymb{\Phi_0}$-propagator $\bsymb{\Delta_{1,0}}$.
All we have to do is to find the BPZ-even left inverse of $\bsymb{Q_{0,1}} = \iu\,Q\ez$ in the restricted subspace
\begin{equation}
\RS^\calZ_0 := \Set{ \bsymb{\Phi_0} | \bsymb{B^{\calZ_{(0)};\hs x_0,\hs y_0}_{2,\hs 1} \Phi_0} = 0 }\,.
\end{equation}
We define
\begin{equation}
\bsymb{K^{\Xi_{(0)};\hs x_0,\hs y_0}_{1,\hs 0}} := \iu\,\frac{\calB_{(0)}\hs\Xi_{(0)}}{\calL_{(0)}}\,,\quad
\bsymb{K^{\calD_{(0)};\hs x_0,\hs y_0}_{1,\hs 0}} := \iu\,\frac{\calB_{(0)}\calD_{(0)}}{\calL^2_{(0)}}\,.
\end{equation}
(Compare these operators with $\bsymb{K^{\xz;\hs x,\hs y}_{1,\hs 0}}$ and $\bsymb{K^{\dz;\hs x,\hs y}_{1,\hs 0}}$ 
in eqs.~\eqref{small K^xi} and \eqref{small K^d}.)
Then the following equation holds:
\begin{align}
\Bigl( 
\bsymb{K^{\calZ_{(0)};\hs x_0,\hs y_0}_{1,\hs 0}}\hs \bsymb{Q_{0,1}}\hs \mrm{bpz}\Bigl( \bsymb{K^{\calZ_{(0)};\hs x_0,\hs y_0}_{1,\hs 0}} \Bigr)
\Bigr)\hs \bsymb{Q_{0,1}}
&= 1 + \bsymb{M^{\calZ_{(0)};\hs x_0,\hs y_0}_{1,\hs 2}\hs B^{\calZ_{(0)};\hs x_0,\hs y_0}_{2,\hs 1}}\,,
\label{KQbpzK}
\end{align}
with
\bs \label{McalD}
\begin{align}
\bsymb{M^{\Xi_{(0)};\hs x_0,\hs y_0}_{1,\hs 2}} &:=
\bbm
-\Xi_{(0)}\ez \frac{Q}{\calL_{(0)}}\ &\  -\ez
\ebm
,\\[1.5ex]
\bsymb{M^{\calD_{(0)};\hs x_0,\hs y_0}_{1,\hs 2}} &:=
\bbm
-\Bigl( 1-\frac{y_0}{x_0 + y_0}\frac{\ez\calD_{(0)}}{\calL_{(0)}}\Bigr) \frac{Q}{\calL_{(0)}}\ 
&
\ -\Bigl( 1-\frac{x_0}{x_0 + y_0}\frac{Q\calB_{(0)}}{\calL_{(0)}}\Bigr) \frac{\ez}{\calL_{(0)}}
\ebm
.
\end{align}
\es
(Note that the matrices in eq.~\eqref{McalD} are similar to those in eqs.~\eqref{Mxz} and \eqref{Mdz}.)
The right-hand side of eq.~\eqref{KQbpzK} acts as the identity operator in $\RS^\calZ_0$.
Thus we conclude that $\bsymb{\Delta_{1,0}}$ is given by
\begin{equation}
\bsymb{\Delta_{1,0}} = 
\bsymb{K^{\calZ_{(0)};\hs x_0,\hs y_0}_{1,\hs 0}}\hs \bsymb{Q_{0,1}}\hs \mrm{bpz}\Bigl( \bsymb{K^{\calZ_{(0)};\hs x_0,\hs y_0}_{1,\hs 0}}\Bigr) 
= \left\{
\begin{aligned}
&-\iu\,\frac{\calB_{(0)}\hs\Xi_{(0)}}{\calL_{(0)}}\;Q\ez\,\frac{\calB^\star_{(0)}\hs\Xi^\star_{(0)}}{\calL^\star_{(0)}} 
\qquad \text{for}\quad \calZ_{(0)} = \Xi_{(0)}\,,\\[1ex]
&-\iu\,\frac{\calB_{(0)}\calD_{(0)}}{\calL^2_{(0)}}\;Q\ez\,\frac{\calB^\star_{(0)}\calD^\star_{(0)}}{\bigl(\calL^\star_{(0)}\bigr)^2}
\qquad \text{for}\quad \calZ_{(0)} = \calD_{(0)}\,.
\end{aligned}
\right.
\end{equation}
This expression is of the same form as in eq.~\eqref{Delta_1}.

\subsubsection{Extended ghost propagators}
\indenths
Next we determine ghost propagators $\bsymb{\Delta_{n+1,n}}$ $\left( n\geq 1\right)$.
In what follows, we denote $n$-independent versions of the operators 
$\calB_{(n)}$, $\Xi_{(n)}$, $\calD_{(n)}$, $\calX_{(n)}$, and $\calL_{(n)}$ by the symbols
$\calB$, $\Xi$, $\calD$, $\calX$, and $\calL$, respectively.
The vector field associated with them will be represented by $v(z)$.
Furthermore, we let untilded (resp.\ tilded) 
matrices with the superscript $\calZ_{(n)}$ denote those obtained from the untilded (resp.\ tilded) matrices with the superscript 
$\zz$ by replacing the zero modes $\bz$, $\xz$, $\dz$, $X_0$, and $L_0$
with $\calB_{(n)}$, $\Xi_{(n)}$, $\calD_{(n)}$, $\calX_{(n)}$, and $\calL_{(n)}$
(resp.\ $\calB^\star_{(n)}$, $\Xi^\star_{(n)}$, $\calD^\star_{(n)}$, $\calX^\star_{(n)}$, and $\calL^\star_{(n)}$).
Recall that the extended gauge-fixing matrices \eqref{ext B} and \eqref{ext tB}
were obtained from $\bsymb{B^{\zz;\hs x_n,\hs y_n}_{n+2,\hs n+1}}$ and $\bsymb{\tB^{\zz;\hs x_n,\hs y_n}_{n+2,\hs n+1}}$
by performing the above-mentioned replacement. Matrices with the superscript $\calZ$ (not $\calZ_{(n)}$)
should be interpreted similarly.

In section~\ref{extended gfc}, we have mentioned that in virtue of the relations \eqref{calB calC Xi} and \eqref{calB calD},
validity of $\calB_{(n)}$-$\calZ_{(n)}$ gauges can be proved in the same manner as that of $\bz$-$\zz$ gauges.
In fact, we find that the matrices
\begin{equation}
\bsymb{K^{\calZ;\hs x,\hs y}_{n+1,\hs n}} 
= \bsymb{P^{\calZ;\hs x,\hs y}_{n+1,\hs n+1}\hs B^{\calZ;\hs x,\hs y}_{n+1,\hs n}} \quad \left( n\geq 0\right),\quad
\bsymb{\tK^{\calZ;\hs x,\hs y}_{n,\hs n+1}} 
= \bsymb{\widetilde{P}^{\calZ;\hs x,\hs y}_{n,\hs n+2}\, \tB^{\calZ;\hs x,\hs y}_{n+2,\hs n+1}} \quad \left( n\geq 1\right),
\label{ext K and tK}
\end{equation}
together with the gauge-fixing matrices \eqref{ext B} and \eqref{ext tB}, satisfy the same relations as
in proposition~\ref{properties of K and tK}. Thus, the ghost propagator $\bsymb{\Delta_{n+1,n}}$ in extended gauges is given by
\begin{equation}
\bsymb{\Delta_{n+1,n}} =
\bsymb{K^{\calZ_{(n)};\hs x_n,\hs y_n}_{n+1,\hs n}\hs Q_{n,n+1}\hs K^{\calZ_{(n-1)};\hs x_{n-1},\hs y_{n-1}}_{n+1,\hs n}}
\quad \left( n\geq 1\right).
\label{ext Delta}
\end{equation}
Compare this result with eqs.~\eqref{Delta} and \eqref{bosonic Delta}.
If the vector fields and the parameters satisfy
\begin{equation}
v^{(n-1)}(z) = v^{(n)}(z) \left( = v(z) \right),\quad
x_{n-1} = x_n \left( = x\right), \quad y_{n-1} = y_n \left( = y\right),
\end{equation}
the above propagator is simplified to
\begin{equation}
\bsymb{\Delta_{n+1,n}} = \bsymb{K^{\calZ;\hs x,\hs y}_{n+1,\hs n}}\,.
\end{equation}
Before closing the present section, we list the concrete forms of $\bsymb{K^{\calZ;\hs x,\hs y}_{n+1,\hs n}}$.
They are the extensions of the matrices shown in section~\ref{some examples}.\\[1ex]
{\bf (I) \boldmath{$\calZ = \Xi$} case} \\*
\indenths
For $x=0$, we have
\bs
\begin{align}
\bsymb{K^{\Xi;\hs 0,\hs y}_{1,\hs 0}} &= \iu\,\frac{\calB\, \Xi}{\calL}\,,\quad
\bsymb{K^{\Xi;\hs 0,\hs y}_{2,\hs 1}} =
\bbm
\frac{\Xi\hs \ez\calB}{\calL} \\[2ex]
\Xi\bigl( 1 - \frac{\calX \ez\calB}{\calL}\bigr)
\ebm
=:
\bbm
\calE \\[2ex]
\calF
\ebm
,\quad
\bsymb{K^{\Xi;\hs 0,\hs y}_{3,\hs 2}} = 
\bbm
\calE &0 \\[.5ex]
\calF &0 \\[.5ex]
(-\calX) \calF &\Xi 
\ebm
,\\[2ex]
\bsymb{K^{\Xi;\hs 0,\hs y}_{n+1,\hs n}} &=
\bbm
\calE &0&0&0&\cdots &0&0 \\[.5ex]
\calF &0&0&0&\cdots &0&0 \\[.5ex]
(-\calX) \calF &\Xi &0 &0&\cdots &0&0 \\[1ex]
(-\calX)^2 \calF &(-\calX)\hs\Xi &\Xi &0&\cdots &0&0 \\[1ex]
(-\calX)^3 \calF &(-\calX)^2\hs\Xi &(-\calX)\hs\Xi &\Xi &\cdots &0&0 \\[.5ex] 
\vdots &\vdots &\vdots &\vdots &\vdots &\vdots &\vdots \\[.5ex]
(-\calX)^{n-2} \calF &(-\calX)^{n-3}\hs\Xi &(-\calX)^{n-4}\hs\Xi &(-\calX)^{n-5}\hs\Xi &\cdots &\Xi &0 \\[1ex]
(-\calX)^{n-1} \calF &(-\calX)^{n-2}\hs\Xi &(-\calX)^{n-3}\hs\Xi &(-\calX)^{n-4}\hs\Xi &\cdots &(-\calX)\hs\Xi &\Xi
\ebm
\quad \left( n\geq 1\right);
\end{align}
\es
and for $y=0$, we have
\bs
\begin{align}
\bsymb{K^{\Xi;\hs x,\hs 0}_{1,\hs 0}} &= \iu\,\frac{\calB\,\Xi}{\calL}\,,\quad
\bsymb{K^{\Xi;\hs x,\hs 0}_{2,\hs 1}} =
\bbm
\frac{\calB}{\calL} \\[1.5ex] 
\frac{\Xi\hs\calB Q}{\calL}
\ebm
,\quad
\bsymb{K^{\Xi;\hs x,\hs 0}_{3,\hs 2}} = 
\bbm
\frac{\calB}{\calL} &0 \\[1ex]
0 &\frac{\calB}{\calL} \\[1.5ex]
0 &\frac{\Xi\hs\calB Q}{\calL}
\ebm
,\\[2ex]
\bsymb{K^{\Xi;\hs x,\hs 0}_{n+1,\hs n}} &= 
\bbm
\frac{\calB}{\calL} &&&&\hsymbu{0} \\[1ex]
&\frac{\calB}{\calL} &&& \\[1ex]
&& \ddots && \\[1ex]
&&&\frac{\calB}{\calL} & \\[1ex]
&&&&\frac{\calB}{\calL} \\[1.5ex]
\hsymbl{0} &&&&\frac{\Xi\hs\calB\hs Q}{\calL}
\ebm
\quad \left( n\geq 1 \right).
\end{align}
\es
{\bf (II) \boldmath{$\calZ =\calD$} case} \\*
\bs
\begin{align}
\bsymb{K^{\calD;\hs x,\hs y}_{1,\hs 0}} &= \iu\,\frac{\calB\hs\calD}{\calL^2}\,,\quad
\bsymb{K^{\calD;\hs x,\hs y}_{2,\hs 1}} =
\bbm
K_\calB \\[1ex]
K_\calD
\ebm
,\quad
\bsymb{K^{\calD;\hs x,\hs y}_{3,\hs 2}} = 
\bbm
K_\calB &0 \\[1ex]
\frac{y\hs\calD}{\calL_{x,y}} &\frac{x\hs\calB}{\calL_{x,y}} \\[1ex]
0 &K_\calD
\ebm
,\\[2ex]
\bsymb{K^{\calD;\hs x,\hs y}_{n+1,\hs n}} &=
\bbm
K_\calB &&&&\hsymbu{0} \\[1.5ex]
\frac{y\hs\calD}{\calL_{x,y}} &\frac{x\hs\calB}{\calL_{x,y}} &&&\\[1.5ex]
&\frac{y\hs\calD}{\calL_{x,y}} &\frac{x\hs\calB}{\calL_{x,y}} &&\\[1.5ex]
&&\ddots &\ddots &\\[1.5ex]
&&&\frac{y\hs\calD}{\calL_{x,y}} &\frac{x\hs\calB}{\calL_{x,y}}\\[1.5ex]
\hsymbl{0} &&&&K_\calD
\ebm
\quad \left( n\geq 1\right),
\end{align}
\es
with
\begin{equation}
K_\calB := \frac{\calB}{\calL} \Bigl( 1-y\frac{\ez\calD}{\calL_{x,y}} \Bigr),\quad
K_\calD := \frac{\calD}{\calL} \Bigl( 1-x\frac{Q\hs\calB}{\calL_{x,y}} \Bigr),\quad
\calL_{x,y} := (x+y)\hs\calL\,.
\end{equation}

\section{Summary and discussion}
\label{summary}
\setcounter{equation}{0}
\indenths
In the present paper, we have made a detailed analysis on validity of gauge-fixing conditions and the structure of propagators
in the WZW-type open superstring field theory.
First, we have proposed a class of gauge-fixing conditions of the form
\bs \label{gfc summary}
\begin{align}
\bsymb{B^{\zz^{(n)};\hs x_n,\hs y_n}_{n+2,\hs n+1} \Phi_{-n} } &=0\quad \left( n\geq 0\right),
\label{gfc B summary} \\*[1ex]
\bsymb{\tB^{\zz^{(n)};\hs x_n,\hs y_n}_{n+2,\hs n+1} \Phi_{n+2}} &=0\quad \left( n\geq 0\right).
\label{gfc tB summary}
\end{align}
\es
Compatibility with the BRST transformation requires that the gauge-fixing matrices satisfy
\begin{equation}
\bsymb{\tB^{\zz^{(n)};\hs x_n,\hs y_n}_{n+2,\hs n+1}} \bpz{\bsymb{B^{\zz^{(n)};\hs x_n,\hs y_n}_{n+2,\hs n+1}}} = 0 
\quad \left( n\geq 0\right).
\label{tBB summary}
\end{equation}
Then, we have demonstrated validity of condition \eqref{gfc summary} by proving reachability and completeness.
In particular, the proof of the completeness was accomplished by showing the existence of the matrices $\bsymb{P^{\zz;\hs x,\hs y}_{n+1,\hs n+1}}$ 
$(n\geq 0)$ and $\bsymb{\widetilde{P}^{\zz;\hs x,\hs y}_{n,\hs n+2}}$ $(n\geq 1)$ such that 
\bs \label{P P anti summary}
\begin{align}
\bsymb{P^{\zz;\hs x,\hs y}_{n+1,\hs n+1}\hs B^{\zz;\hs x,\hs y}_{n+1,\hs n}\hs Q_{n,n+1}}
&=
\bsymb{1_{n+1}}
+
\bsymb{M^{\zz;\hs x\hs y}_{n+1,\hs n+2}\hs B^{\zz;\hs x,\hs y}_{n+2,\hs n+1} }
\quad \left( n\geq 0\right),
\label{P summary} \\[1ex]
\bsymb{\widetilde{P}^{\zz;\hs x,\hs y}_{n,\hs n+2}\hs \tB^{\zz;\hs x,\hs y}_{n+2,\hs n+1} Q_{n+1,n}}
&= \bsymb{1_n} + \bsymb{\tM^{\zz;\hs x,\hs y}_{n,\hs n+1} \tB^{\zz;\hs x,\hs y}_{n+1,\hs n} } \quad \left( n\geq 1 \right).
\label{P anti summary}
\end{align}
\es
We have found that the product of $\bsymb{P}$ and $\bsymb{B}$, and that of $\bsymb{\widetilde{P}}$ and $\bsymb{\tB}$,
\begin{equation}
\bsymb{K^{\zz;\hs x,\hs y}_{n+1,\hs n}} 
= \bsymb{P^{\zz;\hs x,\hs y}_{n+1,\hs n+1}\hs B^{\zz;\hs x,\hs y}_{n+1,\hs n}} \quad \left( n\geq 0\right),\quad
\bsymb{\tK^{\zz;\hs x,\hs y}_{n,\hs n+1}} 
= \bsymb{\widetilde{P}^{\zz;\hs x,\hs y}_{n,\hs n+2}\, \tB^{\zz;\hs x,\hs y}_{n+2,\hs n+1}} \quad \left( n\geq 1\right),
\label{K and tK summary}
\end{equation}
provide the key ingredients of propagators, satisfying the important relations \eqref{KB}-\eqref{tKQ+QtK}.
The $\bsymb{\Phi_0}$-propagator $\bsymb{\Delta_{1,0}}$ and the ghost propagators $\bsymb{\Delta_{n+1,n}}$ $(n\geq 1)$
are given by
\begin{align}
\bsymb{\Delta_{1,0}} &= \bsymb{K^{\zz^{(0)};\hs x_0,\hs y_0}_{1,\hs 0}} \,,\\[1ex]
\bsymb{\Delta_{n+1,n}} &=
\bsymb{K^{\zz^{(n)};\hs x_n,\hs y_n}_{n+1,\hs n}\hs Q_{n,n+1}\hs K^{\zz^{(n-1)};\hs x_{n-1},\hs y_{n-1}}_{n+1,\hs n}}
\quad \left( n\geq 1\right).
\label{Delta summary}
\end{align}
Furthermore, we have also investigated the case in which $\bsymb{B}$ and $\bsymb{\tB}$ are composed of linear combinations of 
world-sheet oscillators, including nonzero modes, and have obtained the corresponding propagators 
\begin{align}
\bsymb{\Delta_{1,0}} &= 
\bsymb{K^{\calZ_{(0)};\hs x_0,\hs y_0}_{1,\hs 0}}\hs \bsymb{Q_{0,1}}\hs \mrm{bpz}\Bigl( \bsymb{K^{\calZ_{(0)};\hs x_0,\hs y_0}_{1,\hs 0}}\Bigr) ,\\[1ex]
\bsymb{\Delta_{n+1,n}} &=
\bsymb{K^{\calZ_{(n)};\hs x_n,\hs y_n}_{n+1,\hs n}\hs Q_{n,n+1}\hs K^{\calZ_{(n-1)};\hs x_{n-1},\hs y_{n-1}}_{n+1,\hs n}}
\quad \left( n\geq 1\right).
\label{ext Delta summary}
\end{align}

In our analysis, gauge-fixing matrices were constructed from the world-sheet
operators $b$ and $\zeta$, with $\zeta$ denoting either $\xi$ or $d = [Q,b\xi]$.
Particularly, if we employ $\bz$-$\dz$ gauges, propagators take simple forms,
in virtue of the relations \eqref{N=2} originating from the twisted $N=2$ superconformal algebra~\cite{Berkovits-Vafa}.
However, as was mentioned in section~\ref{some examples}, these propagators involve the inverse of $L^2_0$ rather than
that of $L_0$.
Although the on-shell four-point tree amplitude was correctly reproduced~\cite{amp} under the gauge choice
\begin{equation}
\bz \Phi_{(0,\hs 0)} = \dz \Phi_{(0,\hs 0)} =0\,,
\label{bzdz}
\end{equation}
we have not yet known much about $\bz$-$\dz$ gauges.
In order to deepen our understanding,
let us now discuss the gauge \eqref{bzdz} from the viewpoint of the massless component field of $\Phi_{(0,\hs 0)}$.

At the massless level, the GSO-even string field $\Phi_{(0,\hs 0)}$ 
and its gauge parameters $\Lambda_{(-1,\hs 0)}$ and $\Lambda_{(-1,\hs 1)}$ are
expanded as\footnote{Here we use real component fields by separating ``$\iu$'' appropriately.
See appendices~\ref{algebraic relations} and~\ref{hc and reality} for our convention.}
\begin{align}
\Phi_{(0,\hs 0)} &= \int \frac{\diff^{10}k}{(2\pi)^{10}}\,\Bigl(
\iu\varphi (k) \ket{0;k} + A_\mu (k) \bigl[\psi^\mu \hs\xi\hs c\bigr]_{z=0}  \ket{-1;k} + \iu B(k) \bigl[\xi\hs\p\xi\hs c\hs\p c\bigr]_{z=0} \ket{-2;k}
\Bigr),
\\[1ex]
\Lambda_{(-1,\hs 0)} &= \int \frac{\diff^{10}k}{(2\pi)^{10}}\,\Bigl(
\frac{\iu}{2}\hs \lambda (k) \bigl[\xi\hs\p\xi\hs c\bigr]_{z=0} \ket{-2;k} \Bigr),\\[1ex]
\Lambda_{(-1,\hs 1)} &= \int \frac{\diff^{10}k}{(2\pi)^{10}}\,\Bigl(
\iu\hs \omega (k)\left. \xi\hs\right|_{z=0} \,\ket{0;k} \Bigr).
\end{align}
Here $\varphi$, $A_\mu$, $B$, $\lambda$, and $\omega$ are real component fields.
(For the reality, see appendix~\ref{hc and reality}.)
The action \eqref{free action} and the gauge transformation \eqref{gdof} are reduced to
\begin{align}
S_0 &= -\frac{\iu}{2} \braket{ \Phi_{(0,\hs 0)}, Q\eta_0\hs\Phi_{(0,\hs 0)}} \nonumber \\*[1ex]
&= \int \frac{\diff^{10} k}{(2\pi)^{10}}\, \left[
-\frac{1}{2}\hs A_\mu (-k) k^2 \eta^{\mu\nu} A_\nu (k) + \iu B(-k) k^\mu A_\mu (k) -\frac{1}{2}\hs B(-k) B(k)
\right] \nonumber \\*[1ex]
&= \int \frac{\diff^{10} k}{(2\pi)^{10}}\, \left[
-\frac{1}{2}\hs A_\mu (-k) \left( k^2 \eta^{\mu\nu} -k^\mu k^\nu \right) A_\nu (k) - \frac{1}{2}\hs \widetilde{B}(-k)\widetilde{B}(k)
\right],
\label{comp action}
\end{align}
\bs
\begin{align}
\delta\varphi(k) &= \omega (k) - \frac{1}{2}\, \lambda (k) \,,\\*[.7ex]
\delta A_\mu (k) &= \iu k_\mu \lambda(k) \,,\\*[.7ex]
\delta B(k) &= -k^2 \lambda (k)\,,\quad \delta \widetilde{B}(k) = 0\,,
\end{align}
\es
with
\begin{equation}
\widetilde{B}(k) := B(k) -\iu k^\mu A_\mu (k)\,.
\end{equation}
To obtain the above action \eqref{comp action}, we have used the normalization convention \eqref{normalization}.
We can express gauge-fixing conditions on $\Phi_{(0,\hs 0)}$ in terms of its component fields as follows:
\begin{align}
\bz \Phi_{(0,\hs 0)} &= 0 \iff B(k) = 0\,,
\label{bzPhi} \\[.5ex]
\xz \Phi_{(0,\hs 0)} &= 0 \iff \varphi (k) = 0\,,
\label{xzPhi} \\[.1ex]
\dz \Phi_{(0,\hs 0)} &= 0 \iff k^2 \varphi (k) -\iu k^\mu A_\mu (k) + \frac{1}{2}\hs B(k) =0\,.
\label{dzPhi}
\end{align}
If we impose condition \eqref{bzPhi}, the resultant action is precisely the one in the Feynman gauge, and
conditions \eqref{xzPhi} and \eqref{dzPhi} determine the value of $\varphi$:
\begin{align}
\xz \Phi_{(0,\hs 0)} = 0 &\ \Longrightarrow \ \varphi (k) = 0\,,
\label{vp0} \\[.5ex]
\dz \Phi_{(0,\hs 0)} = 0 &\ \Longrightarrow \ \varphi (k) = \frac{\iu k^\mu A_\mu (k)}{k^2}\,.
\label{vpA}
\end{align}
Although these two values are different, since the component field $\varphi (k)$ does not appear in the action, 
we conclude that the $\bz$-$\xz$ gauge and the $\bz$-$\dz$ gauge lead to the same gauge-fixed action, 
at least for the massless gauge field $A_\mu$.\footnote{When we consider the interacting theory, the action contains $\varphi$,
which shall be eliminated through condition \eqref{vp0} or \eqref{vpA}.}
Consequently, the propagator of $A_\mu$ takes the regular form even in the $\bz$-$\dz$ gauge.
This can be interpreted as follows.
Eq.~\eqref{d} tells us that the zero mode $\dz$ is composed of $\xz L_0$ and the terms not including $L_0$.
Thus we obtain
\begin{equation}
\frac{\dz}{L_0} = \xz + \frac{1}{L_0}\times\left( \text{the sum of the operators not including $L_0$}\right)\,.
\label{d/L}
\end{equation} 
Our analysis in the present section 
suggests that in the $\bz$-$\dz$ gauge, the factor ${\dz}/{L_0}$ in the propagator $\bsymb{\Delta_{1,0}}$ 
will be effectively the same as $\xz$, and the second term on the right-hand side of eq.~\eqref{d/L} will be negligible. 
Therefore, we expect that $\bz$-$\dz$ gauges, also, will be physically reasonable gauges.

We have dealt with only the free theory throughout the present paper.
In order to complete gauge fixing, however, we have to consider also the interaction terms.
As can be seen from eq.~\eqref{solution}, in the WZW-type superstring field theory,
unlike in bosonic string field theory~\cite{Witten},  the extended action $S$ does not take the same form as the original action $S_0$.
This makes it difficult to obtain completely gauge-fixed action in the interacting theory.
Although we have not yet succeeded in solving the problem, several approaches are proposed in refs.~\cite{paper2, Nathan}.

\section*{Acknowledgments}
\indenths
The author would like to express his gratitude to Nathan Berkovits, Michael Kroyter, Yuji Okawa, Martin Schnabl, and Barton Zwiebach.
This work originated from the collaboration with them. 
The author is also grateful to Mitsuhiro Kato for his helpful comments and would like to thank Toshifumi Noumi for a useful discussion.
This work was supported in part by the Japan Society for the Promotion of Science (JSPS), the Academy of Sciences of the Czech Republic,
and the M\v{S}MT contract No.~LH11106
under the Research Cooperative Program between Japan and the Czech Republic,
and by Research Fellowships of JSPS for Young Scientists.

\appendix
\section{Algebraic relations}
\label{algebraic relations}
\setcounter{equation}{0}
\indenths
We ``bosonize'' the superconformal ghosts $\beta$ and $\gamma$ as\footnote{We omit the normal ordering symbol 
with respect to the SL($2$,$\mathbb{R}$)-invariant vacuum.}
\begin{equation}
\beta = \e^{-\phi} \partial\xi \,,\quad 
\gamma = \eta \e^{\phi} \,.
\label{bosonization}
\end{equation}
Here and in the sequel, we use the convention in which appropriate cocycle factors are implicitly included.
Thus $\e^{l \phi}$ ($l\in$ odd) anticommutes with fermionic operators, such as $b$, $c$, $\xi$, and $\eta$.
The world-sheet ghost number $g$ and the picture number $p$ are assigned as in table~\ref{table}.
\begin{table}[b]
\begin{center}
\caption{The quantum number $(g,p)$ and the conformal weight $h$.}
\ \\[-.5ex]
\begin{tabular}{|c||c|c|c|c|c|c|c|c|}
\hline
operator & $b$ & $c$ & $\xi$ & $\eta$ & $\e^{l\phi}$ & $j_B$ & $d$ &$f$ \\
\hline
$(g,p)$ & $(-1,0)$ & $(1,0)$ & $(-1,1)$ & $(1,-1)$ & $(0,l)$ & $(1,0)$ & $(-1,1)$ & $(1,-1)$ \\
\hline
$h$ & $2$ & $-1$ & $0$ & $1$ & $-\frac{1}{2}\hs l(l+2)$ & $1$ & $2$ & $-1$ \\
\hline
\end{tabular}
\label{table}
\end{center}
\end{table}
The total superconformal field theory on the world-sheet is constructed from the fields
$X^\mu$, $\psi^\mu$, $b$, $c$, $\xi$, $\eta$, and $\phi$.
Their fundamental operator product expansions (OPEs) and mode expansions in the NS sector are given by\footnote{
Basically, we follow the convention in refs.~\cite{Polchinski1, Polchinski2}, setting $\alpha' =2$.}
\bs \label{mode exp}
\begin{align}
X^\mu(z_1,\overline{z_1}) X^\nu(z_2,\overline{z_2}) 
&\sim -\eta^{\mu\nu}\Bigl( \ln \left| z_1 - z_2\right|^2 + \ln \left|z_1 -\overline{z_2}\right|^2\Bigr),\quad
\p X^\mu(z) = -\iu\,\sum_{n\in \mathbb{Z}}\, \frac{\alpha^\mu_n}{z^{n+1}}\,,
\label{X} \\[.5ex]
\psi^\mu (z_1) \psi^\nu (z_2) &\sim \frac{\eta^{\mu\nu}}{z_{12}}\,,\quad 
\psi^\mu (z) = \sum_{r\in\mathbb{Z}+1/2}\,\frac{\psi^\mu_r}{z^{r+1/2}}\,,
\label{psi} \\[.5ex]
b(z_1) c(z_2) &\sim \frac{1}{z_{12}}\,,\quad 
b(z) = \sum_{n\in\mathbb{Z}}\,\frac{b_n}{z^{n+2}}\,,\quad c(z) = \sum_{n\in\mathbb{Z}}\,\frac{c_n}{z^{n-1}}\,,
\label{bc} \\[.5ex]
\xi (z_1) \eta (z_2) &\sim \frac{1}{z_{12}}\,,\quad 
\xi (z) = \sum_{n\in\mathbb{Z}}\,\frac{\xi_n}{z^{n}}\,,\quad \eta (z) = \sum_{n\in\mathbb{Z}}\,\frac{\eta_n}{z^{n+1}}\,,
\label{xi eta} \\[.5ex]
\phi (z_1) \phi (z_2) &\sim -\ln z_{12}\,,\quad \p\phi(z) = - \sum_{n\in\mathbb{Z}}\,\frac{\phi_n}{z^{n+1}}\,,
\label{phi}
\end{align}
\es
with $z_{12} := z_1 -z_2$\hs, and the symbol ``$\sim$'' denoting the equality up to non-singular terms.
Note the factor of $-\iu$ and of $-1$ in the mode expansions in eqs.~\eqref{X} and \eqref{phi}.
In what follows, unless otherwise stated, an operator $\mathcal{O}$ of conformal weight $h$ will be expanded as in
eqs.~\eqref{psi}-\eqref{xi eta}:
\begin{equation}
\mathcal{O}(z) = \sum_{n} \frac{\mathcal{O}_n}{z^{n+h}}\,,
\quad \mathcal{O}_n = \oint_C \frac{\diff z}{2\pi \mrm{i}}\hs z^{n+h-1} \mathcal{O}(z) \,.
\end{equation}
Here we have used the doubling trick~\cite{Polchinski1}, and have denoted by $C$ a counterclockwise contour encircling the origin. 
The BPZ conjugate of $\calO_n$ is defined as
\begin{equation}
\bpz{\calO_n} := \oint_{C'} \frac{\diff z'}{2\pi \mrm{i}}\hs (z')^{n+h-1} \calO' (z')\,,
\label{bpz conj}
\end{equation}
where $\calO'$ is a conformal transform of $\calO$ by the inversion $z \mapsto z' = - \frac{1}{z}$,
and $C'$ is a counterclockwise contour encircling the origin of the $z'$-plane.
In particular, if $\calO$ is a primary operator, we have
\begin{equation}
\calO' (z') = \left( \frac{\diff z'}{\diff z}\right)^{-h} \calO (z)\,,
\end{equation}
so that
\begin{equation}
\bpz{\calO_n} = (-1)^{n+h} \calO_{-n} \,.
\label{bpzO}
\end{equation}
The BPZ conjugate of the product of the operators $A$ and $B$ is given by
\begin{equation}
\bpz{AB} = (-1)^{\epsilon\left(A\right) \epsilon\left(B\right)} \bpz{B} \bpz{A},
\end{equation}
with $\gp{A}$ and $\gp{B}$ denoting the Grassmann parity of $A$ and of $B$.

Under the bosonization \eqref{bosonization}, the BRST current $j_B$ takes the form
\begin{align}
j_B &= c\hs T^\mrm{m} + \gamma\hs G^\mrm{m} + bc\p c +\frac{3}{4}\hs (\p c)\beta\gamma + \frac{1}{4}\hs c(\p\beta)\gamma 
-\frac{3}{4}\hs c\beta\p\gamma -b\gamma^2 + \frac{3}{4}\hs \p^2 c \nonumber \\*[.5ex]
&= c\hs T^\mrm{m} +\eta\e^\phi\hs G^\mrm{m} +bc\p c +\frac{3}{4}\hs\p c\p\phi -\frac{1}{4}\hs c\p^2\phi
-\frac{1}{2}\hs c\p\phi\p\phi -c\eta\p \xi -b\eta\p\eta \e^{2\phi} +\frac{3}{4}\hs\p^2 c \,,
\end{align}
where $T^\mrm{m}$ and $G^\mrm{m}$ are the matter energy-momentum tensor and the matter supercurrent, respectively:
\begin{equation}
T^\mrm{m} = -\frac{1}{2}\hs \p X^\mu \p X_\mu -\frac{1}{2}\hs \psi^\mu \p\psi_\mu \,,\quad
G^\mrm{m} = \iu\hs \psi^\mu \p X_\mu \,.
\label{T^m}
\end{equation}
We have added the total-derivative term $\frac{3}{4}\hs\p^2 c$, in order to make the current primary.
The BRST operator $Q = \oint_C \frac{\diff z}{2\pi\mrm{i}} j_B (z)$ and the zero mode of $\eta$ anticommute with each other,
and thus we have
\begin{equation}
Q^2 = \ez^2 = \{ Q,\ez\} = 0\,.
\end{equation}
By contrast, the anticommutator of $Q$ and $\xi$ produces the picture-changing operator
\begin{equation}
X := \{ Q,\xi\} 
= \e^\phi G^\mrm{m} +c\p\xi + b\p\eta\e^{2\phi} +\p \bigl( b\eta\e^{2\phi}\bigr)\,.
\end{equation}
It obeys the following operator product expansions:
\bs \label{X-OPE}
\begin{align}
X(z_1)\hs b(z_2)
&\sim
- \frac{1}{z_{12}}\hs \partial \xi (z_2) \,,
\\[1ex]
X(z_1)\hs c(z_2)
&\sim
\frac{1}{z^2_{12}}\hs \eta \e^{2\phi} (z_2) 
- \frac{1}{z_{12}}\hs \p\eta \e^{2\phi} (z_2) \,,
\\[1ex]
X(z_1)\hs \xi (z_2)
&\sim
- \frac{2}{z^2_{12}}\hs b \e^{2\phi} (z_2)
- \frac{1}{z_{12}}\hs \p \bigl( b \e^{2\phi} \bigr) (z_2) \,,
\\[1ex]
X(z_1)\hs \eta (z_2)
&\sim
- \frac{1}{z^2_{12}}\hs c (z_2) - \frac{1}{z_{12}}\hs \p c (z_2) \,.
\end{align}
\es
As a result, we obtain
\bs
\begin{align}
[X_m, b_n] &= (m+n)\hs \xi_{m+n}\,,\\[.5ex]
[X_m, \xi_n] &= -(m-n) \bigl( b\e^{2\phi}\bigr)_{m+n} \,,\\[.5ex]
[X_m, \ez] &=0\,.
\end{align}
\es

Next let us see the properties of the primary operator 
\begin{equation}
d := [Q,\hs b\xi]
= T \xi -bX +\p^2\xi \,,
\label{d}
\end{equation}
where $T$ is the total energy-momentum tensor:
\bs \label{Ttotal}
\begin{align}
T &= T^\mrm{m} + T^{bc} + T^{\eta\xi} + T^\phi\,,\quad T(z) = \sum_{n\in\mathbb{Z}}\,\frac{L_n}{z^{n+2}}\,,\\
T^{bc} &= -(\p b)\hs c - 2\hs b\hs\p c\,,\quad
T^{\eta\xi} = -\eta\hs\p\xi\,,\quad
T^\phi = -\frac{1}{2}\hs \p\phi\p\phi - \p^2\phi\,.
\end{align}
\es
OPEs of the operator $d$ are
\bs
\begin{align}
d (z_1)\hs b(z_2) &\sim
\frac{2}{z^2_{12}}\hs \xi b (z_2) + \frac{1}{z_{12}}\hs \p (\xi b) (z_2) \,, 
\label{db} \\[1ex]
d (z_1)\hs c (z_2) &\sim 
- \frac{1}{z^2_{12}} \bigl[ \xi c + b \eta \e^{2\phi} \bigr] (z_2) 
+ \frac{1}{z_{12}} \bigl[ -\e^\phi G^\mrm{m} + \xi \p c  
- (\p b) \eta \e^{2\phi} - \p \bigl( b \eta \e^{2\phi} \bigr) \bigr] (z_2) \,, \\[1ex]
d (z_1)\hs \xi (z_2) &\sim
\frac{1}{z_{12}} \bigl[ \xi \p\xi - b (\p b) \e^{2\phi} \bigr] (z_2) \,, \\[1ex]
d (z_1)\hs \eta (z_2) &\sim
\frac{2}{z^3_{12}} + \frac{1}{z^2_{12}} \bigl[ bc + \xi\eta \bigr] (z_2)
+ \frac{1}{z_{12}} \bigl[ T + \p (bc) + \p (\xi\eta ) \bigr] (z_2) \,,\\[1ex]
d(z_1)\hs d(z_2) &\sim 0\,. 
\end{align}
\es
Thus we obtain
\bs
\begin{align}
\{ d_m, b_n\} &= (m-n)\hs (\xi b)_{m+n} \,,
\label{dm bn} \\[.5ex]
\{ d_m, \ez \} &= L_m \,,
\label{dmez} \\[.5ex]
\{ d_m, d_n \} &= 0\,.
\end{align}
\es
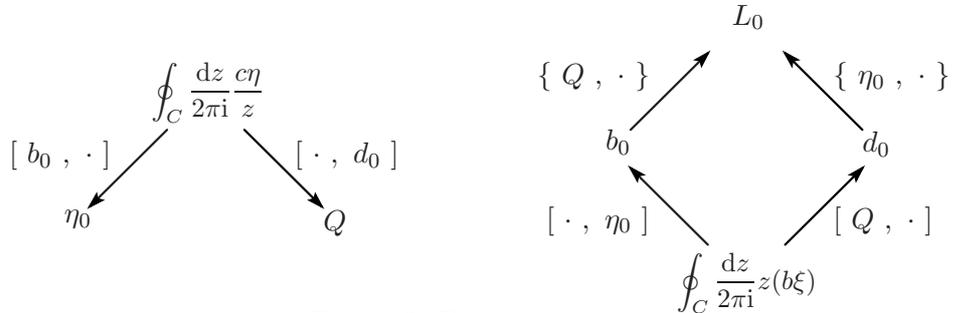
\begin{figure}[b]
\unitlength 0.1in
\begin{picture}( 55.4000, 13.9000)(  2.6000,-25.3500)
\put(24.1000,-16.0500){\makebox(0,0){{\small $\displaystyle \oint_C \frac{{\mathrm d}z}{2\pi{\mathrm i}} \frac{c\eta}{z} $}}}%
\put(18.1000,-22.2000){\makebox(0,0)[rt]{$ \eta_0 $}}%
\put(30.1000,-22.2000){\makebox(0,0)[lt]{$Q$}}%
\put(20.1000,-20.1000){\makebox(0,0)[rb]{$ [\ b_0 \ , \ \cdot \ ]$ \ }}%
\put(28.1000,-20.1000){\makebox(0,0)[lb]{\ $ [\ \cdot \ , \ d_0 \ ]$}}%
%
{\color[named]{Black}{%
\special{pn 13}%
\special{pa 2200 1800}%
\special{pa 1800 2200}%
\special{fp}%
\special{sh 1}%
\special{pa 1800 2200}%
\special{pa 1862 2168}%
\special{pa 1838 2162}%
\special{pa 1834 2140}%
\special{pa 1800 2200}%
\special{fp}%
\special{pa 2600 1800}%
\special{pa 3000 2200}%
\special{fp}%
\special{sh 1}%
\special{pa 3000 2200}%
\special{pa 2968 2140}%
\special{pa 2962 2162}%
\special{pa 2940 2168}%
\special{pa 3000 2200}%
\special{fp}%
}}%
\put(52.1000,-12.1000){\makebox(0,0){$L_0$}}%
%
{\color[named]{Black}{%
\special{pn 13}%
\special{pa 4600 1800}%
\special{pa 5000 1400}%
\special{fp}%
\special{sh 1}%
\special{pa 5000 1400}%
\special{pa 4940 1434}%
\special{pa 4962 1438}%
\special{pa 4968 1462}%
\special{pa 5000 1400}%
\special{fp}%
\special{pa 5800 1800}%
\special{pa 5400 1400}%
\special{fp}%
\special{sh 1}%
\special{pa 5400 1400}%
\special{pa 5434 1462}%
\special{pa 5438 1438}%
\special{pa 5462 1434}%
\special{pa 5400 1400}%
\special{fp}%
}}%
\put(48.0000,-16.0000){\makebox(0,0)[rb]{$\{ \ Q \ , \ \cdot \ \}$ \ }}%
\put(56.0000,-16.0000){\makebox(0,0)[lb]{\ $\{ \ \eta_0 \ , \ \cdot \ \}$}}%
\put(46.0000,-18.0000){\makebox(0,0)[rt]{$b_0$}}%
\put(58.0000,-18.1000){\makebox(0,0)[lt]{$d_0$}}%
%
{\color[named]{Black}{%
\special{pn 13}%
\special{pa 5000 2400}%
\special{pa 4600 2000}%
\special{fp}%
\special{sh 1}%
\special{pa 4600 2000}%
\special{pa 4634 2062}%
\special{pa 4638 2038}%
\special{pa 4662 2034}%
\special{pa 4600 2000}%
\special{fp}%
\special{pa 5400 2400}%
\special{pa 5800 2000}%
\special{fp}%
\special{sh 1}%
\special{pa 5800 2000}%
\special{pa 5740 2034}%
\special{pa 5762 2038}%
\special{pa 5768 2062}%
\special{pa 5800 2000}%
\special{fp}%
}}%
\put(48.0000,-22.0000){\makebox(0,0)[rt]{$[ \ \cdot \ , \ \eta_0 \ ]$ \ }}%
\put(56.0000,-22.0000){\makebox(0,0)[lt]{\ $[\ Q \ , \ \cdot \ ]$}}%
\put(52.0000,-26.0000){\makebox(0,0){{\small $\displaystyle \oint_C \frac{{\mathrm d}z}{2\pi{\mathrm i}} z (b \xi) $}}}%
\end{picture}%
\caption{Diagrams.}
\label{diagrams}
\end{figure}
Note that eq.~\eqref{dmez} is analogous to $\{ b_m, Q \} = L_m\hs$.
In fact, $\ez$ and $\dz$ are the counterparts of $Q$ and $\bz$ 
in the twisted $N=2$ superconformal algebra in ref.~\cite{Berkovits-Vafa}, and the following equations hold:
\bs
\begin{align}
\Bigl[ \,\bz \,,\, \oint_C \frac{\diff z}{2\pi \mrm{i}} \frac{c\eta}{z} \ \Bigr]
&= \eta_0 \,, \\[1ex]
\Bigl[ \, \dz \,,\, \oint_C \frac{\diff z}{2\pi\mrm{i}} \frac{c\eta}{z} \ \Bigr]
&= -Q \,, \\[1ex]
\Bigl[ \, \ez \,,\, \oint_C \frac{\diff z}{2\pi\mrm{i}} z \left( b \xi\right) \ \Bigr]
&= -b_0 \,, \\[1ex]
\left\{ \ez \,,\, \dz \right\}
&= L_0 \,.
\end{align}
\es
These relations form the diagrams in figure~\ref{diagrams}.
Moreover, the weight-$(-1)$ operator
\begin{equation}
f := -c\p c \xi \e^{-2\phi} = -\left[ Q\,,\, c\xi\e^{-2\phi} \right] 
\end{equation}
satisfies 
\begin{equation}
\{ d_m, f_n\} = \delta_{m+n,\hs 0} \,,
\end{equation}
which corresponds to the equation $\{b_m ,c_n\} = \delta_{m+n,\hs 0}$.
\begin{table}[t]
\begin{center}
\caption{Relevant relations among the zero modes. In each box, the value of the graded commutator 
$[ A,B\}$ of an operator $A$ in the first column and $B$ in the first row is shown.
Some nonzero results whose explicit forms are not used in the present paper are omitted.}
\ \\[-.5ex]
\begin{tabular}{|c||c|c|c|c|c|c|c|}
\hline
\backslashbox{A}{B}& $Q$ & $\ez$ & $\bz$ & $\xz$ & $\dz$ & $\cz$ & $X_0$ \\
\hline \hline
$Q$ & $0$ & $0$ & $L_0$ & $X_0$ & $0$ &  & $0$ \\
\hline
$\ez$ & $0$ & $0$ & $0$ & $1$ & $L_0$ & $0$ & $0$ \\
\hline
$\bz$ & $L_0$ & $0$ & $0$ & $0$ & $0$ & $1$ & $0$ \\
\hline
$\xz$ & $X_0$ & $1$ & $0$ & $0$ &  & $0$ & $0$ \\
\hline
$\dz$ & $0$ & $L_0$ & $0$ &  & $0$ &  &  \\
\hline
$\cz$ &  & $0$ & $1$ & $0$ &  & $0$ &  \\
\hline
$X_0$ & $0$ & $0$ & $0$ & $0$ &  &  & $0$ \\
\hline
\end{tabular}
\label{zero modes}
\end{center}
\end{table}

We list the relevant relations among the zero modes in table~\ref{zero modes}.
Throughout the present paper, we often use the symbol $\zz$, which denotes either $\xz$ or $\dz$.
We also introduce the following barred operators in section~\ref{on-shell} and appendix~\ref{reachability}:
\begin{align}
\Lbar &:= \{ \ez ,\zz \} =
\left\{
\begin{aligned}
&1 \quad \bigl( \text{for } \zz = \xz \bigr)\\
&L_0 \quad \bigl( \text{for } \zz = \dz \bigr)
\end{aligned}
\right.
\,,\quad
\Lxy := x L_0 + y\Lbar =
\left\{
\begin{aligned}
&x L_0 + y \quad \bigl( \text{for } \zz = \xz \bigr)\\
&\left(x+y\right) L_0 \quad \bigl( \text{for } \zz = \dz \bigr)
\end{aligned}
\right.
\,,
\label{LbarLxy} \\[2.5ex]
\zzbar &:= 
\left\{
\begin{aligned}
&\xz \quad \left( \text{for } \zz = \xz \right)\\[.5ex]
&\frac{\dz}{L_0} \quad \left( \text{for } \zz = \dz \right)
\end{aligned}
\right.
\,,\quad
\czbar := \left\{
\begin{aligned}
&\cz \quad \left( \text{for } \zz = \xz \right)\\[.5ex]
&\frac{Q}{L_0} \quad \left( \text{for } \zz = \dz \right)
\end{aligned}
\right.
\,,\quad
\ezbar := \left\{
\begin{aligned}
&\eta_0 \quad \left( \text{for } \zz = \xz \right)\\[.5ex]
&\frac{\eta_0}{L_0} \quad \left( \text{for } \zz = \dz \right)
\end{aligned}
\right.
\,,\\[2.5ex]
\Xbar &:= \{ Q\,,\,\zzbar \} =
\left\{
\begin{aligned}
&X_0\quad \bigl(\text{for } \zz = \xz\bigr)\\
&0\quad \bigl(\text{for } \zz = \dz\bigr)
\end{aligned}
\right.\,.
\end{align}
They obey the equations below
\begin{align}
\bz^2 &= \czbar^2 = \zz^2 = \zzbar^2 = \ez^2 = \ezbar^2 = 0\,,\\[.5ex]
\{\bz,\czbar\} &= \{\zz,\ezbar\} = \{\zzbar,\ez\} = 1\,,\\[.5ex]
\{\bz,\zz\} &= \{\bz,\zzbar\} = \{\bz,\ezbar\} = \{\czbar ,\zz\} = \{\czbar,\ezbar\} = 0\,,\\[.5ex]
\{Q,\ezbar\} &= 0\,,\quad \{Q,\zzbar\} = \Xbar\,.
\label{X_0 bar}
\end{align}

\section{Hermitian conjugation and the reality condition}
\label{hc and reality}
\setcounter{equation}{0}
\indenths
We consider the reality of the action and of component fields in the WZW-type open superstring field theory.
First we introduce Hermitian conjugation of mode operators, and then investigate the reality condition~\cite{tensor const} 
in the same manner as in appendix C of ref.~\cite{a-gauge}.

\subsection{Hermitian conjugation of mode operators}
\indenths
We define Hermitian conjugation (hc) of the mode operators $\alpha^\mu_n$, $\psi^\mu_r$, $b_n$, $c_n$, $\xi_n$, and $\eta_n$ 
in eq.~\eqref{mode exp} as
\bs \label{hc mode}
\begin{align}
\hc{\alpha^\mu_n} &= \alpha^\mu_{-n}\,,\quad \hc{\psi^\mu_r} = \psi^\mu_{-r}\,, 
\label{hc alpha psi} \\
\hc{b_n} &= b_{-n}\,,\quad \hc{c_n} = c_{-n}\,,
\label{hc bc} \\
\hc{\xi_n} &= \xi_{-n}\,,\quad \hc{\eta_n} = \eta_{-n}\,.
\label{hc xi eta}
\end{align}
\es
The Hermitian conjugate of the product of the operators $A$ and $B$ is given by
\begin{equation}
\hc{AB} = \hc{B} \hc{A}\,.
\end{equation}
Under the above definitions, the energy-momentum tensors $T^\mrm{m}$, $T^{bc}$, and $T^{\eta\xi}$ in eqs.~\eqref{T^m} and \eqref{Ttotal} 
are Hermitian in the following sense:
\begin{equation}
\hc{L^\mrm{m}_n} = L^\mrm{m}_{-n}\,,\quad \mrm{hc}\bigl(L^{bc}_n\bigr) = L^{bc}_{-n}\,,\quad \mrm{hc}\bigl(L^{\eta\xi}_n\bigr) = L^{\eta\xi}_{-n}\,,
\end{equation}
where $L^\mrm{m}_n$, $L^{bc}_n$, and $L^{\eta\xi}_n$ are mode operators of $T^\mrm{m}$, $T^{bc}$, and $T^{\eta\xi}$, respectively.
We require that the same relation hold also for the mode operators $L^\phi_n$ of the other energy-momentum tensor $T^\phi$: 
$\mrm{hc}\bigl( L^\phi_n\bigr) = L^\phi_{-n}$.
This requirement determines the Hermitian conjugates of $\phi_n$; we find that
\begin{equation}
\hc{\phi_n} = \left\{
\begin{aligned}
&-\phi_{-n}\quad \left( n\neq 0\right),\\
&-\phi_0 -2 \quad\left( n=0\right).
\end{aligned}
\right.
\label{hc phi}
\end{equation}
The transformation property of the zero mode $\phi_0$ reflects the fact that the current $j^\phi = -\p\phi$ is not primary
and the $\phi$-charge is anomalous.
As we shall see in the next subsection, the shift of $-\phi_0$ by $-2$ is consistent with the definition of the inner product \eqref{C_N}.

In order to go further, let us expand also the operator $\e^{l\phi}$ $\left( l\in\mathbb{Z}\right)$ in the following form:
\begin{equation}
\e^{l\phi(z)} = \sum_{r\in\mathbb{Z}+l/2}\, \frac{e^l_r}{z^{r+h_l}}\,,
\label{e mode}
\end{equation} 
where $h_l := -\frac{1}{2}\hs l(l+2)$ is the conformal weight of $\e^{l\phi}$.
The superscript $l$ on the mode operators indicates their $\phi$-charge, and we have
\begin{equation}
\bigl[\phi_0\hs, e^l_r\bigr] = l\, e^l_r\,.
\end{equation}
Note that if $l$ is even (resp.\ odd), the operators $e^l_r$ are bosonic (resp.\ fermionic).
We assume 
\begin{equation}
\mrm{hc}\bigl(e^l_r\bigr) = \varepsilon_l \, e^l_{-r}\quad \left( r\in\mathbb{Z} + \frac{l}{2}\right)
\label{hce}
\end{equation}
with $\varepsilon_l$ a sign factor dependent only on $l$.
We would like to know whether $\varepsilon_l$ is plus one or minus one.
For this purpose, we consider (anti)commutation relations among $e^l_r$. By using the identity
\begin{equation}
\e^{l_1 \phi(z_1)} \e^{l_2 \phi(z_2)} = z^{-l_1 l_2}_{12}\hs \e^{l_1\phi(z_1) + l_2\phi(z_2)}\,,
\end{equation}
we readily obtain 
\begin{align}
\bigl\{ e^{\pm 1}_{r_1} , e^{\pm 1}_{r_2} \bigr\} &= e^{\pm 2}_{r_1 + r_2}\,,
\label{e1e1} \\[1ex]
\bigl[ e^1_{r_1} , e^2_{r_2}\bigr] 
&= -\sum_{n\in\mathbb{Z}}\, \bnol\phi_n e^3_{r_1+r_2-n}\bnor\, + \left( r_1 -\frac{5}{2}\right) e^3_{r_1+r_2}\,,
\label{e1e2} \\*[.5ex]
\bigl[ e^2_{r_2} , e^1_{r_1}\bigr]
&= -2\sum_{n\in\mathbb{Z}}\,\bnol \phi_n e^3_{r_1+r_2-n}\bnor\, + \left( r_2 -5\right) e^3_{r_1+r_2}\,,
\label{e2e1} \\[.5ex]
\bigl[ e^{-1}_{r_1} , e^{-2}_{r_2}\bigr] 
&= \sum_{n\in\mathbb{Z}}\, \bnol\phi_n e^{-3}_{r_1+r_2-n}\bnor\, + \left( r_1 -\frac{1}{2}\right) e^{-3}_{r_1+r_2}\,,
\label{e-1e-2} \\*[.5ex]
\bigl[ e^{-2}_{r_2} , e^{-1}_{r_1}\bigr]
&= 2\sum_{n\in\mathbb{Z}}\,\bnol \phi_n e^{-3}_{r_1+r_2-n}\bnor\, + \left( r_2 -1\right) e^{-3}_{r_1+r_2}\,.
\label{e-2e-1}
\end{align}
Here we have explicitly written the normal ordering symbol \hs $\bnol \cdots \bnor$ with respect to the SL$(2,\mathbb{R})$-invariant vacuum.

First, from eq.~\eqref{e1e1}, we find that $\varepsilon_{\pm 2}$ are plus one because we have
\begin{equation}
\hc{e^{\pm 2}_{r_1 + r_2}} = \hc{\bigl\{ e^{\pm 1}_{r_1} , e^{\pm 1}_{r_2} \bigr\}} 
= \bigl\{ e^{\pm 1}_{-r_1} , e^{\pm 1}_{-r_2} \bigr\} = e^{\pm 2}_{-r_1 - r_2} \,.
\end{equation}
The sign factors $\varepsilon_{\pm 1}$, unlike $\varepsilon_{\pm 2}$, cannot be determined only from the relations \eqref{e1e1}.
However, Hermiticity of the BRST operator $Q$ entails 
\begin{equation}
\varepsilon_1 =-1\,,\quad \mrm{hc}\bigl(e^{1}_r\bigr) = - e^{1}_{-r}\,.
\label{ve1}
\end{equation}
Furthermore, we can determine $\varepsilon_{-1}$ by requiring the consistency with the commutation relation
\begin{equation}
\bigl[ \gamma_s , \beta_r \bigr] = \delta_{r+s,\hs 0}\quad \left(r,s\in\mathbb{Z}+\frac{1}{2}\right),
\label{gamma beta}
\end{equation}
where $\beta_r$ and $\gamma_s$ are mode operators of the superconformal ghosts.
In the bosonized description \eqref{bosonization}, we have
\begin{equation}
\beta_r = -\sum_{n\in\mathbb{Z}}\,n\, e^{-1}_{r-n} \xi_n\,,\quad
\gamma_s = \sum_{n\in\mathbb{Z}}\,\eta_n e^1_{s-n}\,,
\end{equation}
and eqs.~\eqref{hc xi eta} and \eqref{ve1} tell us that $\hc{\gamma_s} = +\gamma_{-s}$.
Thus, in order for the Hermitian conjugation to be consistent with eq.~\eqref{gamma beta},
$\beta_r$ have to transform as $\hc{\beta_r}=-\beta_{-r}$;
we conclude that $\varepsilon_{-1}$ has to be minus one.

Next, let us consider eqs.~\eqref{e1e2} and \eqref{e2e1}. Comparing these two equations, we obtain
\begin{equation}
\sum_{n\in\mathbb{Z}}\, \bnol\phi_n e^3_{r_1+r_2-n}\bnor 
= \frac{1}{3}\left( r_1 + r_2 -\frac{15}{2}\right) e^3_{r_1+r_2}
\label{e3}
\end{equation}
and thus
\begin{equation}
\bigl[ e^1_{r_1} , e^2_{r_2}\bigr] = \frac{1}{3} \left( 2 r_1 - r_2 \right) e^3_{r_1 + r_2}\,.
\label{e1e2e3}
\end{equation}
Similarly, it follows from eqs.~\eqref{e-1e-2} and \eqref{e-2e-1} that
\begin{equation}
\bigl[ e^{-1}_{r_1} , e^{-2}_{r_2}\bigr] = \frac{1}{3} \left( 2 r_1 - r_2 \right) e^{-3}_{r_1 + r_2}\,.
\label{e-1e-2e-3}
\end{equation}
The conjugation \eqref{hce} is compatible with eqs.~\eqref{e1e2e3} and \eqref{e-1e-2e-3} only when $\varepsilon_{\pm 3}$ are minus one.
In this manner, we can determine $\varepsilon_l$, and the result is
\begin{equation}
\varepsilon_l = (-1)^l\,,\quad \mrm{hc}\bigl( e^l_r\bigr) = (-1)^l\,e^l_{-r}\,.
\label{el}
\end{equation}

\subsection{The reality condition}
\indenths
The string field $\Phi_{(0,\hs 0)}$ is expanded in terms of first-quantized states as
\begin{equation}
\Phi_{(0,\hs 0)} = \int \frac{\diff^{10} k}{(2\pi)^{10}}\hs \sum_s \sum_l \varphi^s_l (k) \ket{s(l; k)}.
\label{expansion}
\end{equation}
Here $\ket{s(l; k)}$ denote basis states of the GSO-even NS Fock space of $(g,p)=(0,0)$, and $\varphi^s_l (k)$ denote component fields of momentum $k$.
The states $\ket{s(l; k)}$ are constructed from coherent states of the form
\begin{equation}
\ket{l; k} := \lim_{z\to 0} \e^{l\phi(z)} \e^{\iu k\cdot X(z,\hs \zbar)} \ket{0}
= \lim_{z\to 0} \e^{\iu k\cdot X(z,\hs \zbar)}\ e^l_{\frac{1}{2}\hs l(l+2)} \ket{0}
\quad \left( l\in \mathbb{Z}\right)
\label{ket}
\end{equation} 
and mode operators in eq.~\eqref{mode exp}. (We have utilized the mode expansion \eqref{e mode} in the last equality of eq.~\eqref{ket}.) 
In the above definition \eqref{ket}, the symbol $\ket{0}$ represents the SL($2$,$\mathbb{R}$)-invariant vacuum.
In order to investigate the reality of $\braket{\Phi_{(0,\hs 0)}, Q\ez\hs \Phi_{(0,\hs 0)}}$,
we introduce the Hermitian conjugate $\langle\bra{l;k} = \mrm{hc}\left(\, \ket{l;k}\,\right)$ satisfying the equation
\begin{equation}
\langle\bra{l;k} \xz c_{-1} \cz c_1 \ket{l';k'} = \left( -1\right)^l C_N \hs\delta^{10} (k-k') \hs\delta_{l+l',-2}\,.
\label{C_N}
\end{equation}
Here $C_N$ is a normalization constant independent of the $\phi$-charge $l$ and the momentum $k$,
and the factor $(-1)^l$ originates from the relation \eqref{el}.
The above normalization is consistent with the transformation property of $\phi_0$ given in eq.~\eqref{hc phi}.
Indeed we have
\begin{align}
l' \,\langle\bra{l;k} \xz c_{-1} \cz c_1 \ket{l';k'} &= \langle\bra{l;k} \xz c_{-1} \cz c_1 \bigl(\phi_0 \ket{l';k'}\hs\bigr) \nonumber \\*
&= \bigl(\,\langle\bra{l;k} \phi_0 \hs\bigr) \xz c_{-1} \cz c_1 \ket{l';k'} = -\left( l+2\right) \langle\bra{l;k} \xz c_{-1} \cz c_1 \ket{l';k'}\,.
\end{align}
In the last equality, we have used
\begin{equation}
\langle\bra{l;k} \phi_0 = \mrm{hc}\bigl(\, \hc{\phi_0} \ket{l;k}\,\bigr) = \mrm{hc}\bigl(\, (-\phi_0 -2) \ket{l;k}\,\bigr)
= -\left(l+2\right) \langle\bra{l;k}\,.
\end{equation}
Note that we may relate $\langle \bra{l;k}$ to the BPZ conjugate $\bra{l;-k} = \bpz{\,\ket{l;-k}\hs}$ as
\begin{equation}
\langle \bra{l;k} = \left(-1\right)^l \bra{l;-k} \,.
\label{hc bpz}
\end{equation}
The Kronecker delta $\delta_{l+l',-2}$ in eq.~\eqref{C_N} owes its origin to the $\phi$-charge anomaly.
The BPZ conjugate of $\phi_0$ takes the same form as the Hermitian conjugate:
\begin{equation}
\bpz{\phi_0} = -\phi_0 -2\,.
\end{equation}
From eqs.~\eqref{hc mode} and \eqref{C_N}, it follows that
\begin{align}
\left(-1\right)^l C_N \hs\delta^{10} (k-k') \hs\delta_{l+l',-2}
&= \langle\bra{l';k'} \xz c_{-1} \cz c_1 \ket{l;k} \nonumber \\*
&= - \mrm{hc}\bigl(\, \langle\bra{l;k} \xz c_{-1} \cz c_1 \ket{l';k'} \,\bigr) 
= - \left(-1\right)^l \,\overline{C_N}\, \hs\delta^{10} (k-k') \hs\delta_{l+l',-2}\,.
\end{align}
Thus $C_N$ has to be pure imaginary.
Let us set $C_N = \frac{\iu}{2}\,(2\pi)^{10}$, in which case we have
\begin{equation}
\bra{l;k} \xi\hs c\hs \p c\hs \p^2 c \ket{l';k'} = -\iu \left(2\pi\right)^{10} \delta^{10} (k+k') \hs\delta_{l+l',-2}\,.
\label{normalization}
\end{equation}
Eq.~\eqref{hc bpz} tells us that the Hermitian conjugate of $\ket{s(l;k)}$ coincides with the BPZ conjugate of $\ket{s(l;-k)}$ 
except for some sign factor $\varepsilon_s$:
\begin{equation}
\langle\bra{s(l;k)} = \varepsilon_s \bra{s(l;-k)} \,.
\label{s s}
\end{equation}
The factor $\varepsilon_s$ depends not on the coherent state \eqref{ket} but only on the mode operators.
Using eqs.~\eqref{expansion} and \eqref{s s}, we obtain
\begin{align}
\braket{\Phi_{(0,\hs 0)}, Q\ez\hs \Phi_{(0,\hs 0)}}
&= \int \frac{\diff^{10} k}{(2\pi)^{10}}\hs \frac{\diff^{10} k'}{(2\pi)^{10}}\hs \sum_{s,s'} \sum_{l,l'} 
\varphi^{s'}_{l'} (k') \bra{s'(l'; k')} Q\ez \ket{s(l; k)} \varphi^s_l (k) \nonumber \\*
&= \int \frac{\diff^{10} k}{(2\pi)^{10}}\hs \frac{\diff^{10} k'}{(2\pi)^{10}}\hs \sum_{s,s'} \sum_{l,l'} 
\varepsilon_{s'}\, \varphi^{s'}_{l'} (k') \, \langle\bra{s'(l'; -k')} Q\ez \ket{s(l; k)} \varphi^s_l (k) \nonumber \\*
&= \int \frac{\diff^{10} k}{(2\pi)^{10}}\hs \frac{\diff^{10} k'}{(2\pi)^{10}}\hs \sum_{s,s'} \sum_{l,l'} 
\varepsilon_{s}\, \varphi^{s}_{l} (-k) \, \langle\bra{s(l; k)} Q\ez \ket{s'(l'; k')} \varphi^{s'}_{l'} (k') \,,
\label{action} \\[1ex]
\overline{\braket{\Phi_{(0,\hs 0)}, Q\ez\hs \Phi_{(0,\hs 0)}}}
&= - \int \frac{\diff^{10} k}{(2\pi)^{10}}\hs \frac{\diff^{10} k'}{(2\pi)^{10}}\hs \sum_{s,s'} \sum_{l,l'} 
\overline{\varphi^{s}_{l} (k)} \, \langle\bra{s(l; k)} Q\ez \ket{s'(l'; -k')} \varepsilon_{s'}\, \overline{\varphi^{s'}_{l'} (k')}
\nonumber \\*
&= - \int \frac{\diff^{10} k}{(2\pi)^{10}}\hs \frac{\diff^{10} k'}{(2\pi)^{10}}\hs \sum_{s,s'} \sum_{l,l'} 
\overline{\varphi^{s}_{l} (k)} \, \langle\bra{s(l; k)} Q\ez \ket{s'(l'; k')} \varepsilon_{s'}\, \overline{\varphi^{s'}_{l'} (-k')}\,.
\label{bar action}
\end{align}
Comparing the above two equations, we find that $\braket{\Phi_{(0,\hs 0)}, Q\ez\hs \Phi_{(0,\hs 0)}}$ \emph{multiplied by an imaginary unit}
is real if component fields satisfy
\begin{equation}
\overline{\varphi^{s}_{l} (k)} = - \varepsilon_{s}\, \varphi^{s}_{l} (-k) \,,
\label{comp reality}
\end{equation}
or equivalently 
\begin{equation}
\overline{\varphi^{s}_{l} (x)} = - \varepsilon_{s}\, \varphi^{s}_{l} (x) 
\end{equation}
in the coordinate representation. Condition \eqref{comp reality} can be expressed in terms of the string field as
\begin{equation}
\Phi^\ddagger_{(0,\hs 0)} = - \Phi_{(0,\hs 0)} \,,
\label{ddagger}
\end{equation}
where $\ddagger$ denotes the composition of Hermitian conjugation and the inverse BPZ conjugation ($\mrm{bpz}^{-1}$):
\begin{equation}
\Phi^\ddagger_{(0,\hs 0)} := \mrm{bpz}^{-1}\circ \mrm{hc}\left(\Phi_{(0,\hs 0)}\right).
\end{equation}
It should be mentioned that the other possibility $\Phi_{(0,\hs 0)}^\ddagger = + \Phi_{(0,\hs 0)}$ 
($\Longleftrightarrow \overline{\varphi^{s}_{l} (k)} = + \varepsilon_{s}\, \varphi^{s}_{l} (-k)$)
is excluded when we consider the interacting theory: only condition \eqref{ddagger} is compatible with the equation of motion
\begin{equation}
\ez \bigl( \e^{-\Phi_{(0, 0)}}\, Q \e^{\Phi_{(0, 0)}}\bigr) = 0
\end{equation}
because we have
\begin{equation}
\ez \bigl( \e^{-\Phi_{(0, 0)}}\, Q \e^{\Phi_{(0, 0)}}\bigr) = 0\ \iff\ 
0 = \Bigl( \ez \bigl( \e^{-\Phi_{(0, 0)}}\, Q \e^{\Phi_{(0, 0)}}\bigr) \Bigr)^\ddagger 
= -\ez \Bigl( \bigl( Q\hs\e^{\Phi_{(0,0)}^\ddagger}\bigr) \e^{-\Phi_{(0,0)}^\ddagger}\Bigr)\,.
\end{equation}
Here we have used the relations
\begin{equation}
\left( Q\hs \Phi\right)^\ddagger = -(-1)^{\epsilon\left(\Phi\right)}\hs Q\hs\Phi^\ddagger\,,\quad 
\left( \ez \Phi\right)^\ddagger = -(-1)^{\epsilon\left(\Phi\right)}\hs \ez \Phi^\ddagger\,,\quad
\left( \Phi\hs\Psi\right)^\ddagger = \Psi^\ddagger\hs \Phi^\ddagger\,,
\label{properties of ddagger}
\end{equation}
where $\Phi$ and $\Psi$ are string fields, and $\gp{\Phi}$ denotes the Grassmann parity of $\Phi$.

In summary, in order for the action to be real, we have to consider not $\braket{\Phi_{(0,\hs 0)}, Q\ez\hs \Phi_{(0,\hs 0)}}$ 
but $\iu \times \braket{\Phi_{(0,\hs 0)}, Q\ez\hs \Phi_{(0,\hs 0)}}$ with $\Phi_{(0,\hs 0)}$ satisfying the reality condition \eqref{ddagger}.
One should note that in our convention, in which $Q$ and $\ez$ are Hermitian,
the operator $Q\ez$ is not Hermitian but $\iu \times Q\ez$ is.
The sign factor $\varepsilon_s$ for the state
\begin{align}
\ket{s(l;k)} = 
\bigl( \alpha^{\mu_1}_{-k_1}\cdots\alpha^{\mu_i}_{-k_i} \bigr) 
&\bigl( \psi^{\nu_1}_{-r_1}\cdots\psi^{\nu_j}_{-r_j} \bigr)
\bigl( b_{-\ell_1}\cdots b_{-\ell_q} \bigr) \bigl( c_{-\ell'_1}\cdots c_{-\ell'_{q'}} \bigr) \nonumber \\*
&\times \bigl( \xi_{-m_1} \cdots \xi_{-m_t} \bigr) \bigl( \eta_{-m'_1} \cdots \eta_{-m'_{t'}} \bigr)
\bigl( \phi_{-n_1} \cdots \phi_{-n_a} \bigr) \ket{l;k} 
\label{state}
\end{align}
with
\bs
\begin{align}
1 &\leq k_1 \leq k_2 \leq \cdots \leq k_i\,,\quad \frac{1}{2}\leq r_1 < r_2 < \cdots < r_j\,,\\
2 &\leq \ell_1 < \ell_2 < \cdots < \ell_q\,,\quad -1 \leq \ell'_1 < \ell'_2 < \cdots < \ell'_{q'}\,,\\
0 &\leq m_1 < m_2 < \cdots < m_t\,,\quad 1\leq m'_1 < m'_2 < \cdots < m'_{t'}\,,\quad 1\leq n_1 \leq n_2 \leq \cdots \leq n_a\,,
\end{align}
\es
\begin{equation}
j+l \in 2\mathbb{Z}\,,\quad -q+q'-t+t' = 0\,,\quad t-t' +l = 0
\label{GSOeven}
\end{equation}
is explicitly computed as
\begin{equation}
\varepsilon_s = (-1)^{-N + i + \frac{j^2}{2}}\,,\quad 
N:= \sum_a k_a + \sum_b r_b + \sum_c \ell_c + \sum_{c'} \ell'_{c'} + \sum_d m_d + \sum_{d'} m'_{d'} + \sum_e n_e\,.
\end{equation}
The conditions \eqref{GSOeven} ensure that the state \eqref{state} is GSO even and is of $(g,p)=(0,0)$.

Consistency with the reality condition \eqref{ddagger} sets constraints also on the parameters of the gauge transformation \eqref{gdof}:
\begin{equation}
\Lambda^\ddagger_{(-1,\hs 0)} = -\Lambda_{(-1,\hs 0)}\,,\quad
\Lambda^\ddagger_{(-1,\hs 1)} = -\Lambda_{(-1,\hs 1)}\,.
\end{equation}
These constraints follow immediately from the first two relations in eq.~\eqref{properties of ddagger}.

\section{Proofs of propositions \ref{reachability bxi}\ and \ref{reachability bxi anti}}
\label{reachability}
\setcounter{equation}{0}
\indenths In the present appendix, we give proofs of propositions~\ref{reachability bxi} and~\ref{reachability bxi anti},
which guarantee the reachability of conditions \eqref{gfc for ghosts} and \eqref{gfc for antighosts}.
In what follows, we utilize the relations \eqref{LbarLxy}-\eqref{X_0 bar}.
\begin{Propbis}{reachability bxi}
For any set of string fields of the form $\{ \Phi_{(-n,\hs m)} \mid 0 \leq m \leq n \}$ ($n\geq 0$ is fixed),
there exists a set of string fields 
$\{ \Lambda_{(-(n+1),\hs m)} \mid 0 \leq m \leq n+1 \}$ such that
\begin{equation}
\bsymb{B^{\zz;\hs x,\hs y}_{n+2,\hs n+1}} \Bigl( \bsymb{\Phi_{-n}} + \bsymb{Q_{n+1,n+2}\hs \Lambda_{-(n+1)}} \Bigr) =0
\quad \left( n\geq 0\right).
\tag{\ref{bxi gfc} bis}
\end{equation}
\end{Propbis}
\vspace{1ex}
{\bf \underline{Proof.}}\quad Eq.~\eqref{bxi gfc} is equivalent to the following set of equations:
\bs \label{xy eq}
\begin{align}
&\bz\bigl( \Phi_{(-n,\hs 0)} + Q\Lambda_{(-(n+1),\hs 0)}\bigr) +\bz\eta_0 \Lambda_{(-(n+1),\hs 1)} =0
\quad \left( n\geq 0\right),\\[1.5ex]
&y\hs\zz\Bigl(\bigl( \Phi_{(-n,\hs m)} + Q\Lambda_{(-(n+1),\hs m)} \bigr) +\eta_0\Lambda_{(-(n+1),\hs m+1)}\Bigr)
+x\hs\bz \bigl( \Phi_{(-n,\hs m+1)} +Q\Lambda_{(-(n+1),\hs m+1)} \bigr) \nonumber \\*[1ex]
&+x\hs\bz\eta_0 \Lambda_{(-(n+1),\hs m+2)} =0
\quad \left( 0\leq m \leq n-1 \right),\\[1.5ex]
&\zz\Bigl(\bigl( \Phi_{(-n,\hs n)} + Q\Lambda_{(-(n+1),\hs n)} \bigr) +\eta_0\Lambda_{(-(n+1),\hs n+1)}\Bigr) =0
\quad \left( n\geq 0 \right).
\end{align}
\es
{\bf (I) \boldmath{$x=0$} (and \boldmath{$y\neq 0$}) case} \\*
\indenths
For the purpose of showing the existence of the solution to eq.~\eqref{xy eq}, 
it is sufficient to find $\Lambda_{(-(n+1),\hs m)}$ ($0\leq n$, $0\leq m\leq n+1$)
such that
\bs \label{x=0 eq}
\begin{align}
&\bz\bigl( \Phi_{(-n,\hs 0)} + Q\Lambda_{(-(n+1),\hs 0)}\bigr) =0\,,\quad \bz\eta_0 \Lambda_{(-(n+1),\hs 1)} =0
\quad \left( n\geq 0\right),\\[1ex]
&\zz\Bigl(\bigl( \Phi_{(-n,\hs m)} + Q\Lambda_{(-(n+1),\hs m)} \bigr) +\eta_0\Lambda_{(-(n+1),\hs m+1)}\Bigr) =0
\quad \left( 0\leq m \leq n \right).
\end{align}
\es
These equations are solved by
\bs \label{x=0 sol}
\begin{align}
\Lambda_{(-(n+1),\hs 0)} &= -\frac{\bz}{L_0}\hs \Phi_{(-n,\hs 0)}
\quad \left( n\geq 0\right),\\[1ex]
\Lambda_{(-(n+1),\hs m+1)} &= -\zzbar \bigl( \Phi_{(-n,\hs m)} + Q\Lambda_{(-(n+1),\hs m)} \bigr)
\quad \left( 0\leq m \leq n \right)
\end{align}
\es
with
\begin{equation}
\zzbar := 
\left\{
\begin{aligned}
&\xz \quad \left( \text{for } \zz = \xz \right),\\[.5ex]
&\frac{\dz}{L_0} \quad \left( \text{for } \zz = \dz \right).
\end{aligned}
\right.
\label{zzbar}
\end{equation}
Indeed, given eq.~\eqref{x=0 sol}, we have
\bs
\begin{align}
& \Phi_{(-n,\hs 0)} + Q\Lambda_{(-(n+1),\hs 0)} = \frac{\bz Q}{L_0}\hs \Phi_{(-n,\hs 0)}
\quad \left( n\geq 0\right),\\[.5ex]
& \Lambda_{(-(n+1),\hs 1)} = -\frac{\zzbar\bz Q}{L_0}\hs \Phi_{(-n,\hs 0)}\,,\quad
\bz\eta_0 \Lambda_{(-(n+1),\hs 1)} =0
\quad \left( n\geq 0\right),\\[1ex]
\bigl( &\Phi_{(-n,\hs m)} + Q\Lambda_{(-(n+1),\hs m)} \bigr) +\eta_0\Lambda_{(-(n+1),\hs m+1)}
=\zzbar\eta_0 \bigl( \Phi_{(-n,\hs m)} + Q\Lambda_{(-(n+1),\hs m)} \bigr)
\quad \left( 0\leq m\leq n\right).
\end{align}
\es
{\bf (II) \boldmath{$x\neq 0$} case} \\*
\indenths
It is sufficient to show the existence of $\Lambda_{(-(n+1),\hs m)}$
($0\leq n$, $0\leq m\leq n+1$) such that
\bs \label{x neq 0 eq}
\begin{align}
&\bsymb{\mrm{{\bf Eq}}_{-(n+1),\hs 0}}:\quad
\bz \bigl( \Phi_{(-n,\hs 0)} + Q\Lambda_{(-(n+1),\hs 0)} \bigr) =0
\qquad \left( n\geq 0 \right),\\[1.5ex]
&\bsymb{\mrm{{\bf Eq}}_{-(n+1),\hs m+1}}:\quad \nonumber \\*[1ex]
&\left\{
\begin{aligned}
&\bz\eta_0 \Lambda_{(-(n+1),\hs m+1)} =0 \\[.5ex]
&y\hs\zz\Bigl(\bigl( \Phi_{(-n,\hs m)} + Q\Lambda_{(-(n+1),\hs m)} \bigr) +\eta_0\Lambda_{(-(n+1),\hs m+1)}\Bigr)
+x\hs\bz \bigl( \Phi_{(-n,\hs m+1)} +Q\Lambda_{(-(n+1),\hs m+1)} \bigr) =0  
\end{aligned}
\right.
\nonumber \\*[1ex]
&\quad \left( 0\leq m \leq n-1 \right),\\[1.5ex]
&\bsymb{\mrm{{\bf Eq}}_{-(n+1),\hs n+1}}:\quad
\left\{
\begin{aligned}
&\bz\eta_0 \Lambda_{(-(n+1),\hs n+1)} =0 \\[.5ex]
&\zz\Bigl(\bigl( \Phi_{(-n,\hs n)} + Q\Lambda_{(-(n+1),\hs n)} \bigr) +\eta_0\Lambda_{(-(n+1),\hs n+1)}\Bigr) =0
\end{aligned}
\right.
\qquad \left( n\geq 0\right).
\end{align}
\es
If $n=0$, eq.~\eqref{x neq 0 eq} is the same as eq.~\eqref{x=0 eq}.
Therefore we concentrate on the case in which $n \geq 1$.
Noting that $\bsymb{\mrm{{\bf Eq}}_{-(n+1),\hs 0}}$ $\left( n\geq 1 \right)$ is solved by the choice
\begin{equation}
\Lambda_{(-(n+1),\hs 0)} = -\frac{\bz}{L_0}\hs \Phi_{(-n,\hs 0)}
\qquad \left( n\geq 1\right),
\end{equation}
we find that the following lemma ensures the existence of the solution to eq.~\eqref{x neq 0 eq}.
\qquad $\Box$
\\
\begin{Lem}
\label{Lem1}
Let $n\geq 1$ be fixed, and let $x$ be nonzero.
If \hs $\bsymb{\mrm{{\bf Eq}}_{-(n+1),\hs m}}$ holds for some $m$ $\left(0\leq \exists m\leq n\right)$,
then there exists $\Lambda_{(-(n+1),\hs m+1)}$ which solves $\bsymb{\mrm{{\bf Eq}}_{-(n+1),\hs m+1}}$.
\end{Lem}
\vspace{1ex}
{\bf \underline{Proof.}}\quad
By the use of the relation
\begin{equation}
\left\{\bz, \czbar \right\} = \left\{\zz , \ezbar \right\} = 1
\end{equation}
with
\begin{equation}
\czbar := \left\{
\begin{aligned}
&\cz \quad \left( \text{for } \zz = \xz \right)\\[.5ex]
&\frac{Q}{L_0} \quad \left( \text{for } \zz = \dz \right)
\end{aligned}
\right.
\,,\quad
\ezbar := \left\{
\begin{aligned}
&\eta_0 \quad \left( \text{for } \zz = \xz \right)\\[.5ex]
&\frac{\eta_0}{L_0} \quad \left( \text{for } \zz = \dz \right)
\end{aligned}
\right.
\,,
\end{equation}
one can decompose any string field $\Phi_{(g,\hs p)}$
of world-sheet ghost number $g$ and picture number $p$ as\footnote{Properties of the barred operators are given
at the end of appendix~\ref{algebraic relations}.}
\begin{equation}
\Phi_{(g,\hs p)} = 
\czbar \zz \, \Pcz_{(g,\hs p)}
+ \czbar \, \Pc_{(g,\hs p)}
+ \zz \, \Pz_{(g,\hs p)}
+ \Pn_{(g,\hs p)}
\label{decomposition}
\end{equation}
with
\begin{equation}
\Pcz_{(g,\hs p)} = \ezbar \bz \Phi_{(g,\hs p)} \,,\quad 
\Pc_{(g,\hs p)} = \ezbar \zz \bz \Phi_{(g,\hs p)} \,,\quad 
\Pz_{(g,\hs p)} = \ezbar \bz \czbar \Phi_{(g,\hs p)} \,,\quad
\Pn_{(g,\hs p)} = \ezbar \zz \bz \czbar \Phi_{(g,\hs p)} \,.
\end{equation}
It should be noted that the subscripts on $\Pcz_{(g,\hs p)}$, $\Pc_{(g,\hs p)}$, $\Pz_{(g,\hs p)}$, and $\Pn_{(g,\hs p)}$
are simply carried over from $\Phi_{(g,\hs p)}$, and thus they do not necessarily indicate the quantum numbers of the string fields.
Indeed, whereas $\Pn_{(g,\hs p)}$ is of quantum number $(g,p)$, 
the other fields $\Pcz_{(g,\hs p)}$, $\Pc_{(g,\hs p)}$, and $\Pz_{(g,\hs p)}$ \emph{are not}:
their quantum numbers are $(g,p-1)$, $(g-1,p)$, and $(g+1,p-1)$, respectively.
All of the string fields $\Pcz_{(g,\hs p)}$, $\Pc_{(g,\hs p)}$, $\Pz_{(g,\hs p)}$, and $\Pn_{(g,\hs p)}$
are annihilated by both $\bz$ and $\eta_0$.
\\[1ex]
\indenths
The equations $\bsymb{\mrm{{\bf Eq}}_{-(n+1),\hs m}}$ 
and \hs $\bz \eta_0 \Lambda_{(-(n+1),\hs m+1)} = 0$ \hs 
$\left(0\leq m\leq n\right)$ \hs 
respectively imply
\begin{equation}  
\tPc_{(-n,\hs m)} = 0\hs \quad
(\text{this follows from}\; \ 
\zz\bz \tP_{(-n,\hs m)} =0)
\label{tP -n m}
\end{equation}
and
\begin{equation}
\Lcz_{(-(n+1),\hs m+1)} = 0\,.
\label{Lambda}
\end{equation}
Here we have defined $\tP_{(-n,\hs m)}$ $\left( 0\leq m \leq n\right)$ as
\begin{equation}
\tP_{(-n,\hs m)} := \Phi_{(-n,\hs m)} + Q\Lambda_{(-(n+1),\hs m)} \,.
\end{equation}
Our proof of the lemma is composed of two parts.
\\[1.5ex]
{\bf (I)}
$\bsymb{\mrm{{\bf Eq}}_{-(n+1),\hs m}} \quad \left(0\leq \exists m\leq n-1\right) \ \Longrightarrow \ 
\exists \Lambda_{(-(n+1),\hs m+1)}$, s.t.\ $\bsymb{\mrm{{\bf Eq}}_{-(n+1),\hs m+1}}$ holds.
\\*[.5ex]
\indenths
We seek $\Lambda_{(-(n+1),\hs m+1)}$ such that
\begin{equation}
y\hs\zz \left( \tP_{(-n,\hs m)} + \eta_0 \Lambda_{(-(n+1),\hs m+1)} \right) 
+ x\hs\bz \left( \Phi_{(-n,\hs m+1)} + Q \Lambda_{(-(n+1),\hs m+1)} \right) 
=0\,,
\label{L -(n+1) m+1}
\end{equation}
under conditions \eqref{tP -n m} and \eqref{Lambda}.
This suffices to prove claim (I).
Substituting eqs.~\eqref{tP -n m} and \eqref{Lambda} into eq.~\eqref{L -(n+1) m+1},
we obtain
\begin{align}
&\zz \Bigl[ \bigl( y\hs\tPn_{(-n,\hs m)} + x\hs\Pcz_{(-n,\hs m+1)} \bigr) + \Lxy \Lz_{(-(n+1),\hs m+1)} \Bigr]
+ x \Bigl[ \Pc_{(-n,\hs m+1)} +  L_0 \Ln_{(-(n+1),\hs m+1)} \Bigr] \nonumber \\*[1ex]
&- x \left( Q -\czbar L_0 \right) \Lc_{(-(n+1),\hs m+1)}
=0\,.
\end{align}
Thus if we set
\bs
\begin{align}
\Lcz_{(-(n+1),\hs m+1)} &= \Lc_{(-(n+1),\hs m+1)} =0\,,\\[1.5ex]
\Lz_{(-(n+1),\hs m+1)} 
&= -\frac{1}{\Lxy} \bigl( y\hs\tPn_{(-n,\hs m)} + x\hs\Pcz_{(-n,\hs m+1)} \bigr) \nonumber \\*[1ex]
&= -\frac{1}{\Lxy} \bigl( y\hs\ezbar \zz \bz \czbar \tP_{(-n,\hs m)} + x\hs\ezbar \bz \Phi_{(-n,\hs m+1)} \bigr) ,
\\[1.5ex]
\Ln_{(-(n+1),\hs m+1)} 
&= -\frac{1}{L_0}\hs \Pc_{(-n,\hs m+1)} = - \frac{\ezbar \zz \bz}{L_0}\hs \Phi_{(-n,\hs m+1)} \,,
\end{align}
\es
that is, if we set
\begin{equation}
\Lambda_{(-(n+1),\hs m+1)}
= -\frac{\ezbar \zz \bz}{L_0}\hs \Phi_{(-n,\hs m+1)}
- \frac{\zz}{\Lxy} \bigl( y\hs\bz \czbar \tP_{(-n,\hs m)} + x\hs \ezbar \bz \Phi_{(-n,\hs m+1)} \bigr),
\end{equation}
eq.~\eqref{L -(n+1) m+1} is indeed satisfied.
\\[1.5ex]
{\bf (II)}
$\bsymb{\mrm{{\bf Eq}}_{-(n+1),\hs n}} \ \Longrightarrow \ 
\exists \Lambda_{(-(n+1),\hs n+1)}$, s.t.\ $\bsymb{\mrm{{\bf Eq}}_{-(n+1),\hs n+1}}$ holds.
\\*[.5ex]
\indenths
The proof of claim (II) is completed if we find $\Lambda_{(-(n+1),\hs n+1)}$ such that
\begin{equation}
\zz \bigl( \tP_{(-n,\hs n)} + \eta_0 \Lambda_{(-(n+1),\hs n+1)} \bigr) = 0\,,
\label{L -(n+1) n+1}
\end{equation}
under conditions \eqref{tP -n m} and \eqref{Lambda} with $m=n$.
Putting eqs.~\eqref{tP -n m} and \eqref{Lambda} with $m=n$ into eq.~\eqref{L -(n+1) n+1}
leads to
\begin{equation}
\zz \bigl( \tPn_{(-n,\hs n)} + \Lbar \Lz_{(-(n+1),\hs n+1)} \bigr) = 0\,.
\end{equation}
Thus the choice
\begin{align}
\Lcz_{(-(n+1),\hs n+1)} &= \Lc_{(-(n+1),\hs n+1)} = \Ln_{(-(n+1),\hs n+1)} = 0\,,\quad
\Lz_{(-(n+1),\hs n+1)} = -\frac{1}{\Lbar}\hs\tPn_{(-n,\hs n)} \,,\quad \text{i.e.,}
\nonumber \\*[1ex]
\Lambda_{(-(n+1),\hs n+1)} &= -\zzbar \bz \czbar \tP_{(-n,\hs n)} 
\end{align}
solves eq.~\eqref{L -(n+1) n+1}.
\qquad $\Box$
\\
\begin{Propbis}{reachability bxi anti}
\ \\
{\bf {\upshape (A)}}
For any string field $\Phi_{(2,\hs -1)}$ of indicated quantum number,
there exists a string field $\Lambda_{(0,\hs 0)}$ such that
\begin{equation}
\bz\zz \left( \Phi_{(2,\hs -1)} + Q\eta_0\hs \Lambda_{(0,\hs 0)} \right) =0 \,.
\tag{\ref{bxi gfc 2,-1} bis}
\end{equation}
{\bf {\upshape (B)}}
Let $n\geq 1$ be fixed.
For any set of string fields of the form $\{ \Phi_{(n+2,\hs -m)} \mid 1 \leq m \leq n+1 \}$,
there exists a set of string fields $\{ \Lambda_{(n+1,\hs  -m)} \mid 1 \leq m \leq n \}$ such that
\begin{equation}
\bsymb{\tB^{\zz;\hs x,\hs y}_{n+2,\hs n+1}} \Bigl( \bsymb{\Phi_{n+2}} + \bsymb{Q_{n+1,n}\hs \Lambda_{n+1}} \Bigr) =0
\quad \left( n\geq 1\right).
\tag{\ref{bxi gfc anti} bis}
\end{equation}
\end{Propbis}
\vspace{1ex}
{\bf \underline{Proof of (A).}}\quad
Eq.~\eqref{bxi gfc 2,-1} is solved by
\begin{equation}
\Lambda_{(0,\hs 0)} = \frac{\bz\zzbar}{L_0}\hs \Phi_{(2,\hs -1)}
\end{equation}
because we have
\begin{equation}
1+ Q\eta_0 \left( \frac{\bz\zzbar}{L_0}\right)
= 1- \left( \frac{Q\bz}{L_0} \right) \eta_0 \zzbar
= 1- \left( 1- \frac{\bz Q}{L_0}\right) \left( 1-\zzbar\eta_0 \right)
= \frac{\bz Q}{L_0} + \zzbar\eta_0 - \frac{\bz Q}{L_0}\hs \zzbar\eta_0 \,.
\end{equation}
\ \\
{\bf \underline{Proof of (B).}}\quad
We prove the proposition case by case:
the $x=0$ case, the $y=0$ case, and the $x,\, y \neq 0$ case.
(Note the remark below eq.~\eqref{xy}.)
\\[1ex]
{\bf (I) \boldmath{$x=0$} (and \boldmath{$y\neq 0$}) case} \\*
\indenths
Eq.~\eqref{bxi gfc anti} is reduced to the following set of equations:
\bs
\begin{align}
&\bz\zz \bigl( \Phi_{(n+2,\hs -1)} + Q\Lambda_{(n+1,\hs -1)} \bigr) =0\quad \left( n\geq 1\right),\\[1ex]
&\zz \bigl( \hP_{(n+2,\hs -(m+1))} + \eta_0 \Lambda_{(n+1,\hs -m)}\bigr) =0\quad \left( 1\leq m\leq n \right),
\end{align}
\es
with
\begin{equation}
\hP_{(n+2,\hs -(m+1))} :=
\left\{
\begin{aligned}
&\Phi_{(n+2,\hs -(m+1))} +Q\Lambda_{(n+1,\hs -(m+1))} \quad \left( 1\leq m\leq n-1\right), \\[1ex]
&\Phi_{(n+2,\hs -(n+1))} \quad \left( m=n\right). 
\end{aligned}
\right.
\end{equation}
We can solve this by setting 
\bs
\begin{align}
&\Lambda_{(n+1,\hs -m)} = -\zzbar \hP_{(n+2,\hs -(m+1))} \quad \left( 2\leq m\leq n\right),\\[1ex]
&\Lambda_{(n+1,\hs -1)} = -\zzbar \hP_{(n+2,\hs -2)} + \frac{\bz}{L_0}\hs \eta_0\zzbar
\bigl( -\Phi_{(n+2,\hs -1)} + \Xbar \hP_{(n+2,\hs -2)} \bigr) \quad \left(n\geq 1\right),
\end{align}
\es
with
\begin{equation}
\Xbar := \{ Q,\zzbar \} =
\left\{
\begin{aligned}
&X_0\quad \bigl(\text{for } \zz = \xz\bigr),\\
&0\quad \bigl(\text{for } \zz = \dz\bigr).
\end{aligned}
\right.
\end{equation}
\\
{\bf (II) \boldmath{$y=0$} (and \boldmath{$x\neq 0$}) case} \\*
\indenths
Eq.~\eqref{bxi gfc anti} is reduced to
\bs
\begin{align}
&\bz\bigl( \hP_{(n+2,\hs -m)} + Q\Lambda_{(n+1,\hs -m)}\bigr) =0\quad \left( 1\leq m \leq n\right),\\[1ex]
&\bz\zz \bigl( \Phi_{(n+2,\hs -(n+1))} +\eta_0 \Lambda_{(n+1,\hs -n)} \bigr) =0\quad \left( n\geq 1\right),
\end{align}
\es
with
\begin{equation}
\hP_{(n+2,\hs -m)} :=
\left\{
\begin{aligned}
&\Phi_{(n+2,\hs -1)}\quad \left( m=1\right), \\[1ex]
&\Phi_{(n+2,\hs -m)} +\eta_0 \Lambda_{(n+1,\hs -(m-1))} \quad \left( 2\leq m\leq n\right) .
\end{aligned}
\right.
\end{equation}
The above equations are solved by
\bs
\begin{align}
&\Lambda_{(n+1,\hs -m)} = -\frac{\bz}{L_0}\hs \hP_{(n+2,\hs -m)} \quad \left( 1\leq m\leq n\right),\\[1ex]
&\Lambda_{(n+1,\hs -n)} = -\frac{\bz}{L_0}\hs \hP_{(n+2,\hs -n)} - \frac{Q\bz}{L_0}\hs\zzbar \Phi_{(n+2,\hs -(n+1))} \quad \left( n\geq 1\right).
\end{align}
\es
\\
{\bf (III) \boldmath{$x\neq 0$}, \boldmath{$y\neq 0$} case} \\*
\indenths
According as $n=1$ or $n\geq 2$, the set of constituent equations of \eqref{bxi gfc anti} takes the different form:
for $n=1$, we have
\begin{equation}
x\hs\bz\bigl( \Phi_{(3,\hs -1)} + Q\Lambda_{(2,\hs -1)}\bigr)
+y\hs\zz\bigl( \Phi_{(3,\hs -2)} + \eta_0 \Lambda_{(2,\hs -1)} \bigr) =0\,,
\label{n=1}
\end{equation}
whereas for $n\geq 2$, we have the three types of constituent equations
\bs \label{n neq 1}
\begin{align}
&x\hs\bz \bigl( \Phi_{(n+2,\hs -1)} + Q\Lambda_{(n+1,\hs -1)} \bigr)
+y\hs\zz\bigl( \Phi_{(n+2,\hs -2)} +\eta_0 \Lambda_{(n+1,\hs -1)} +Q\Lambda_{(n+1,\hs -2)} \bigr)
=0\quad \left( n\geq 2\right),\\[1.5ex]
&x\hs\bz \bigl( \Phi_{(n+2,\hs -m)} + \eta_0 \Lambda_{(n+1,\hs -(m-1))} + Q\Lambda_{(n+1,\hs -m)} \bigr) \nonumber \\*[.8ex]
&+y\hs\zz\bigl( \Phi_{(n+2,\hs -(m+1))} +\eta_0 \Lambda_{(n+1,\hs -m)} +Q\Lambda_{(n+1,\hs -(m+1))} \bigr)
=0\quad \left( 2 \leq m \leq n-1 \right),\\[1.5ex]
&x\hs\bz \bigl( \Phi_{(n+2,\hs -n)} + \eta_0 \Lambda_{(n+1,\hs -(n-1))} + Q\Lambda_{(n+1,\hs -n)} \bigr)
+y\hs\zz\bigl( \Phi_{(n+2,\hs -(n+1))} +\eta_0 \Lambda_{(n+1,\hs -n)} \bigr)
=0\quad \left( n\geq 2 \right).
\end{align}
\es
However, we can prove the proposition by considering only a single set of equations:
it is sufficient to show the existence of a solution to the set of equations below.
\bs
\begin{align}
&\bsymb{\mrm{{\bf Eq}}_{n+1,\hs -n}}:\quad
x\hs\bz \bigl( \Phi_{(n+2,\hs -n)} + Q\Lambda_{(n+1,\hs -n)} \bigr)
+y\hs\zz\bigl( \Phi_{(n+2,\hs -(n+1))} +\eta_0 \Lambda_{(n+1,\hs -n)} \bigr)
=0\quad \left( n\geq 1 \right),\\[-.2ex]
&\bsymb{\mrm{{\bf Eq}}_{n+1,\hs -m}}: \nonumber \\*[1ex]
&\left\{
\begin{aligned}
&\bz \eta_0 \Lambda_{(n+1,\hs -m)} = 0 \\[.5ex]
&x\hs\bz \bigl( \Phi_{(n+2,\hs -m)} + Q\Lambda_{(n+1,\hs -m)} \bigr) 
+y\hs\zz \Bigl( \bigl( \Phi_{(n+2,\hs -(m+1))} +Q\Lambda_{(n+1,\hs -(m+1))} \bigr) +\eta_0 \Lambda_{(n+1,\hs -m)} \Bigr)
=0
\end{aligned}
\right.
\nonumber \\*[1ex]
&\quad
\left( 1\leq m\leq n-1 \right).
\end{align}
\es
The following lemma completes the proof.
\qquad $\Box$
\\
\begin{Lem}
Let $n\geq 1$ be fixed, and let $x$ and $y$ be nonzero.\\
{\bf {\upshape (A)}}
There exists a solution $\Lambda_{(n+1,\hs -n)}$ to the equation $\bsymb{\mrm{{\bf Eq}}_{n+1,\hs -n}}$.\\
{\bf {\upshape (B)}}
If $\bsymb{\mrm{{\bf Eq}}_{n+1,\hs -(m+1)}}$ holds for some $m$ $\left(1\leq \exists m\leq n-1 \right)$,
then there exists $\Lambda_{(n+1,\hs -m)}$ which solves $\bsymb{\mrm{{\bf Eq}}_{n+1,\hs -m}}$.
\end{Lem}
\vspace{1ex}
{\bf \underline{Proof of (A).}}\quad
Using the zero-mode decomposition \eqref{decomposition}, we find that $\bsymb{\mrm{{\bf Eq}}_{n+1,\hs -n}}$ is equivalent to
\bs
\begin{align}
&y\hs \czbar \zz \Bigl[ -\Pc_{(n+2,\hs -(n+1))} + \Lbar \Lcz_{(n+1,\hs -n)} \Bigr] \nonumber \\*[1ex]
&+\zz \Bigl[ 
x\hs\Pcz_{(n+2,\hs -n)} + y\hs\Pn_{(n+2,\hs -(n+1))} +\Lxy \Lz_{(n+1,\hs -n)} +x\bigl( Q- \czbar L_0 -\eta_0 \Xbar \bigr) \Lcz_{(n+1,\hs -n)}
\Bigr] \nonumber \\*[1ex]
&+x\Bigl[
\Pc_{(n+2,\hs -n)} +L_0 \Ln_{(n+1,\hs -n)} -\Lbar\Xbar \Lcz_{(n+1,\hs -n)} -\bigl( Q -\czbar L_0 - \eta_0 \Xbar \bigr) \Lc_{(n+1,\hs -n)} 
\Bigr] =0\,.
\end{align}
\es
Thus, if we set
\bs
\begin{align}
\Lcz_{(n+1,\hs -n)} &= \frac{1}{\Lbar}\hs \Pc_{(n+2,\hs -(n+1))}\,,\\[1ex]
\Lc_{(n+1,\hs -n)} &= 0\,,\\[1ex]
\Lz_{(n+1,\hs -n)} &= -\frac{1}{\Lxy} \Bigl( x\hs\Pcz_{(n+2,\hs -n)} + y\hs\Pn_{(n+2,\hs -(n+1))}
+ x\bigl( Q- \czbar L_0 -\eta_0 \Xbar \bigr) \Lcz_{(n+1,\hs -n)} \Bigr)\,,\\[1ex]
\Ln_{(n+1,\hs -n)} &= -\frac{1}{L_0} \bigl( \Pc_{(n+2,\hs -n)} -\Lbar\Xbar \Lcz_{(n+1,\hs -n)} \bigr)\,,
\end{align}
\es
that is, if we set
\begin{align}
\Lambda_{(n+1,\hs -n)}
=&\ -\zzbar\Phi_{(n+2,\hs -(n+1))}
+ \frac{x\hs\zzbar\bz}{\Lxy} \Bigl( \eta_0 \Phi_{(n+2,\hs -n)} +\bigl( Q-\eta_0 \Xbar \bigr) \Phi_{(n+2,\hs -(n+1))}\Bigr)
\nonumber \\*[.5ex]
&-\frac{\eta_0 \zzbar\bz}{L_0} \bigl( \Phi_{(n+2,\hs -n)} - \Xbar \Phi_{(n+2,\hs -(n+1))} \bigr)\,,
\end{align}
the equation $\bsymb{\mrm{{\bf Eq}}_{n+1,\hs -n}}$ is satisfied.\\[2ex]
{\bf \underline{Proof of (B).}}\quad
The equations $\bsymb{\mrm{{\bf Eq}}_{n+1,\hs -(m+1)}}$ and \hs $\bz \eta_0 \Lambda_{(n+1,\hs -m)} = 0$ \hs respectively imply
\begin{equation}
\tPc_{(n+2,\hs -(m+1))} = 0\quad
(\text{this follows from}\; \ \zz\bz \tP_{(n+2,\hs -(m+1))} = 0)
\label{tP n+2 -(m+1)}
\end{equation}
and
\begin{equation}
\Lcz_{(n+1,\hs -m)} = 0\,.
\label{Lcx}
\end{equation}
Here we have defined $\tP_{(n+2,\hs -(m+1))}$ as
\begin{equation}
\tP_{(n+2,\hs -(m+1))} := \Phi_{(n+2,\hs -(m+1))} + Q\Lambda_{(n+1,\hs -(m+1))} \,.
\end{equation}
Therefore, for the purpose of proving the claim, it is sufficient to show the existence of $\Lambda_{(n+1,\hs -m)}$ such that
\begin{equation}
x\hs\bz \bigl( \Phi_{(n+2,\hs -m)} + Q\Lambda_{(n+1,\hs -m)} \bigr) 
+y\hs\zz \bigl( \tP_{(n+2,\hs -(m+1))} +\eta_0 \Lambda_{(n+1,\hs -m)} \bigr)
=0\,,
\label{L n+1 -m}
\end{equation}
under conditions \eqref{tP n+2 -(m+1)} and \eqref{Lcx}.
Because eqs.~\eqref{tP n+2 -(m+1)}, \eqref{Lcx}, and \eqref{L n+1 -m} correspond to
eqs.~\eqref{tP -n m}, \eqref{Lambda}, and \eqref{L -(n+1) m+1},
existence of such a $\Lambda_{(n+1,\hs -m)}$ can be shown in the manner
used when we proved claim (I) in the proof of lemma~\ref{Lem1}.
\qquad $\Box$

\section{Solutions to equations \eqref{P} and \eqref{P anti}}
\label{solutions to PtP}
\setcounter{equation}{0}
\indenths
In the present appendix, we list the solutions to eqs.~\eqref{P} and \eqref{P anti}.

\subsection{Solutions for the \texorpdfstring{$\zz =\xz$}{zeta0 = xi0} case}
{\bf (I) The form of $\bsymb{P}$}\\*
\indenths
Small-size matrices are given by
\bs
\begin{align}
&\bsymb{P^{\xz;\hs x,\hs y}_{1,\hs 1}} = \frac{\iu}{L_0}\,,\quad
\bsymb{P^{\xz;\hs x,\hs y}_{2,\hs 2}}
=
\bbm
\frac{1}{L_0} & \frac{y\hs\eta_0 \bz}{L_0 \Lxy} \\[1.5ex]
-\frac{\xi_0 Q}{L_0}P_\xi & 1 
\ebm
,\\[2ex]
&\bsymb{P^{\xz;\hs x,\hs y}_{3,\hs 3}}
=
\begin{bmatrix}
\frac{1}{L_0} & \frac{\eta_0 \bz}{L_0 \Lxy } & 0 \\[1.5ex]
\frac{P_X}{L_0}P_\xi & \frac{1}{\Lxy} & 0 \\[1ex]
\hline\\[-2ex]
-\xz Q \frac{P_X}{L_0} & -\xz Q \frac{\PXb}{\Lxy} & 1 
\end{bmatrix}
,\quad
\bsymb{P^{\xz;\hs x,\hs y}_{4,\hs 4}}
=
\begin{bmatrix}
\frac{1}{L_0} & \frac{\eta_0 \bz}{L_0 \Lxy } & 0 &0 \\[1.5ex]
\frac{P_X}{L_0}P_\xi & \frac{1}{\Lxy} & 0 &0 \\[1ex]
\hline\\[-2ex]
\frac{P^2_X}{L_0} & \frac{P_X}{\Lxy} \PXb & \frac{1}{\Lxy} &0 \\[1ex]
\hline\\[-2ex]
-\xz Q \frac{P^2_X}{L_0} & -\xz Q \frac{P_X}{\Lxy} \PXb & -\xz Q \frac{1}{\Lxy} & 1 
\end{bmatrix} 
,\\[3ex]
&\bsymb{P^{\xz:\hs x,\hs y}_{5,\hs 5}}
=
\begin{bmatrix}
\frac{1}{L_0} & \frac{\eta_0 \bz}{L_0 \Lxy } & 0 &0 &0 \\[1.5ex]
\frac{P_X}{L_0}P_\xi & \frac{1}{\Lxy} & 0 &0 &0 \\[1ex]
\hline\\[-2ex]
\frac{P^2_X}{L_0} & \frac{P_X}{\Lxy} \PXb & \frac{1}{\Lxy} &0 &0 \\[1.5ex]
\frac{P^3_X}{L_0} & \frac{P^2_X}{\Lxy} \PXb & \frac{P_X}{\Lxy} &\frac{1}{\Lxy} &0 \\[1ex]
\hline\\[-2ex]
-\xz Q\frac{P^3_X}{L_0} & -\xz Q\frac{P^2_X}{\Lxy} \PXb & -\xz Q\frac{P_X}{\Lxy} &-\xz Q\frac{1}{\Lxy} &1
\end{bmatrix}
,
\end{align}
\es
with
\begin{equation}
P_X := -\frac{y X_0}{\Lxy}\,,\quad
P_\xi :=1-y\frac{\eta_0\xz}{\Lxy}\,,\quad
\PXb := 1+ P_X \frac{\ez\bz}{L_0} = 1-y\frac{X_0 \eta_0 \bz}{L_0 \Lxy}\,.
\label{P_X}
\end{equation}
The symbol $X_0$ denotes $\{Q,\xz\}$, which is precisely the zero mode of the picture-changing operator $X(z) = \{Q,\xi(z)\}$.
For an arbitrary $n\left( \geq 1\right)$, we have
\begin{align}
& \bsymb{P^{\xz;\hs x,\hs y}_{n+2,\hs n+2}} \ \left(n\geq 1\right) \nonumber \\*[2ex]
&=
\begin{bmatrix}
\frac{1}{L_0} &\frac{\eta_0 \bz}{L_0 \Lxy } &0 &0 &0 &\cdots &0 &0 &0 \\[2ex]
\frac{P_X}{L_0}P_\xi &\frac{1}{\Lxy} &0 &0 &0 &\cdots &0 &0 &0 \\[2ex]
\hline\\[-2ex]
\frac{P^2_X}{L_0} &\frac{P_X}{\Lxy} \PXb &\frac{1}{\Lxy} &0 &0 &\cdots &0 &0 &0 \\[2ex]
\frac{P^3_X}{L_0} &\frac{P^2_X}{\Lxy} \PXb &\frac{P_X}{\Lxy} &\frac{1}{\Lxy} &0 &\cdots &0 &0 &0 \\[2ex]
\frac{P^4_X}{L_0} &\frac{P^3_X}{\Lxy} \PXb &\frac{P^2_X}{\Lxy} &\frac{P_X}{\Lxy} &\frac{1}{\Lxy} &\cdots &\vdots &\vdots &\vdots \\[1ex]
\vdots &\vdots &\vdots &\vdots &\vdots &\cdots &\vdots &\vdots &\vdots \\
\vdots &\vdots &\vdots &\vdots &\vdots &\cdots & 0 &0 &0 \\[1ex]
\frac{P^{n-1}_X}{L_0} &\frac{P^{n-2}_X}{\Lxy} \PXb &\frac{P^{n-3}_X}{\Lxy} &\frac{P^{n-4}_X}{\Lxy} &\frac{P^{n-5}_X}{\Lxy}
&\cdots &\frac{1}{\Lxy} &0 &0 \\[2ex]
\frac{P^{n}_X}{L_0} &\frac{P^{n-1}_X}{\Lxy} \PXb &\frac{P^{n-2}_X}{\Lxy} &\frac{P^{n-3}_X}{\Lxy} &\frac{P^{n-4}_X}{\Lxy}
&\cdots &\frac{P_X}{\Lxy} &\frac{1}{\Lxy} &0 \\[2ex]
\hline\\[-2ex]
-\xz Q \frac{P^{n}_X}{L_0} &-\xz Q\frac{P^{n-1}_X}{\Lxy} \PXb &-\xz Q\frac{P^{n-2}_X}{\Lxy} &-\xz Q\frac{P^{n-3}_X}{\Lxy} &-\xz Q\frac{P^{n-4}_X}{\Lxy}
&\cdots &-\xz Q\frac{P_X}{\Lxy} &-\frac{\xz Q}{\Lxy} &1 
\end{bmatrix}
.
\end{align}
To avoid confusion, we write down the $(i,\hs j)$-components $p^{\xz;\hs i,\hs j}_{n+2}$ of 
$\bsymb{P^{\xz;\hs x,\hs y}_{n+2,\hs n+2}}$ $(n\geq 1)$ below:
\bs
\begin{align}
&\ \quad p^{\xz;\hs 1,\hs 1}_{n+2} = \frac{1}{L_0} \,,\quad 
p^{\xz;\hs 1,\hs 2}_{n+2} = \frac{\eta_0 \bz}{L_0\Lxy} \,,\quad 
p^{\xz;\hs 1,\hs j}_{n+2} = 0 \quad \left( 3\leq j \leq n+2 \right),
\\[1ex]
&\ \quad p^{\xz;\hs 2,\hs 1}_{n+2} = \frac{P_X}{L_0} P_\xi \,,\quad
p^{\xz;\hs 2,\hs 2}_{n+2} = \frac{1}{\Lxy} \,,\quad 
p^{\xz;\hs 2,\hs j}_{n+2} = 0 \quad \left( 3\leq j \leq n+2 \right),
\\[1ex]
&\ \quad p^{\xz;\hs i,\hs 1}_{n+2} = \frac{P^{i-1}_X}{L_0} \,,\quad 
p^{\xz;\hs i,\hs 2}_{n+2} = \frac{P^{i-2}_X}{\Lxy} \PXb \,,\quad
p^{\xz;\hs i,\hs j}_{n+2} = \frac{P^{i-j}_X}{\Lxy} \,,\quad 
p^{\xz;\hs i,\hs k}_{n+2} = 0 \quad \left(3 \leq j \leq i < k \leq n+2 \right),
\\
&\left\{
\begin{aligned}
p^{\xz;\hs n+2,\hs 1}_{n+2} &= -\xz Q \frac{P^{n}_X}{L_0} \,,\quad 
p^{\xz;\hs n+2,\hs 2}_{n+2} = -\xz Q \frac{P^{n-1}_X}{\Lxy} \PXb \quad \left( n\geq 1\right),\\[.5ex]
p^{\xz;\hs n+2,\hs j}_{n+2} &= -\xz Q \frac{P^{n+1-j}_X}{\Lxy} \quad \left( 3 \leq j \leq n+1 \right),\\[.5ex]
p^{\xz;\hs n+2,\hs n+2}_{n+2} &= 1 \quad \left( n\geq 1\right).
\end{aligned}
\right.
\end{align}
\es
\\
{\bf (II) The form of $\bsymb{M}$}
\bs
\begin{align}
&\bsymb{M^{\xz;\hs x,\hs y}_{1,\hs 2}}
=
\bbm
-\xz\ez \frac{Q}{L_0} & -\eta_0
\ebm
,\label{Mxz} \\[1ex]
&\bsymb{M^{\xz;\hs x,\hs y}_{2,\hs 3}}
= 
\bbm
-P_\xi \frac{Q}{L_0} & -\frac{\eta_0}{\Lxy} &0 \\[1ex]
\hline\\[-2ex]
\xz Q P_\xi \frac{Q}{L_0} & \xz Q\frac{\eta_0}{\Lxy} &-\eta_0
\ebm
,\quad
\bsymb{M^{\xz;\hs x,\hs y}_{3,\hs 4}}
= 
\bbm
-P_\xi \frac{Q}{L_0} & -\frac{\eta_0}{\Lxy} &0 &0 \\[1.5ex]
-P_X P_\xi \frac{Q}{L_0} & -\frac{M_Q}{\Lxy} &-\frac{\eta_0}{\Lxy} &0 \\[1ex]
\hline\\[-2ex]
\xz Q P_X P_\xi \frac{Q}{L_0} & \xz Q \frac{M_Q}{\Lxy} &\xz Q \frac{\eta_0}{\Lxy} &-\eta_0
\ebm
,
\end{align}
\es
\begin{align}
& \bsymb{M^{\xz;\hs x,\hs y}_{n+2,\hs n+3}} \ \left(n\geq 0\right) \nonumber \\*[2ex]
&=
\begin{bmatrix}
-P_\xi \frac{Q}{L_0} & -\frac{\eta_0}{\Lxy} &0 &0 &\cdots &0 &0 &0 \\[2ex]
-P_X P_\xi \frac{Q}{L_0} & -\frac{M_Q}{\Lxy} &-\frac{\eta_0}{\Lxy} &0 &\cdots &0 &0 &0 \\[2ex]
-P^2_X P_\xi \frac{Q}{L_0} & -P_X \frac{M_Q}{\Lxy} &-\frac{M_Q}{\Lxy} &-\frac{\eta_0}{\Lxy} &\cdots &0 &0 &0 \\[2ex]
\vdots &\vdots &\vdots &\vdots &\cdots &\vdots &\vdots &\vdots \\
\vdots &\vdots &\vdots &\vdots &\cdots & 0 &0 &0 \\[1ex]
-P^{n-1}_X P_\xi \frac{Q}{L_0} & -P^{n-2}_X \frac{M_Q}{\Lxy} &-P^{n-3}_X \frac{M_Q}{\Lxy} &-P^{n-4}_X \frac{M_Q}{\Lxy} 
&\cdots &-\frac{\eta_0}{\Lxy} &0 &0 \\[2ex]
-P^{n}_X P_\xi \frac{Q}{L_0} & -P^{n-1}_X \frac{M_Q}{\Lxy} &-P^{n-2}_X \frac{M_Q}{\Lxy} &-P^{n-3}_X \frac{M_Q}{\Lxy} 
&\cdots &-\frac{M_Q}{\Lxy} &-\frac{\eta_0}{\Lxy} &0 \\[2ex]
\hline\\[-2ex]
\xz Q P^{n}_X P_\xi \frac{Q}{L_0} & \xz Q P^{n-1}_X \frac{M_Q}{\Lxy} &\xz Q P^{n-2}_X \frac{M_Q}{\Lxy} &\xz Q P^{n-3}_X \frac{M_Q}{\Lxy} 
&\cdots &\xz Q \frac{M_Q}{\Lxy} &\xz Q \frac{\eta_0}{\Lxy} &-\eta_0 \\[2ex] 
\end{bmatrix}
,
\end{align}
with
\begin{equation}
M_Q := Q + P_X \eta_0 = Q - y\frac{X_0\eta_0}{\Lxy}\,.
\label{M_Q}
\end{equation}
The operators $P_X$ and $P_\xi$ are defined in eq.~\eqref{P_X}.
The $(i,\hs j)$-components $m^{\xz;\hs i,\hs j}_{n+2,\hs n+3}$ of $\bsymb{M^{\xz;\hs x,\hs y}_{n+2,\hs n+3}}$ $(n\geq 0)$ are given by
\bs
\begin{align}
&\left\{
\begin{aligned}
m^{\xz;\hs i,\hs 1}_{n+2,\hs n+3} &= -P^{i-1}_X P_\xi \frac{Q}{L_0} \quad \left( 1\leq i\leq n+1\right),\\[.5ex]
m^{\xz;\hs i,\hs j}_{n+2,\hs n+3} &= -P^{i-j}_X \frac{M_Q}{\Lxy} \quad \left( 2\leq j\leq i\leq n+1 \right),\\[.5ex]
m^{\xz;\hs i,\hs i+1}_{n+2,\hs n+3} &= -\frac{\eta_0}{\Lxy}\,,\quad
m^{\xz;\hs i,\hs j}_{n+2,\hs n+3} =0 \quad \left( 1\leq i\leq j-2\leq n+1 \right),
\end{aligned}
\right.\\[1ex]
&\left\{
\begin{aligned}
m^{\xz;\hs n+2,\hs j}_{n+2,\hs n+3} &= -\xz Q\, m^{\xz;\hs n+1,\hs j}_{n+2,\hs n+3} \quad \left( 1\leq j\leq n+2 \right),\\[.5ex]
m^{\xz;\hs n+2,\hs n+3}_{n+2,\hs n+3} &= -\eta_0 \quad \left( n\geq 0\right). 
\end{aligned}
\right.
\end{align}
\es
\\
{\bf (III) The form of $\bsymb{\widetilde{P}}$}
\bs
\begin{align}
&\bsymb{\widetilde{P}^{\xz;\hs x,\hs y}_{1,\hs 3}}
= 
\bbm
-\frac{y\ez}{L_0\Lxy} &\frac{1}{\Lxy}\tPb &\frac{xQ}{\Lxy}
\ebm
,\quad
\bsymb{\widetilde{P}^{\xz;\hs x,\hs y}_{2,\hs 4}}
= 
\bbm
-\frac{y\ez}{L_0\Lxy} &\frac{1}{\Lxy}\PXbstar &\frac{P_X}{\Lxy}\tPb &x\frac{P_X}{\Lxy}Q \\[2ex]
\hline \\[-2ex]
0 &0 &\frac{1}{\Lxy} &\frac{xQ}{\Lxy}
\ebm
,\\[2ex]
&\bsymb{\widetilde{P}^{\xz;\hs x,\hs y}_{3,\hs 5}}
= 
\bbm
-\frac{y\ez}{L_0\Lxy} &\frac{1}{\Lxy}\PXbstar &\frac{P_X}{\Lxy}\PXbstar &\frac{P_X^2}{\Lxy}\tPb &x\frac{P^2_X}{\Lxy}Q \\[2ex]
\hline \\[-2ex]
0 &0 &\frac{1}{\Lxy} &\frac{P_X}{\Lxy} &x\frac{P_X}{\Lxy}Q \\[2ex]
0 &0 &0 &\frac{1}{\Lxy} &\frac{xQ}{\Lxy}
\ebm
,
\end{align}
\es
\begin{align}
&\bsymb{\widetilde{P}^{\xz;\hs x,\hs y}_{n,\hs n+2}} \ \left(n\geq 1\right) \nonumber \\*[2ex]
&= 
\bbm
-\frac{y\ez}{L_0\Lxy} & \frac{1}{\Lxy} \PXbstar & \frac{P_X}{\Lxy}\PXbstar & \frac{P_X^2}{\Lxy}\PXbstar &\frac{P_X^3}{\Lxy}\PXbstar &\cdots
&\frac{P_X^{n-3}}{\Lxy}\PXbstar &\frac{P_X^{n-2}}{\Lxy}\PXbstar & \frac{P_X^{n-1}}{\Lxy}\tPb &x\frac{P_X^{n-1}}{\Lxy}Q \\[2ex]
\hline \\[-2ex]
0 &0 &\frac{1}{\Lxy} &\frac{P_X}{\Lxy} &\frac{P_X^2}{\Lxy} &\cdots 
&\frac{P_X^{n-4}}{\Lxy} &\frac{P_X^{n-3}}{\Lxy} &\frac{P_X^{n-2}}{\Lxy} &x\frac{P_X^{n-2}}{\Lxy} Q \\[2ex]
0 &0 &0 &\frac{1}{\Lxy} &\frac{P_X}{\Lxy} &\cdots 
&\frac{P_X^{n-5}}{\Lxy} &\frac{P_X^{n-4}}{\Lxy} &\frac{P_X^{n-3}}{\Lxy} &x\frac{P_X^{n-3}}{\Lxy} Q \\[2ex]
0 &0 &0 &0 &\frac{1}{\Lxy} &\cdots &\frac{P_X^{n-6}}{\Lxy} &\frac{P_X^{n-5}}{\Lxy} &\frac{P_X^{n-4}}{\Lxy} &x\frac{P_X^{n-4}}{\Lxy} Q \\[1.5ex]
\vdots &\vdots &\vdots &\vdots &\vdots &\cdots &\vdots &\vdots &\vdots &\vdots \\[1.5ex]
0 &0 &0 &0 &0 &\cdots &0 &\frac{1}{\Lxy} &\frac{P_X}{\Lxy} &x\frac{P_X}{\Lxy}Q \\[2ex]
0 &0 &0 &0 &0 &\cdots &0 &0 &\frac{1}{\Lxy} &x\frac{1}{\Lxy}Q
\ebm
.
\end{align}
Here we have denoted the BPZ conjugation by $\star$ and have defined $\tPb$ as 
\begin{equation}
\tPb := 1 + \frac{X_0 \ez\bz}{L_0}\,.
\label{tPb}
\end{equation}
For the operators $P_X$ and $\PXb$, see eq.~\eqref{P_X}. 
The $(i,\hs j)$-components $\tp^{\xz;\hs i,\hs j}_{n,\hs n+2}$ of $\bsymb{\widetilde{P}^{\xz;\hs x,\hs y}_{n,\hs n+2}}$ $(n\geq 1)$ are given by
\bs
\begin{align}
&\left\{
\begin{aligned}
\tp^{\xz;\hs 1,\hs 1}_{n,\hs n+2} &= -\frac{y\ez}{L_0\Lxy} \,,\quad 
\tp^{\xz;\hs 1,\hs j}_{n,\hs n+2} = \frac{P_X^{j-2}}{\Lxy}\PXbstar \quad \left( 2\leq j\leq n\right),\\[.5ex]
\tp^{\xz;\hs 1,\hs n+1}_{n,\hs n+2} &= \frac{P_X^{n-1}}{\Lxy}\tPb \,,\quad 
\tp^{\xz;\hs 1,\hs n+2}_{n,\hs n+2} = x\frac{P_X^{n-1}}{\Lxy}Q \quad \left( n\geq 1\right),
\end{aligned}
\right.
\\[1ex]
&\left\{
\begin{aligned}
\tp^{\xz;\hs i,\hs j}_{n,\hs n+2} &= 0 \quad \left( 1\leq j\leq i,\ 2\leq i\leq n \right), \\[.5ex]
\tp^{\xz;\hs i,\hs j}_{n,\hs n+2} &= \frac{P^{j-i-1}_X}{\Lxy} \,,\quad
\tp^{\xz;\hs i,\hs n+2}_{n,\hs n+2} = x\frac{P^{n-i}_X}{\Lxy}Q \quad \left( 2\leq i < j\leq n+1\right).
\end{aligned}
\right. 
\end{align}
\es
\\
{\bf (IV) The form of $\bsymb{\tM}$}
\bs
\begin{align}
&\bsymb{\tM^{\xz;\hs x,\hs y}_{1,\hs 2}}
=
\bbm
\frac{yQ\ez}{L_0\Lxy} &\frac{xQ\ez}{L_0\Lxy}
\ebm
,\\[1ex]
&\bsymb{\tM^{\xz;\hs x,\hs y}_{2,\hs 3}}
= 
\bbm
\frac{yQ\ez}{L_0\Lxy} &-\frac{Q}{\Lxy}\tPb &0 \\[2ex]
\hline \\[-2ex]
0 &-\frac{\ez}{\Lxy} &\frac{xQ\ez}{\Lxy}
\ebm
,\quad
\bsymb{\tM^{\xz;\hs x,\hs y}_{3,\hs 4}}
= 
\bbm
\frac{yQ\ez}{L_0\Lxy} &-\frac{Q}{\Lxy}\PXbstar &-\frac{Q P_X}{\Lxy}\tPb &0 \\[2ex]
\hline \\[-2ex]
0 &-\frac{\ez}{\Lxy} &-\frac{1}{\Lxy}M_Q &x\frac{P_X}{\Lxy}Q\ez \\[2ex]
0 &0 &-\frac{\ez}{\Lxy} &\frac{xQ\ez}{\Lxy}
\ebm
,
\end{align}
\es
\begin{align}
&\bsymb{\tM^{\xz;\hs x,\hs y}_{n,\hs n+1}}\ \left(n\geq 2\right) \nonumber \\*[2ex]
&=
\bbm
\frac{yQ\ez}{L_0\Lxy} &-\frac{Q}{\Lxy}\PXbstar &-\frac{Q P_X}{\Lxy}\PXbstar &-\frac{Q P_X^2}{\Lxy}\PXbstar &\cdots
&-\frac{Q P_X^{n-3}}{\Lxy}\PXbstar &-\frac{Q P_X^{n-2}}{\Lxy}\tPb &0\\[2ex]
\hline \\[-2ex]
0 &-\frac{\ez}{\Lxy} &-\frac{1}{\Lxy}M_Q & -\frac{P_X}{\Lxy}M_Q &\cdots 
& -\frac{P_X^{n-4}}{\Lxy}M_Q &-\frac{P_X^{n-3}}{\Lxy}M_Q &x\frac{P_X^{n-2}}{\Lxy} Q\ez \\[2ex]
0 &0 &-\frac{\ez}{\Lxy} &-\frac{1}{\Lxy}M_Q &\cdots & -\frac{P_X^{n-5}}{\Lxy}M_Q &-\frac{P_X^{n-4}}{\Lxy}M_Q &x\frac{P_X^{n-3}}{\Lxy} Q\ez \\[1.5ex]
\vdots &\vdots &\vdots &\vdots &\cdots &\vdots &\vdots &\vdots \\[1.5ex]
0 &0 &0 &0 &\cdots &-\frac{\ez}{\Lxy} &-\frac{1}{\Lxy}M_Q &x\frac{P_X}{\Lxy}Q\ez \\[2ex]
0 &0 &0 &0 &\cdots &0 &-\frac{\ez}{\Lxy} &x\frac{1}{\Lxy}Q\ez
\ebm
,
\end{align}
where we have used the operators in eqs.~\eqref{P_X}, \eqref{M_Q}, and \eqref{tPb}. 
The $(i,\hs j)$-components $\tm^{\xz;\hs i,\hs j}_{n,\hs n+1}$ of $\bsymb{\tM^{\xz;\hs x,\hs y}_{n,\hs n+1}}$ $(n\geq 2)$ are given by
\bs
\begin{align}
&\left\{
\begin{aligned}
\tm^{\xz;\hs 1,\hs 1}_{n,\hs n+1} &= \frac{yQ\ez}{L_0\Lxy} \quad \left( n\geq 2\right),\\[.5ex]
\tm^{\xz;\hs 1,\hs j}_{n,\hs n+1} &= -\frac{Q P_X^{j-2}}{\Lxy} \PXbstar \quad \left( 2\leq j\leq n-1 \right), \\[.5ex]
\tm^{\xz;\hs 1,\hs n}_{n,\hs n+1} &= -\frac{Q P_X^{n-2}}{\Lxy} \tPb \,,\quad
\tm^{\xz;\hs 1,\hs n+1}_{n,\hs n+1} = 0 \quad \left( n\geq 2 \right),
\end{aligned}
\right. \\[2ex]
&\left\{
\begin{aligned}
\tm^{\xz;\hs i,\hs j}_{n,\hs n+1} &= 0 \quad \left( 1\leq j < i \leq n\right), \\[1ex]
\tm^{\xz;\hs i,\hs i}_{n,\hs n+1} &= -\frac{\ez}{\Lxy} \quad \left( 2\leq i\leq n\right), \\[.5ex]
\tm^{\xz;\hs i,\hs j}_{n,\hs n+1} &= -\frac{P_X^{j-i-1}}{\Lxy} M_Q \quad \left( 2\leq i < j \leq n\right), \\[.5ex]
\tm^{\xz;\hs i,\hs n+1}_{n,\hs n+1} &= x\frac{P_X^{n-i}}{\Lxy}Q\ez \quad \left( 2\leq i\leq n \right).
\end{aligned}
\right.
\end{align}
\es

\subsection{Solutions for the \texorpdfstring{$\zz =\dz$}{zeta0 = d0} case}
{\bf (I) The form of $\bsymb{P}$}\\*
\indenths
Small-size matrices are given by
\bs
\begin{align}
&\bsymb{P^{\dz;\hs x,\hs y}_{1,\hs 1}} = \frac{\iu}{L_0^2}\,,\quad
\bsymb{P^{\dz;\hs x,\hs y}_{2,\hs 2}} =
\bbm
\frac{1}{L_0} & yP_b \\[1ex]
xP_d & \frac{1}{L_0}
\ebm
,\\[2ex]
&\bsymb{P^{\dz;\hs x,\hs y}_{3,\hs 3}} =
\bbm
\frac{1}{L_0} & P_b & 0 \\[1ex]
0 & \frac{1}{\Lxy} & 0 \\[1ex]
0 & P_d & \frac{1}{L_0}
\ebm
,\quad
\bsymb{P^{\dz;\hs x,\hs y}_{4,\hs 4}} =
\bbm
\frac{1}{L_0} & P_b & 0 &0 \\[1ex]
0 & \frac{1}{\Lxy} & 0 &0 \\[1ex]
0 &0 &\frac{1}{\Lxy} &0 \\[1ex]
0 &0 & P_d & \frac{1}{L_0}
\ebm
,
\end{align}
\es
with
\begin{equation}
P_b:= \frac{\eta_0 \bz}{L_0\Lxy}\,,\quad P_d := \frac{Q\dz}{L_0\Lxy}\,.
\end{equation}
For an arbitrary $n \left(\geq 1\right)$, we have
\begin{equation}
\bsymb{P^{\dz;\hs x,\hs y}_{n+2,\hs n+2}} =
\bbm
\frac{1}{L_0} & P_b &&&&& \hsymbu{0} \\[1ex]
& \frac{1}{\Lxy} &&&&& \\[1ex]
&&\frac{1}{\Lxy} &&&& \\[1ex]
&&& \ddots &&& \\[1ex]
&&&& \frac{1}{\Lxy} && \\[1ex]
&&&&& \frac{1}{\Lxy} & \\[1.5ex]
\hsymbl{0} &&&&& P_d & \frac{1}{L_0}
\ebm
\quad \left(n\geq 1\right).
\label{P^d}
\end{equation}
\\
{\bf (II) The form of $\bsymb{M}$}
\begin{align}
&\bsymb{M^{\dz;\hs x,\hs y}_{1,\hs 2}}
=
\bbm
M_d & M_b
\ebm
,\quad
\bsymb{M^{\dz;\hs x,\hs y}_{2,\hs 3}} = 
\bbm
M_d & -\frac{\eta_0}{\Lxy} & 0 \\[1.5ex]
0 & -\frac{Q}{\Lxy} & M_b
\ebm
,\label{Mdz} \\[2.5ex]
&\bsymb{M^{\dz;\hs x,\hs y}_{n+2,\hs n+3}} =
\bbm
M_d & -\frac{\eta_0}{\Lxy} &&&\hsymbu{0} \\[1.5ex]
& -\frac{Q}{\Lxy} & \ddots &&\\[1.5ex]
&& \ddots & -\frac{\eta_0}{\Lxy} &\\[1.5ex]
\hsymbl{0}&&& -\frac{Q}{\Lxy} & M_b
\ebm
\quad \left(n\geq 0\right),
\label{M^d}
\end{align}
with
\begin{equation}
M_d := -\left( 1-y\frac{\eta_0\dz}{\Lxy} \right) \frac{Q}{L_0}\,,\quad
M_b := -\left( 1- x\frac{Q\bz}{\Lxy} \right) \frac{\eta_0}{L_0}\,.
\label{MbMd}
\end{equation}
\\
{\bf (III) The form of $\bsymb{\widetilde{P}}$}
\begin{align}
&\bsymb{\widetilde{P}^{\dz;\hs x,\hs y}_{1,\hs 3}}
=
\bbm
-\frac{y\ez}{L_0\Lxy} &\frac{1}{\Lxy} &\frac{xQ}{L_0\Lxy}
\ebm
,\quad
\bsymb{\widetilde{P}^{\dz;\hs x,\hs y}_{2,\hs 4}}
=
\bbm
-\frac{y\ez}{L_0\Lxy} & \frac{1}{\Lxy} &0 &0 \\[1ex]
0 &0 &\frac{1}{\Lxy} &\frac{xQ}{L_0\Lxy}
\ebm
,\\[2ex]
&\bsymb{\widetilde{P}^{\dz;\hs x,\hs y}_{n,\hs n+2}}
=
\bbm
-\frac{y\ez}{L_0\Lxy} & \frac{1}{\Lxy} &&&&&\hsymbu{0}\\[1ex]
&&\frac{1}{\Lxy} &&&&\\[1ex]
&&&\ddots &&&\\[1ex]
&&&&\frac{1}{\Lxy} &&\\[1ex]
\hsymbl{0} &&&&&\frac{1}{\Lxy} &\frac{xQ}{L_0\Lxy}
\ebm
\quad \left( n\geq 1\right).
\end{align}
\\
{\bf (IV) The form of $\bsymb{\tM}$}
\begin{align}
&\bsymb{\tM^{\dz;\hs x,\hs y}_{1,\hs 2}}
=
\bbm
\frac{yQ\ez}{L_0\Lxy} &\frac{xQ\ez}{L_0\Lxy} 
\ebm
,\quad
\bsymb{\tM^{\dz;\hs x,\hs y}_{2,\hs 3}}
=
\bbm
\frac{yQ\ez}{L_0\Lxy} &-\frac{Q}{\Lxy} &0 \\[1.5ex]
0 &-\frac{\ez}{\Lxy} &\frac{xQ\ez}{L_0\Lxy} 
\ebm
,\\[2ex]
&\bsymb{\tM^{\dz;\hs x,\hs y}_{n,\hs n+1}}
=
\bbm
\frac{yQ\ez}{L_0\Lxy} &-\frac{Q}{\Lxy} &&&&\hsymbu{0} \\[1.5ex]
&-\frac{\ez}{\Lxy} &-\frac{Q}{\Lxy} &&& \\[1ex]
&&-\frac{\ez}{\Lxy} &\ddots && \\[1ex]
&&&\ddots &-\frac{Q}{\Lxy} & \\[1.5ex]
\hsymbl{0} &&&&-\frac{\ez}{\Lxy} &\frac{xQ\ez}{L_0\Lxy} 
\ebm
\quad \left( n\geq 1\right).
\end{align}

\section{Concrete forms of \texorpdfstring{$\bsymb{K}$ and $\bsymb{\tK}$}{K and tilded K}}
\label{concrete forms}
\setcounter{equation}{0}
\indenths
After listing some small-size matrices, we give $\bsymb{K}$ and $\bsymb{\tK}$ for general $n$.
Moreover, we specify the components of the matrices when $\zz$ is equal to $\xz$.

\subsection{\texorpdfstring{$\bsymb{K}$ and $\bsymb{\tK}$ for the $\zz =\xz$}{K and tilded K for the zeta0 = xi0} case}
{\bf (I) The form of $\bsymb{K}$}
\bs
\begin{align}
&\bsymb{K^{\xz;\hs x,\hs y}_{1,\hs 0}}
= \iu\,\frac{\bz\xz}{L_0}\,,
\qquad
\bsymb{K^{\xz;\hs x,\hs y}_{2,\hs 1}}
=
\bbm
\frac{P_\xi \bz}{L_0} \\[2ex]
-\xz Q\frac{P_\xi \bz}{L_0} +\xz
\ebm
,\label{small K^xi} \\[1ex]
&\bsymb{K^{\xz;\hs x,\hs y}_{3,\hs 2}}
=
\bbm
\frac{P_\xi \bz}{L_0} &0 \\[2ex]
\frac{P_X P_\xi \bz}{L_0} + \frac{y\xz}{\Lxy} &\frac{x\bz}{\Lxy} \\[2ex]
\hline \\[-2ex]
- \xz Q \KXb &\xz\bigl( 1-\frac{xQ\bz}{\Lxy}\bigr)
\ebm
,\quad
\bsymb{K^{\xz;\hs x,\hs y}_{4,\hs 3}} 
=
\bbm
\frac{P_\xi \bz}{L_0} &0 &0 \\[2ex]
\frac{P_X P_\xi \bz}{L_0} + \frac{y\xz}{\Lxy} &\frac{x\bz}{\Lxy} &0 \\[2ex]
\hline \\[-2ex]
P_X \KXb &\frac{1}{\Lxy} K_\xi &\frac{x\bz}{\Lxy} \\[2ex]
\hline \\[-2ex]
-\xz Q P_X \KXb &- \frac{\xz Q}{\Lxy} K_\xi &\xz\bigl( 1-\frac{xQ\bz}{\Lxy}\bigr)
\ebm
,\\[1ex]
&\bsymb{K^{\xz;\hs x,\hs y}_{5,\hs 4}}
=
\bbm
\frac{P_\xi \bz}{L_0} &0 &0 &0 \\[2ex]
\frac{P_X P_\xi \bz}{L_0} + \frac{y\xz}{\Lxy} &\frac{x\bz}{\Lxy} &0 &0 \\[2ex]
\hline \\[-2ex]
P_X \KXb &\frac{1}{\Lxy} K_\xi &\frac{x\bz}{\Lxy} &0 \\[2ex]
P_X^2 \KXb &\frac{P_X}{\Lxy} K_\xi &\frac{1}{\Lxy} K_\xi &\frac{x\bz}{\Lxy} \\[2ex]
\hline \\[-2ex]
-\xz Q P_X^2 \KXb &-\xz Q \frac{P_X}{\Lxy} K_\xi &-\frac{\xz Q}{\Lxy} K_\xi &\xz\bigl( 1-\frac{xQ\bz}{\Lxy}\bigr)
\ebm
,
\end{align}
\es
\vspace{0ex}
\begin{align}
&\bsymb{K^{\xz;\hs x,\hs y}_{n+1,\hs n}} \ \left( n\geq 3\right) \nonumber \\*[2ex]
&=
\bbm
\frac{P_\xi \bz}{L_0} &0 &0 &0 &\cdots &0 &0\\[2ex]
\frac{P_X P_\xi \bz}{L_0} + \frac{y\xz}{\Lxy} &\frac{x\bz}{\Lxy} &0 &0 &\cdots &0 &0\\[2ex]
\hline \\[-2ex]
P_X \KXb &\frac{1}{\Lxy} K_\xi &\frac{x\bz}{\Lxy} &0 &\cdots &0 &0\\[2ex]
P_X^2 \KXb &\frac{P_X}{\Lxy} K_\xi &\frac{1}{\Lxy} K_\xi &\frac{x\bz}{\Lxy} &\cdots &0 &0\\[2ex]
P_X^3 \KXb &\frac{P_X^2}{\Lxy} K_\xi &\frac{P_X}{\Lxy} K_\xi &\frac{1}{\Lxy} K_\xi &\cdots &0 &0\\[1.5ex]
\vdots &\vdots &\vdots &\vdots &\cdots &\vdots &\vdots \\[1.5ex]
P_X^{n-3} \KXb &\frac{P_X^{n-4}}{\Lxy} K_\xi &\frac{P_X^{n-5}}{\Lxy} K_\xi &\frac{P_X^{n-6}}{\Lxy} K_\xi &\cdots &\frac{x\bz}{\Lxy} &0\\[2ex]
P_X^{n-2} \KXb &\frac{P_X^{n-3}}{\Lxy} K_\xi &\frac{P_X^{n-4}}{\Lxy} K_\xi &\frac{P_X^{n-5}}{\Lxy} K_\xi &\cdots 
&\frac{1}{\Lxy} K_\xi &\frac{x\bz}{\Lxy}\\[2ex]
\hline \\[-2ex]
-\xz Q P_X^{n-2} \KXb &-\xz Q \frac{P_X^{n-3}}{\Lxy} K_\xi &-\xz Q \frac{P_X^{n-4}}{\Lxy} K_\xi &-\xz Q \frac{P_X^{n-5}}{\Lxy} K_\xi &\cdots 
&- \frac{\xz Q}{\Lxy} K_\xi &\xz\bigl( 1-\frac{xQ\bz}{\Lxy}\bigr)
\ebm
,
\end{align}
with
\begin{align}
\KXb := \frac{P_X \bz}{L_0} + \frac{y}{\Lxy} \PXb\xz \,,\quad
K_\xi := xP_X \bz +y\xz\,.
\label{KXb}
\end{align}
The operators $P_X$, $P_\xi$, and $\PXb$ are defined in eq.~\eqref{P_X}.
The $(i,\hs j)$-components $k^{\xz;\hs i,\hs j}_{n+1,\hs n}$ of $\bsymb{K^{\xz;\hs x,\hs y}_{n+1,\hs n}}$ $\left( n\geq 3\right)$ are given by
\bs
\begin{align}
&\ \quad k^{\xz;\hs 1,\hs 1}_{n+1,\hs n} = \frac{P_\xi \bz}{L_0} \,,\quad
k^{\xz;\hs 1,\hs j}_{n+1,\hs n} = 0\quad \left( 2\leq j\leq n\right),\\[1ex]
&\ \quad k^{\xz;\hs 2,\hs 1}_{n+1,\hs n} = \frac{P_X P_\xi \bz}{L_0} + \frac{y\xz}{\Lxy} \,,\quad
k^{\xz;\hs 2,\hs 2}_{n+1,\hs n} = \frac{x\bz}{\Lxy}\,,\quad
k^{\xz;\hs 2,\hs j}_{n+1,\hs n} =0 \quad \left( 3\leq j\leq n\right),\\[1ex]
&\left\{
\begin{aligned}
k^{\xz;\hs i,\hs 1}_{n+1,\hs n} &= P_X^{i-2} \KXb \,,\quad
k^{\xz;\hs i,\hs j}_{n+1,\hs n} = \frac{P_X^{i-j-1}}{\Lxy} K_\xi \,,\quad
k^{\xz;\hs i,\hs i}_{n+1,\hs n} = \frac{x\bz}{\Lxy} \quad \left( 2\leq j<i\leq n\right),\\[.5ex]
k^{\xz;\hs i,\hs j}_{n+1,\hs n} &= 0 \quad \left( 3\leq i<j\leq n\right),
\end{aligned}
\right.\\[1ex]
&\left\{
\begin{aligned}
k^{\xz;\hs n+1,\hs 1}_{n+1,\hs n} &= -\xz Q P_X^{n-2} \KXb \,,\quad
k^{\xz;\hs n+1,\hs j}_{n+1,\hs n} = -\xz Q \frac{P_X^{n-j-1}}{\Lxy} K_\xi \\[.5ex]
k^{\xz;\hs n+1,\hs n}_{n+1,\hs n} &= \xz\Bigl( 1-\frac{xQ\bz}{\Lxy}\Bigr) 
\end{aligned}
\right.
\quad \left( 2\leq j\leq n-1\right).
\end{align}
\es
\\
{\bf (II) The form of $\bsymb{\tK}$ }
\bs
\begin{align}
&\bsymb{\tK^{\xz;\hs x,\hs y}_{0,\hs 1}} := \bpz{\bsymb{K^{\xz;\hs x,\hs y}_{1,\hs 0}}} = \iu\,\frac{\bz\xz}{L_0}\,,\quad
\bsymb{\tK^{\xz;\hs x,\hs y}_{1,\hs 2}}
=
\bbm
\frac{\Pxistar \bz}{L_0} &\frac{1}{\Lxy}\bigl( y\tPb +xQ\bz\bigr) \xz
\ebm
,\\[2ex]
&\bsymb{\tK^{\xz;\hs x,\hs y}_{2,\hs 3}}
=
\bbm
\frac{\Pxistar\bz}{L_0} &\frac{1}{\Lxy}\tKXb &\frac{P_X}{\Lxy}\bigl(y\tPb +xQ\bz\bigr)\xz \\[2ex]
\hline \\[-2ex]
0 &\frac{x\bz}{\Lxy} &\frac{1}{\Lxy}\bigl( y+xQ\bz\bigr)\xz
\ebm
,\\[3ex]
&\bsymb{\tK^{\xz;\hs x,\hs y}_{3,\hs 4}}
=
\bbm
\frac{\Pxistar\bz}{L_0} &\frac{1}{\Lxy}\tKXb &\frac{P_X}{\Lxy}\tKXb &\frac{P_X^2}{\Lxy}\bigl(y\tPb +xQ\bz\bigr)\xz \\[2ex]
\hline \\[-2ex]
0 &\frac{x\bz}{\Lxy} &\frac{1}{\Lxy}K_\xi &\frac{P_X}{\Lxy}\bigl( y+xQ\bz\bigr)\xz \\[2ex]
0 &0 &\frac{x\bz}{\Lxy} &\frac{1}{\Lxy}\bigl( y+xQ\bz\bigr)\xz
\ebm
,
\end{align}
\es
\vspace{0ex}
\begin{align}
&\bsymb{\tK^{\xz;\hs x,\hs y}_{n,\hs n+1}}\ \left( n\geq 1\right) \nonumber \\*[2ex]
&=
\bbm
\frac{\Pxistar\bz}{L_0} &\frac{1}{\Lxy}\tKXb &\frac{P_X}{\Lxy}\tKXb &\frac{P_X^2}{\Lxy}\tKXb &\cdots
&\frac{P_X^{n-3}}{\Lxy}\tKXb &\frac{P_X^{n-2}}{\Lxy}\tKXb &\frac{P_X^{n-1}}{\Lxy}\bigl(y\tPb +xQ\bz\bigr)\xz \\[2ex]
\hline \\[-2ex]
0 &\frac{x\bz}{\Lxy} &\frac{1}{\Lxy}K_\xi &\frac{P_X}{\Lxy}K_\xi &\cdots 
&\frac{P_X^{n-4}}{\Lxy}K_\xi &\frac{P_X^{n-3}}{\Lxy}K_\xi &\frac{P_X^{n-2}}{\Lxy}\bigl( y+xQ\bz\bigr)\xz \\[2ex]
0 &0 &\frac{x\bz}{\Lxy} &\frac{1}{\Lxy}K_\xi &\cdots 
&\frac{P_X^{n-5}}{\Lxy}K_\xi &\frac{P_X^{n-4}}{\Lxy}K_\xi &\frac{P_X^{n-3}}{\Lxy}\bigl( y+xQ\bz\bigr)\xz \\[1.5ex]
\vdots &\vdots &\vdots &\vdots &\cdots &\vdots &\vdots &\vdots \\[1.5ex]
0 &0 &0 &0 &\cdots &\frac{x\bz}{\Lxy} &\frac{1}{\Lxy}K_\xi &\frac{P_X}{\Lxy}\bigl( y+xQ\bz\bigr)\xz \\[2ex]
0 &0 &0 &0 &\cdots &0 &\frac{x\bz}{\Lxy} &\frac{1}{\Lxy}\bigl( y+xQ\bz\bigr)\xz
\ebm
,
\end{align}
where we have denoted the BPZ conjugation by $\star$ and have defined $\tKXb$ as
\begin{equation}
\tKXb := xP_X \bz + y\PXbstar \xz\,.
\end{equation}
See also the definitions \eqref{P_X}, \eqref{tPb}, and \eqref{KXb}.
The $(i,\hs j)$-components $\tk^{\xz;\hs i,\hs j}_{n,\hs n+1}$ of $\bsymb{\tK^{\xz;\hs x,\hs y}_{n,\hs n+1}}$ $\left( n\geq 1\right)$ are given by
\bs
\begin{align}
&\left\{
\begin{aligned}
\tk^{\xz;\hs 1,\hs 1}_{n,\hs n+1} &= \frac{\Pxistar \bz}{L_0}\,,\quad
\tk^{\xz;\hs 1,\hs j}_{n,\hs n+1} = \frac{P_X^{j-2}}{\Lxy}\tKXb \quad \left( 2\leq j\leq n\right),\\[.5ex]
\tk^{\xz;\hs 1,\hs n+1}_{n,\hs n+1} &= \frac{P_X^{n-1}}{\Lxy}\bigl( y\tPb +xQ\bz\bigr) \xz \quad \left( n\geq 1\right),
\end{aligned}
\right.\\[1ex]
&\left\{
\begin{aligned}
\tk^{\xz;\hs i,\hs j}_{n,\hs n+1} &= 0\quad \left( 1\leq j <i\leq n\right),\\[.5ex]
\tk^{\xz;\hs i,\hs i}_{n,\hs n+1} &= \frac{x\bz}{\Lxy}\,,\quad
\tk^{\xz;\hs i,\hs j}_{n,\hs n+1} = \frac{P_X^{j-i-1}}{\Lxy} K_\xi \quad \left( 2\leq i<j\leq n\right),\\[.5ex]
\tk^{\xz;\hs i,\hs n+1}_{n,\hs n+1} &= \frac{P_X^{n-i}}{\Lxy} \bigl( y+xQ\bz \bigr) \xz \quad \left( 2\leq i\leq n\right).
\end{aligned}
\right.
\end{align}
\es

\subsection{\texorpdfstring{$\bsymb{K}$ and $\bsymb{\tK}$ for the $\zz =\dz$}{K and tilded K for the zeta0 = d0} case}
{\bf (I) The form of $\bsymb{K}$}
\begin{align}
&\bsymb{K^{\dz;\hs x,\hs y}_{1,\hs 0}}
= \iu\,\frac{\bz\dz}{L_0^2}\,,
\quad
\bsymb{K^{\dz;\hs x,\hs y}_{2,\hs 1}}
=
\bbm
K_b \\[1ex]
K_d
\ebm
,\quad
\bsymb{K^{\dz;\hs x,\hs y}_{3,\hs 2}}
=
\bbm
K_b &0 \\[1ex]
\frac{y\dz}{\Lxy} &\frac{x\bz}{\Lxy} \\[1ex]
0 &K_d
\ebm
,\label{small K^d} \\[3ex]
&\bsymb{K^{\dz;\hs x,\hs y}_{n+1,\hs n}}
=
\bbm
K_b &&&&\hsymbu{0} \\[1.5ex]
\frac{y\dz}{\Lxy} &\frac{x\bz}{\Lxy} &&&\\[1.5ex]
&\frac{y\dz}{\Lxy} &\frac{x\bz}{\Lxy} &&\\[1.5ex]
&&\ddots &\ddots &\\[1.5ex]
&&&\frac{y\dz}{\Lxy} &\frac{x\bz}{\Lxy}\\[1.5ex]
\hsymbl{0} &&&&K_d
\ebm
\quad \left(n\geq 1\right),
\end{align}
with
\begin{equation}
K_b := \frac{\bz}{L_0} \Bigl( 1-y\frac{\ez\dz}{\Lxy} \Bigr),\quad
K_d := \frac{\dz}{L_0} \Bigl( 1-x\frac{Q\bz}{\Lxy} \Bigr).
\end{equation}
\\
{\bf (II) The form of $\bsymb{\tK}$ }
\begin{align}
&\bsymb{\tK^{\dz;\hs x,\hs y}_{0,\hs 1}} := \bpz{\bsymb{K^{\dz;\hs x,\hs y}_{1,\hs 0}}} = \iu\,\frac{\bz\dz}{L_0^2}\,,\quad
\bsymb{\tK^{\dz;\hs x,\hs y}_{1,\hs 2}}
=
\bbm
\tK_b &\tK_d
\ebm
,\quad
\bsymb{\tK^{\dz;\hs x,\hs y}_{2,\hs 3}}
=
\bbm
\tK_b &\frac{y\dz}{\Lxy} &0 \\[1.5ex]
0 &\frac{x\bz}{\Lxy} &\tK_d
\ebm
,\\[-2ex]
&\bsymb{\tK^{\dz;\hs x,\hs y}_{n,\hs n+1}}
=
\bbm
\tK_b &\frac{y\dz}{\Lxy} &&&&\hsymbu{0}\\[1.5ex]
&\frac{x\bz}{\Lxy} &\frac{y\dz}{\Lxy} &&& \\[1.5ex]
&&\frac{x\bz}{\Lxy} &\ddots && \\[1.5ex]
&&&\ddots &\frac{y\dz}{\Lxy} & \\[1.5ex]
\hsymbl{0} &&&&\frac{x\bz}{\Lxy} &\tK_d
\ebm
\quad \left( n\geq 1\right),
\end{align}
with
\begin{equation}
\tK_b := \Bigl( x+y\frac{\ez\dz}{L_0} \Bigr)\frac{\bz}{\Lxy}\,,\quad
\tK_d := \Bigl( y+x\frac{Q\bz}{L_0} \Bigr)\frac{\dz}{\Lxy}\,.
\end{equation}


\end{document}